\documentstyle[12pt,twoside]{report}
\def\eqalign#1{\null\vcenter{\def\\{\cr}\openup\jot
  \ialign{\strut$\displaystyle{##}$\hfil&$\displaystyle{{}##}$\hfil
      \crcr#1\crcr}}\,}
\newcounter{eqnval} 
\def\eqlabel#1{\refstepcounter{equation}\label{#1}%
     \addtocounter{equation}{-1}}%
\def\a{\alpha}      \def\b{\beta}     \def\c{\gamma}      \def\d{\delta}
\def\e{\epsilon}        \def\l{\lambda}     \def\m{\mu}
\def\n{\nu}         \def\o{\omega}             
\def\s{\sigma}      \def\t{\tau}      
\def\vare{\varepsilon} 

\def\adag{a^\dagger}   \def\sbar{\bar{s}}
\def\bdag{b^\dagger}   \def\ubar{\bar{u}}
\def\cdag{c^\dagger}   \def\vbar{\bar{v}}
\def\ddag{d^\dagger}   \def\wbar{\bar{w}}

\def\hats{\mbox{$\hat{\s}$}} 
\def\hatone{\mbox{$\hat{1}$}} 
\def\hatl{\mbox{$\hat{\l}$}} 
\def\calu{\mbox{$\cal U$}}

\newcommand{\sssty}{\scriptscriptstyle}
\def\fff#1{\mbox{\boldmath$#1$}}

\def\beq{\begin{equation}}            \def\eeq{\end{equation}}
\def\baq{\begin{eqnarray}}            \def\eaq{\end{eqnarray}}  
\def\baqn{\begin{eqnarray*}}          \def\eaqn{\end{eqnarray*}}

\def\mm{$M$} 
\def\pfm{\mbox{$F(M)$}} 
\def\pgm{\mbox{$G(M)$}} 
\def\phm{\mbox{$H(M)$}} 
\def\pkm{\mbox{$K(M)$}} 
\def\pvm{\mbox{$V(M)$}} 
\def\psm{\mbox{$S(M)$}} 

\def\real{\mbox{$\mbox{I}\!\mbox{R}$}}
\def\complex{\mbox{$\,\,\mbox{\rule{0.06em}{1.48ex}}\!\!\mbox{C}$}}
\def\One{\mbox{$1\!\!\!\;\mbox{l}$}}
\def\Ones{\leavevmode\hbox{\footnotesize1\kern-3.9pt\small1}} 
\def\square{\mbox{\framebox[2.5mm]{}}}

\def\glfourr{\mbox{$\mbox{GL}(4,\real)$}}
\def\glfourc{\mbox{$\mbox{GL}(4,\complex)$}}
\def\sltwoc{\mbox{$\mbox{SL}(2,\complex)$}}
\def\glonec{\mbox{$\mbox{GL}(1,\complex)$}}
\def\uone{\mbox{$\mbox{U}(1)$}}
\def\cspin{\mbox{$\complex\mbox{Spin}$}}

\def\lplus{\mbox{$L^+_\uparrow$}} 
\def\cl{\mbox{$\complex L$}}
\def\clplus{\mbox{$\complex L^+$}}

\def\ctm{\mbox{$\complex\otimes TM$}}
\def\lplusm{\mbox{$L^+_\uparrow(M)$}}
\def\clplusm{\mbox{$\complex L^+(M)$}}
\def\cspinm{\mbox{$\complex\mbox{Spin}(M)$}}
\def\spinm{\mbox{$\mbox{Spin}(M)$}} 

\def\ctimes{\mbox{$\complex^\times$}} 
\def\cxm{\mbox{$\complex^\times(M)$}}
\def\liesltwoc{\mbox{$\fff{sl}(2,\complex)$}}
 
\def\lieglfourc{\mbox{$\fff{gl}(4,\complex)$}} 

\def\Con#1#2#3{\Gamma^#1{}_{\! #2 #3}}
\def\con#1#2#3{\Gamma_{#1 #2 #3}}
\def\Chr#1#2#3{\big\{{}^#1{}_{#2 #3}\big\}}
\def\chr#1#2#3{\big\{{}_{#1 #2 #3}\big\}}

\def\PSI{\overline{\psi}}
\def\CHI{\overline{\chi}} 
\def\planck{l_{\sssty 0}}

\begin{document}

\thispagestyle{empty} 

\begin{center} 
\mbox{}\\[2cm]
{\Large Geometric Interpretation of Electromagnetism}\\[3mm] 
{\Large in a Gravitational Theory with Torsion and Spinorial Matter}\\[6cm] 
{\large Dissertation}\\[1cm] 
zur Erlangung des Grades\\ 
''Doktor der Naturwissenschaften''\\
am Fachbereich Physik\\ 
der Johannes Gutenberg--Universit\"at in Mainz\\[2cm] 
Kenichi Horie\\ 
geboren in Baltimore\\[1cm] 
Mainz 1995 
\end{center} 
\newpage 
\mbox{}\\[6cm]
1.\ Berichterstatter: Prof.\ Dr.\ M.\ Kretzschmar\\
2.\ Berichterstatter: Prof.\ Dr.\ N.\ A.\ Papadopoulos\\
3.\ Berichterstatter: Prof.\ Dr.\ G.\ Mack\\[1cm]
M\"undliche Pr\"ufung: 25.\ August 1995.

\newpage

\pagenumbering{roman} 
\tableofcontents

                         \chapter{Introduction}

\pagenumbering{arabic}

Einstein's general relativistic theory of gravitation (see e.g.\ 
\cite{mtw} and 
references therein) is based on a semi-Riemannian geometry. This spacetime 
geometry is characterized by a pseudo-Riemannian metric $g_{\mu\nu}$ and a 
linear connection $\Con\a\m\b$, which is compatible with the metric,
\beq\label{metric}
  \nabla_{\!\m}g_{\a\b} = \partial_\m g_{\a\b}-\Con\c\m\a g_{\c\b}
                                              -\Con\c\m\b g_{\a\c} = 0\;,
\eeq
and has vanishing torsion
\beq\label{zerot}
  T^\a{}_{\!\m\b} := \Con\a\m\b-\Con\a\b\m = 0\;.
\eeq
These two requirements uniquely determine a special connection, the 
Levi--Civita connection
\beq\label{lc}
  \Con\a\m\b = \Chr\a\m\b := 
  \frac{1}{2}g^{\a\e}\left( \partial_\m g_{\e\b}
                           +\partial_\b g_{\e\m}
                           -\partial_\e g_{\m\b} \right)\;.
\eeq
Within this semi-Riemannian geometry, the mass-energy of matter influences 
the spacetime via the Einstein's field equation
\beq\label{einstein}
  \frac{1}{k}G^\ast_{\a\b} = T^m_{\a\b}\;,
\eeq
where $G^\ast_{\a\b}$ is the Einstein--tensor, $k=8\pi G/c^4$, and 
$T^m_{\a\b}$ is the energy-momentum tensor of matter. Since $G^\ast_{\a\b}$ 
is a tensor built from the Riemann curvature, the Einstein equation describes 
how the mass-energy of matter curves the spacetime. As far as macroscopic 
bulk matter is considered, the physical property of the matter is 
sufficiently characterized by this 
energy-momentum equation. However, on the microscopic level, the elementary 
particles are described by quantum mechanics and are not only characterized 
by mass, but also by spin. Therefore, to 
consider gravitational phenomena also on the microscopic level and to make 
general relativity more compatible with quantum mechanics, it seems 
necessary to take into account the influence of spin on the 
geometry of spacetime.

This aim is achieved in the so-called Einstein--Cartan theory 
by the use of an extended geometry. The crucial feature of this geometry 
is the non-vanishing torsion of the linear connection. Torsion was originally 
introduced by E.\ Cartan \cite{car22,car23}, who also developed a general 
relativistic theory with torsion, which contained the rudiments of the 
Einstein--Cartan theory. Surprisingly, although the spin of elementary 
particles was unknown at that time, he expected a connection between torsion 
and the intrinsic angular-momentum properties of matter.%
\footnote{Besides this connection between torsion and elementary spin in the 
framework of general relativity, the geometrical concept of torsion is also 
employed in the solid state physics for the description of dislocations in 
solids \cite{bil55,kro64,kro81,kat92}. Furthermore, there is an interesting 
link between both types of torsion-theories \cite{heh65a,heh65b,heh66}.} 
The Einstein--Cartan theory in its final form was developed by many authors 
\cite{kib61,sci62,heh66,tra71,heh76}. For a review see \cite{heh76}. The 
geometry of the spacetime is now described by the so-called 
Riemann--Cartan geometry, in which the connection $\Con\a\m\b$ is only 
required to be compatible with the metric (\ref{metric}), but is allowed 
to have non-vanishing torsion contrary to the torsionless Levi--Civita 
connection. Due to the metricity condition (\ref{metric}) the connection 
now becomes \cite{heh76}
\beq\label{lct}
  \Con\a\m\b = \Chr\a\m\b
              +\frac{1}{2}( T_\m{}^\a{}_{\!\b}
                           +T_\b{}^\a{}_{\!\m}
                           +T^\a{}_{\!\m\b})\;,
\eeq
where the second expression on the right side is called the contorsion tensor. 
This generalization of the connection not only enables the spacetime to 
respond to mass as before in the general relativity, but also to spin, where 
spinning matter produces 
torsion and thus generates a non-vanishing contorsion in (\ref{lct}).

To illustrate the new features of the Einstein--Cartan theory, let us consider 
a Dirac spinor $\psi$. It is coupled to the full connection (\ref{lct}) and, 
especially to torsion, by means of a covariant spinor derivative. By 
employing the variational principle to an appropriate Lagrangian density the 
following field equation for the torsion is obtained \cite{heh71} 
\beq\label{ectorsion}
  T_{\a\b\c} = -\frac{1}{2}\planck^2\,\eta_{\a\b\c\d}\,\PSI\c^5\c^\d\psi\;. 
\eeq
Here $\planck^2=\hbar ck$ is the square of the Planck length and 
$\eta_{\a\b\c\d}$ is the volume form, see (\ref{volume}). Note that the 
right-hand side of (\ref{ectorsion}) is proportional to the canonical spin 
density of a Dirac particle, see \cite{rom69,itz80}. Due to the presence of 
torsion, the energy-momentum equation, which is obtained by varying the 
Lagrangian density with respect to the metric, now becomes
\beq\label{ecenergy}
  \frac{1}{k}G^\ast_{\a\b} = T^m_{\a\b}+\frac{3}{16k}\planck^4 g_{\a\b}
                  (\PSI\c^5\c^\d\psi)(\PSI\c^5\c_\d\psi)\;,
\eeq
where $T^m_{\a\b}$ is the usual energy-momentum tensor of Dirac 
particles already present in general relativity. The second term 
on the right side of (\ref{ecenergy}) describes a spin-spin self-interaction 
induced by torsion, which was absent in the energy-momentum equation 
(\ref{einstein}) of general relativity. Since this interaction occures only 
when matter fields overlap with each other, it is called a contact 
interaction. It does not only influence the curvature via (\ref{ecenergy}),
but also creates a cubic self-interaction in the Dirac equation \cite{heh71}
\beq\label{ecdirac}
  i\c^\m\nabla^\ast_{\!\m}\psi-\frac{mc}{\hbar}\psi
  +\frac{3}{8}\planck^2(\PSI\c^5\c^\d\psi)\c^5\c_\d\psi = 0\;,
\eeq
where $\nabla^\ast_{\!\m}$ is the covariant spinor derivative with respect 
to the Levi--Civita connection, see (\ref{usualsd}).

Besides this well-known aspect of torsion in the framework of 
Einstein--Cartan theory, 
another physical role for it has been suggested in several works on the 
unification of gravity and electromagnetism. These works originated from 
Einstein's own approach to an unified field theory \cite{ein55}, in which 
he considered an arbitrary linear connection and a non-symmetric metric
${\tilde{g}}_{\a\b}(\neq {\tilde{g}}_{\b\a})$, of which the antisymmetric 
part was identified with the dual of the electromagnetic field strength.%
\footnote{Similar attempts at an unification of gravity and electromagnetism 
were considered by many other authors, see e.g.\ \cite{edd21,sch54,ton55,%
kur52,kur74}.} 
To remedy the serious drawbacks \cite{inf50,cal53} (see also \cite{pau58}) 
of Einstein's field theory several authors have suggested to identify the 
torsion trace (or torsion vector)
\beq\label{torsiontrace}
  T_\mu = T^\a{}_{\!\m\a} = \Con\a\m\a-\Con\a\a\m
\eeq
of an arbitrary linear connection with the electromagnetic vector potential 
in an ad hoc manner \cite{bor76a,mof77,kun79}
\beq\label{t-a}
  T_\mu \sim A_\mu\;,
\eeq
still considering a non-symmetric metric. McKellar \cite{mck79} and 
Jakubiec and Kijowski \cite{jak85} deduced this relation (\ref{t-a}) 
very naturally 
using only the variational principle and avoiding any ad hoc assumptions. 
Also, the somewhat unnatural concept of a non-symmetric metric was withdrawn.

McKellar considers the usual metric and an arbitrary general linear 
connection $\Con\a\m\b$, which is neither compatible with the metric nor 
torsionless. As the result of the field equations, the connection is 
restricted to be of the form \cite{mck79} 
\beq\label{mckellar} 
  \Con\a\m\b = \Chr\a\m\b+\frac{1}{3}\d^\a{}_{\!\b}\,T_\m\;.
\eeq
Furthermore, his field equations resemble precisely the source-free 
Einstein--Maxwell equations, provided that (\ref{t-a}) is assumed.

Ferraris and Kijowski \cite{fer82} arrive at the same field equations as 
McKellar, but they do not conclude (\ref{t-a}), but consider $\Con\a\m\a 
= \Chr\a\m\a + \frac{4}{3}T_\m$, 
which is not a vector, as the electromagnetic potential and develop a 
\uone\ gauge theory differing from the usual understanding.

Jakubiec and Kijowski consider in \cite{jak85} the same theory as Ferraris 
and Kijowski \cite{fer82}, but now the relation $T_\mu \sim A_\mu$ is 
adopted implicitly. Although Dirac spinors are included in their unified 
theory \cite{jak85}, the employed spinor derivative is mainly built from 
the Levi--Civita connection, and from the general linear connection 
$\Con\a\m\b$, merely its trace $\Con\a\m\a$ couples to spinors properly. 
Since the torsion of $\Con\a\m\b$ does not couple to spinors, the 
spin-torsion aspect established in Einstein--Cartan theory is missing in 
their theory. 

The drawbacks of the above mentioned unified theories are twofold: 

First, the identification $T_\mu \sim A_\mu$ (\ref{t-a}) lacks a clear 
geometric and physical meaning, because the torsion trace is an ordinary  
vector but not a gauge potential. It remains invariant under \uone, while 
$A_\mu$ transforms in the well-known inhomogeneous way as a potential. 
This inconsistency can not be remedied by introducing a so-called 
$\lambda$--transformation, first introduced by Einstein in another 
context \cite{ein55}, by which $T_\mu$ can formally be transformed like a 
potential \cite{mck79}. What is really missing here is a clear fibre bundle 
geometric conception, from which a consistent \uone\ theory can be deduced. 
Another related problem with unified field theories is the missing physical 
interpretation of the resulting connection (\ref{mckellar}): Since it is 
not compatible with the metric, $\nabla_{\!\m}g_{\a\b} = 
-\frac{2}{3}T_\mu\cdot g_{\a\b}\neq 0$, 
it must not be applied for the parallel transports of signals on the 
spacetime: Otherwise, this would lead to the dependence of physical 
invariants 
upon their histories like in Weyl's unified theory \cite{wey22}. Therefore, 
it is necessary to decompose the whole connection (\ref{mckellar}) into a 
metric part and a ``non-metric'' part. But this can be done in several 
ways, for example, as 
\baq\label{decompose1} 
  \Con\a\m\b &=& \big[\Chr\a\m\b\big] 
                +\big[\frac{1}{3}\d^\a{}_{\!\b}T_\m\big] 
  \quad\mbox{or} 
\\ 
  \Con\a\m\b &=& \big[ \Chr\a\m\b 
                      +\frac{1}{6}(\d^\a{}_{\!\b}T_\m-T^\a g_{\m\b})\big] 
                +\big[ \frac{1}{6}(\d^\a{}_{\!\b}T_\m+T^\a g_{\m\b})\big]
  \;.\label{decompose2} 
\eaq 
In both examples the first bracket $[\ldots]$ represents a connection
compatible with the metric. Although the field equations seem to suggest 
that the metric part of (\ref{mckellar}) is provided by the Christoffel 
symbol alone, there is no rigorous geometric justification for this 
assumption. 

Secondly, although the linear connection used in these unifications is much 
more general than the Lorentzian connection (\ref{lct}) of Einstein--Cartan 
theory, the important spin-torsion coupling is missing either 
because spinning matter is not considered \cite{mck79,fer82}, or because 
the treatment of Dirac spinors is somewhat inappropriate \cite{jak85}.

In my diploma thesis \cite{diplom,paper} I have proposed a new theory 
of gravity and electromagnetism, which incorporates both aforementioned 
aspects of torsion. To achieve this purpose it is necessary to further expand 
the spacetime geometry by introducing a complex rather than a real linear 
connection and an {\em extended spinor derivative based on this 
connection\/}. Contrary to \cite{jak85}, this new spinor derivative not only 
couples the trace part, but also other components including the contorsion
of the linear connection to Dirac spinors. As a consequence of this``tight'' 
coupling, the resulting field equation for the connection can not be solved 
in the real numbers but require complex degrees of freedom. Thus, it is 
necessary to consider a complex linear connection. Through the consideration 
of spinorial matter it is possible to fully clarify the underlying fibre 
bundle geometry of this theory, and, as a consequence, especially its 
\uone\ structure. Due to the new spinor derivative, both aspects of torsion 
must be revised: First, the long-standing and unsatisfactory relation
(\ref{t-a}) turns out to be merely a formal remnant of the new fibre bundle 
geometry. Instead, the electromagnetic potential $A_\mu$ is truly related 
to another vector part $S_\mu$ via (\ref{s-a}). Secondly, the torsion-induced
spin-spin contact interaction now only occurs between distinct particles. The 
missing of the self-interaction leads to the vanishing of the second term 
on the right side of (\ref{ecenergy}) and also of the cubic spinor term in 
(\ref{ecdirac}), if a one-particle system is considered.

The field equations are derived directly from the variational principle 
and do not require any ad hoc assumptions. Formally, they resemble 
precisely the well-known equations of Einstein--Maxwell theory with charged 
Dirac particles. But this physical interpretation is now fully justified by 
the structure of the underlying fibre bundle geometry, according to which the 
resulting complex connection can be {\em decomposed} into a gravitational 
Lorentzian (that is, compatible with the metric) connection and an 
electromagnetic vector potential. This splitting of the connection together 
with a characteristic length scale in the theory suggests that gravity and 
electromagnetism have the same geometrical origin.

Although the main part of this theory was developed in the diploma thesis 
\cite{diplom}, there are still many features of the theory, which were not 
clarified rigorously and therefore deserve detailed considerations: 

First of all, the exact role of the torsion trace and its connection to the 
``true'' underlying bundle structure were not analysed exhaustively. It was 
stated in \cite{diplom} that the true electromagnetic vector potential is not 
given by the torsion but by some another vector part, $S_\m$, of the 
connection. But {\em formally\/}, the torsion trace $T_\m$ is still related 
to $A_\m$ and seems to play a role in electromagnetic phenomena. This point 
was not clarified in the diploma thesis. In this work I will show rigorously 
that torsion is connected to electromagnetism {\em not physically} but only 
{\em formally\/}. For this purpose, the electromagnetic vector potential in 
the resultant complex connection of the theory will be detached from the 
tangent frame bundle of the spacetime manifold. Since torsion is a tensor 
defined on the tangent bundle of the spacetime, torsion will be disconnected 
from electromagnetic phenomena in this way. This will also help to clarify 
the gauge transformation aspect of the electromagnetic potential and its 
connection to torsion. 

We may say that the long-standing relation $T_\m \sim A_\m$ is {\em no 
leading principle} for an unification of gravity and electromagnetism, but 
rather a {\em formal first hint} that both physical phenomena can be 
explained through the geometry of spacetime. 

To understand how the vector potential originates from the intrinsic 
spacetime geometry, we must consider the underlying fibre bundle background 
geometry of our theory very carefully. This makes necessary to reconsider 
this fibre bundle structure developed in the diploma thesis, since there 
some essential points were skipped. The decomposition of the resulting 
complex linear connection into its metric part and an electromagnetic part, 
and the corresponding decomposition principle of the extended covariant 
spinor derivative, which are vital to the understanding of the unifying 
principle of our theory, will be treated in detail in this work. In so doing, 
we will notice why it is {\em not} possible in our theory to consider 
arbitrary \uone\ principal bundles for electromagnetism but only the trivial 
bundle $M\times\uone$. Also, we will correct an error occured in the 
derivation of the connections in the diploma thesis. 

In the diploma thesis, I have employed a real orthonormal tetrad field to 
pull back a complex connection 1-form from the complex frame bundle $F_c(M)$ 
onto the spacetime manifold $M$ without further explanation. In this work  
I will explain and justify why this real valued structure is used in an 
otherwise entirely complex geometrical structure.

Finally, the spin-spin contact interaction of the new theory will now be 
investigated in detail by considering the energy eigenvalues of Dirac test
particles in a background torsion field and also by quantizing the 
interaction Hamiltonian in the first Born approximation. 

The organization of this work is as follows: 

In the second chapter we represent the new unified field theory of gravity 
and electromagnetism. Although details of the computations can be found in 
the diploma thesis \cite{diplom} and therefore will not be repeated again, 
the presentation in this work is kept fairly self-contained. In addition, 
the essential structures of the field equations are now clarified, so that 
they can be understood quite easily without going into details. More 
importantly, the physical content of the theory, which was outlined in the 
diploma thesis, is now explained in great detail. We clarify the basic 
building principle of our theory and its physical consequences. Also, the 
above mentioned formal aspect of torsion and its link to the basic 
geometrical background are explained. 
                                                                
In the third chapter the fibre bundle geometry of the theory is examined in 
every detail. First, some special topics from differential geometry are 
provided, which will be needed to explain the various construction steps of 
our fibre geometrical background: Although the basic concepts of the 
differential geometry like principal fibre bundle, connection 1-forms, and 
covariant derivatives are by now fairly well-known, there are special topics 
of differential geometry, which, in my opinion, are less familiar: For 
example, the local representation of the fibre geometry based on cross 
sections, mappings of connections, and the real and complex spin geometries. 
After these preliminaries, the bundle geometry of our theory is 
constructed step by step, and the beforementioned points on the geometrical 
background of the theory will be discussed. 

In the next chapter, we consider the spin-spin contact interaction of the new 
theory and discuss its differences to the interaction of the ordinary 
Einstein--Cartan theory. First, we study the classical Dirac equation of a 
test particle in a background torsion field caused by a classical plane 
wave field. Contrary to the contact interaction of the Einstein--Cartan 
theory, which is universal \cite{ker75}, that is, does not depend upon the 
interacting particle types, this is no longer the case for the new contact 
interaction: Now the interaction between two particles or two anti-particles 
differs from that between a particle and an anti-particle. However, if both 
types of contact interactions are quantized, and if identical particles are 
considered, then both interactions turn out to be non-universal. 

The final chapter gives a summary of the results and an outlook on future 
research.

                 \chapter{The Theory of Gravity and Electromagnetism}
                 \label{maintheory} 

                 \section{Lagrangian density}
                 \subsection{Metric and tetrads}

As mentioned in the introduction, the theory \cite{diplom} employs a 
complex linear connection. Further field variables are a metric or 
orthonormal tetrads, and Dirac spinors. Note that tetrad fields are 
needed to define Dirac spinors appropriately, see e.\ g.\ \cite{heh71}.

To introduce these field variables and their Lagrangian density, from which 
the field equations will be computed using the variational principle, let us 
consider a real 4-dimensional spacetime manifold, denoted by $M$. Let 
$F(M)$ be its frame bundle, which is a \glfourr\ principal bundle consisting 
of all tangent frames. We assume that $M$ is endowed with a pseudo-Riemannian 
metric $g_{\m\n}$, so that $F(M)$ can be reduced to the Lorentz subbundle 
consisting of orthonormal tangent frames only. Assuming further that $M$ is 
space- and time-orientable with respect to $g_{\m\n}$, this Lorentz 
subbundle has exactly 4 connected components, by the choice of one of 
which we introduce a definite space- and time-orientation on $M$ 
\cite{ble81,bau81}. This chosen subbundle is called the special Lorentz 
bundle \lplusm, which is a principal bundle consisting of orthonormal 
tangent frames such that the structure group is given by the special 
orthochronous Lorentz group \lplus\ with Lie algebra \fff{l}, 

\beq\label{lorentz} 
\eqalign{ 
  \mbox{\lplus} &:=   
  \left\{ \Lambda\in\mbox{Mat}(4,\!\real)\,| \,\Lambda^T\eta\Lambda=\eta
         \:,\:\det\Lambda=1\:,\:\Lambda^0{}_{\!0}\ge1 \right\} \;;     \\
  \fff{l} &\phantom{:}=   
  \left\{ \Lambda\in\mbox{Mat}(4,\!\real)\,| \,\Lambda^T\eta+\eta\Lambda=0
    \right\} \; , 
    }
\eeq 
where $\eta=(\eta_{ab})=(\eta^{ab})=\mbox{diag}(1,-1,-1,-1)$. A tetrad
$\sigma=(e_a{}^{\!\mu}\partial_\mu)$ is a local cross section in \lplusm, 
where latin indices, running from 0 to 3, are anholonomic indices and will be 
lowered and raised with $\eta_{ab}$ and $\eta^{ab}$, respectively. Greek 
indices run also from 0 to 3 and refer to local coordinates. They are 
lowered and raised with $g_{\mu\nu}$ and $g^{\mu\nu}$, respectively, 
the latter being the inverse of $g_{\mu\nu}$. Let $(e^a{}_{\!\mu}dx^\mu)$ 
denote the reciprocal tetrad satisfying $e_a{}^{\!\mu}e^a{}_{\!\nu}=
\delta^\mu{}_{\!\nu}$ and $e_a{}^{\!\mu}e^b{}_{\!\mu}=\delta^b{}_{\!a}$. 
Since the tetrad $\s = (e_a{}^{\!\m}\partial_\m)$ is orthonormal, it 
satisfies 
\beq\label{tetradhelp} 
  g_{\m\n}e_a{}^{\!\m}e_b{}^{\!\n} = \eta_{ab} \;, 
\eeq 
which results in the following relations: 
\beq\label{tetrad}
  g^{\mu\nu}=e_a{}^{\!\mu}e^{a\nu}\;,\quad
  g_{\mu\nu}=e_{a\mu}e^a{}_{\!\nu}\;,\quad
  g:=|\det(e^a{}_{\!\mu})|=\sqrt{-\det(g_{\mu\nu})} \;.
\eeq
The components $e_a{}^{\!\mu}$ and $e^a{}_{\!\mu}$ will be used to 
convert coordinate indices to anholonomic ones and vice versa. With the 
help of the determinant $g$, the volume element on $M$ reads
\beq\label{volume}
  \eta_{\a\b\c\d}:=g\cdot\e_{\a\b\c\d}\;,
\eeq
where $\e_{\a\b\c\d}$ is totally antisymmetric in its indices and 
$\e_{0123}=+1$.

                  \subsection{Complex linear connection}

Let $\complex\otimes TM$ be the complexified tangent bundle of $M$. 
Contrary to $F(M)$, the complex frame bundle $F_c(M)$ consists of all 
complex tangent frames of $\complex\otimes TM$ and is a \glfourc\ principal 
bundle naturally containing $F(M)$ as a canonical subbundle. Since \lplusm\
is naturally contained in $F(M)$, a tetrad $\s = (e_a{}^{\!\m}\partial_\m)$ 
is in particular a cross section in $F(M)$ and thus also in $F_c(M)$.%
\footnote{For more information on differential geometry see the next 
chapter.} 

A {\em complex linear connection} $\o$ is a \glfourc\ connection on 
$F_c(M)$, which can be pulled back to $M$ locally with the tetrad $\s$, 
yielding a $\fff{gl}(4,\!\complex)$-valued 1-form 
\beq\label{con}
  (\s^\ast\o)^a{}_{\!b}=:\Con{a}{\mu}{b}\,\mbox{d}x^\mu\;,
\eeq
which we call also a complex linear connection. 

In the first instance, the real-valuedness of the tetrad $\s$ seems to be 
confusing with respect to the complex structures introduced. We 
remark that the whole theory remains valid if we allow also for complex 
tetrads, which are cross sections of the special complex Lorentz bundle 
\clplusm\ containing not only real, but also all complex orthonormal 
tangent frames in $\complex\otimes TM$, see \ref{complexspingeometry}.
This is due to the fact that the metric, and hence all expressions 
derived from it (like the Levi--Civita connection, Einstein--tensor etc.) 
and also the matter currents $\PSI\c^\m\psi$ and $\PSI\c^\m\c^5\psi$ are 
gauge-invariant expressions and thus remain 
real valued in any case, guaranteeing the same field equations and their 
physical interpretations as in the real tetrad case.  

In order to show this invariance explicitly, let 
$(f_a{}^{\!\m}\partial_\m)$ denote 
a complex tetrad field. Then, since any two tetrads, regarded as cross  
sections into the complex Lorentz bundle \clplusm, are connected by a 
gauge transformation (see \ref{gaugetransformations}) of the complex 
Lorentz group \clplus, there exists a \clplus-valued function $\Lambda^b
{}_{\!a}$ with 
\beq 
  f_a{}^{\!\m} = e_b{}^{\!\m}\Lambda^b{}_{\! a}\;, 
\eeq 
where $e_b{}^{\!\m}$ are the components of a real tetrad, which is  
viewed here as a special complex tetrad. Since, by definition of \clplus\ 
(see (\ref{lorentzcomplex}) or simply the complexified version of 
(\ref{lorentz})), 
\beq 
  \Lambda^c{}_{\!a}\eta_{cd}\Lambda^d{}_{\!b} = \eta_{ab}\;, 
\eeq 
we have for the inverse metric the desired invariance: 
\beq 
  g^{\m\n} = f_a{}^{\!\m}\eta_{ab} f_b{}^{\!\n} 
           = e_a{}^{\!\m}\eta_{ab} e_b{}^{\!\n}\;. 
\eeq 
Thus in particular, the metric $g^{\m\n}$ and also $g_{\m\n}$ are always 
real valued quantities. The real valuedness of the currents $\PSI\c^\m\psi$ 
and $\PSI\c^5 \c^\m\psi$ can be verified in a similar fashion, using a spin 
gauge transformation instead of the Lorentz transformation. 

The reason why we have restricted the tetrads to be real valued is that we 
want to avoid confusion concerning their physical meaning as orthonormal 
reference frames, this being necessary for example to describe the physics  
studied in an observer's laboratory \cite{mtw}. 

The connection was introduced in (\ref{con}) in its anholonomic, tetrad 
components. Its corresponding coordinate components are obtained by the rule
\beq\label{coordinatecon}
  \Con\a\m\b= e_a{}^{\!\a}\,e^b{}_{\!\b}\Con{a}{\m}{b}
             +e_c{}^{\!\a}\partial_\m e^c{}_{\!\b}\;.
\eeq
These components transform in the well-known inhomogeneous way under 
coordinate changes. The curvature tensor, Ricci tensor, curvature scalar, 
and the curvature trace of the complex linear connection are defined in its 
anholonomic components as follows%
\footnote{We remark that the holonomic, i.e.\ coordinate, components of 
the various tensor quantities in 
(\ref{ricci}) can be obtained simply by employing the tetrad components. 
For example, $R^\a{}_{\!\b\m\n} = R^a{}_{\!b\m\n} e_a{}^{\!\a}e^b{}_{\!\b}$.} 
\eqlabel{ricci} 
\begin{eqroman} 
\baq
  R^a{}_{\!b\m\n} &=&
  \partial_\m\Con{a}\n{b}+\Con{a}\m{c}\Con{c}\n{b}
  -\partial_\n\Con{a}\m{b}-\Con{a}\n{c}\Con{c}\m{b}\;; 
\label{riccia}\\
  R_{\m\n} &=&
  R^a{}_{\!b\a\n}\cdot e_a{}^{\!\a}\,e^b{}_{\!\m}\;; 
\label{riccib}\\
  R &=&
  R^a{}_{\!b\m\n}\cdot e_a{}^{\!\m}\,e^{b\n}\;; 
\label{riccic}\\
  Y_{\m\n} &=&
  R^a{}_{\!a\m\n}=
  \partial_\m\Con{a}\n{a}-\partial_\n\Con{a}\m{a} \;.
\label{riccid}
\eaq%
\end{eqroman}%
It is important to note that since $\Con{a}{\m}{b}$ is an arbitrary 
complex linear connection, it is not compatible with the metric in general. 
Using the above coordinate expression (\ref{coordinatecon}) it is easy to 
show the following equivalence
\baqn 
  \nabla_{\!\m}g_{\a\b} &=& 
  \partial_\m g_{\a\b} - \Con\e\m\a g_{\e\b} - \Con\e\m\b g_{\a\e} 
  \;=\; 
  \partial_\m g_{\a\b} - \con\b\m\a - \con\a\m\b 
\\ 
  &=& 
  \partial_\m(e^c{}_{\!\a}e_{c\b}) 
  -\big( e^b{}_{\!\b}e^a{}_{\!\a}\con{b}\m{a}+e_{c\b}\partial_\m e^c{}_{\!\a} 
        +e^a{}_{\!\a}e^b{}_{\!\b}\con{a}\m{b}+e_{c\a}\partial_\m e^c{}_{\!\b}  
   \big) 
\\ 
  &=& 
  -e^a{}_{\!\a}e^b{}_{\!\b}\big( \con{b}\m{a}+\con{a}\m{b} \big) \;. 
\eaqn 
Thus we obtain a simple {\em metricity condition} in terms of the anholonomic 
connection components, 
\beq\label{metricity}
  \nabla_{\!\m} g_{\a\b}=0 \quad\Leftrightarrow\quad 
  \con{a}{\m}{b}+\con{b}{\m}{a}=0\;,
\eeq
where the right equation is precisely the condition of the Lie algebra 
\fff{l} (\ref{lorentz}) of the Lorentz group \lplus\ (to be more precise, of 
its complexified version), see (\ref{lorentzcondition}). Therefore, a 
connection which is compatible with the metric will be called henceforth 
a {\em Lorentzian connection}. 

Since in our theory 
$\Con{a}{\m}{b}$ does not satisfy (\ref{metricity}), its trace 
$\Con{a}{\m}{a}$ does not vanish in (\ref{riccid}) in general. Note that 
$\Con{a}{\n}{a}$ is a vector, contrary to $\Con\a\n\a$.

               \subsection{Extended spinor derivative}

It is well-known that spinor derivatives can be constructed not only from 
the Christoffel symbol (see, for example, \cite{dew64}) 
but also from any Lorentzian connection with non-vanishing contorsion 
\cite{heh71}. Such a connection $\Con{a}{\mu}{b}$ 
compatible with the metric defines the following spinor derivative (for 
details, see \ref{covsd}) 
\beq\label{spinorderivative} 
\nabla_{\!\mu}\psi = \partial_\mu\psi 
                    -\frac{1}{4}\Gamma_{a\mu b}\c^b\c^a\psi\;, 
\eeq  
where the $\c$--matrices satisfy $\c^a\c^b+\c^b\c^a=2\eta^{ab}\One$. In 
(\ref{spinorderivative}) the product $\c^b\c^a$ has been employed 
instead of the commonly used Lorentz generators%
\footnote{In physics, Lorentz generators are usually defined to be $i$ 
times our $\s^{ba}$, which then are hermitian matrices. But for our 
purposes it is more convenient to work without the factor $i$.} 
\beq\label{sigmaba} 
  \s^{ba}=\frac{1}{2}(\c^b\c^a-\c^a\c^b)  
\eeq   
in virtue of the metricity condition $\Gamma_{a\m b} = -\Gamma_{b\m a}$ 
\cite{heh71,heh91,law89,ber91}. If we now omit this 
condition and use our complex linear connection instead, its 
non-vanishing trace part $\Gamma_{a\mu b}\cdot\frac{1}{2} 
(\c^b\c^a+\c^a\c^b) = \Gamma_{a\m b}\eta^{ba}=\Gamma^a{}_{\!\mu a}$ 
also contributes to the spinor derivative, 
\beq\label{esd}
  \nabla_{\!\m}\psi = \partial_\m\psi 
                     -\frac{1}{4}\Gamma_{a\m b}\c^b\c^a\psi 
                    = \partial_\mu\psi 
                     -\frac{1}{4}\Gamma_{a\m b}\s^{ba}\psi
                     -\frac{1}{4}\Gamma^a{}_{\!\m a}\psi\;.  
\eeq  
Note that at this stage (\ref{esd}) is rather a formal extension since it is 
only \lplus\ covariant but of course not with respect to \glfourc. Its full 
geometric meaning is expounded in the next chapter. We remark that this 
extension is not unique, since $\s^{ba}$ could have been replaced equally 
well by $-\c^a\c^b$ or, more generally, by $\s^{ba}+\vare\cdot\eta^{ba}$. 
Due to this freedom, spinors with any multiple of the elementary charge, 
$\vare e$, can be treated, see \ref{theextension}.

               \subsection{Lagrangian density}

We introduce the adjoint spinor $\PSI:=\psi^\dagger\c^0$, $\c^\mu:=\c^a\,
e_a{}^{\!\mu}$, the mass of the spinor particle $m$, $k=8\pi G / c^4$, the 
Planck length $\planck:=\sqrt{\hbar ck} \approx 8\cdot 10^{-35}\mbox{m}$ and 
a length scale $l$ to be determined later. Using the extended spinor 
derivative and the curvature expressions derived from the complex linear 
connection in (\ref{ricci}) we write down the following Lagrangian density
\eqlabel{lagrangian} 
\begin{eqroman} 
\baq
  {\cal L} &=& {\cal L}_m + {\cal L}_G + {\cal L}_Y 
\label{lagrangiana}\\
           &=:&
  g\cdot \hbar c
  \left[ i\PSI\c^\mu\nabla_{\!\mu}\psi
        -\frac{mc}{\hbar}\PSI\psi \right]
   -\frac{g}{2k}R + \frac{g}{4k}l^2 Y_{\mu\nu}Y^{\mu\nu}
\label{lagrangianb}\\
           & =&
  \frac{g}{k}\cdot\left[
  i\planck^2\,\PSI\c^\mu\nabla_{\!\mu}\psi - mc^2 k\PSI\psi
  -\frac{1}{2}R +\frac{1}{4}l^2 Y_{\mu\nu}Y^{\mu\nu}\right]\;.
\label{lagrangianc}
\eaq%
\end{eqroman}%
Note that this Lagrangian is complex valued. We must consider this 
whole complex expression and not only its real part, since otherwise 
the contribution of the full complex connection would be taken away 
from the theory, making it meaningless.%
\footnote{For another interesting complex Lagrangian theory of gravity we 
refer to Ashtekar's formulation of general relativity \cite{ash91}, which 
might be related to the complex structures developed in this chapter, cf.\
\cite{mag87,gam93}.} 
Obviously, the three parts 
${\cal L}_m$, ${\cal L}_G$, and ${\cal L}_Y$ resemble the usual 
Lagrangian densities of spinorial matter, gravity, and the electromagnetic 
field, respectively. But now they are all complex valued and, whereas 
expressions similar to ${\cal L}_G$ and ${\cal L}_Y$ for a real connection 
were already used in \cite{mck79,jak85}, the matter Lagrangian
${\cal L}_m$ including the extended spinor derivative (\ref{esd})
is new and plays a key role in our unification. 

Note also that the introduction of the characteristic length $l$ is necessary 
from purely dimensional arguments.%
\footnote{The partial derivatives $\partial_\mu$ and the connection have 
the dimension of inverse length.} 
In (\ref{lagrangian}) it is possible to relate this length scale $l$ with the 
already given Planck length $\planck$. Regarding the last term of 
(\ref{lagrangianc}), we recognize $l^2$ as the self-coupling constant of the 
connection and, looking at the first term, $\planck^2$ as the coupling 
strength between the connection and the Dirac particles. Since Dirac fields 
are vector fields on a certain spinor bundle, which in turn is closely 
related to the intrinsic spacetime structure,%
\footnote{Spinors can be introduced on a spacetime manifold $M$ provided 
that it possesses a spin-structure with respect to a given metric, see 
\ref{spingeometry}. The existence of such a structure is a topological 
property of $M$ \cite{bau81,law89}. Since in most cases, a spacetime $M$ 
posseses a spin structure \cite{ger68,ger70}, we may say that Dirac spinors 
are natural geometric objects on $M$ like ordinary vector fields on the 
tangent bundle.} 
it is legitimate to consider Dirac spinors as intrinsical geometrical 
objects of the spacetime, at least on the non-quantum level. 
Thus, $\planck$ and $l$ both describe couplings between objects belonging to 
the same intrinsic spacetime geometry and should therefore be of the same 
order of magnitude. If, on the other hand, the unknown length $l$ turns out 
to be drastically different from the Planck length $\planck$, we may say 
that the Lagrangian (\ref{lagrangian}) does not provide a physically sensible 
theory. 

%
                     \section{Field equations}
%

The field equations are obtained by the action principle based on the 
Lagrangian constructed in (\ref{lagrangian}). The variation acts 
independently on the field quantities $\Con{a}{\mu}{b}$, $\psi$, $\PSI$ 
and $e_a{}^{\!\mu}$. Since in our Lagrangian no second derivative is 
present, the Euler--Lagrange equations will be of the simple form
\beq\label{euler} 
  0 = \frac{\d {\cal L}}{\d v}
   := \frac{\partial {\cal L}}{\partial v}
     -\partial_\nu\frac{\partial {\cal L}}{\partial\partial_\nu v}\;,
\eeq
where $v$ is an arbitrary field component. In the following only an outline 
of the calculations is presented. 
For the full computations we refer to \cite{diplom}.

         \subsection{Field equation for the connection}

To simplify the computations, we subtract the Levi--Civita connection
(\ref{lc}) from the complex connection (in its coordinate components, see 
(\ref{coordinatecon}))
\beq\label{sigma}
  \Sigma^\a{}_{\!\m\b}:=\Con\a\m\b-\Chr\a\m\b\;.
\eeq 
Being the difference of two linear connections, $\Sigma^\a{}_{\!\m\b}$ is a 
third rank tensor, which is complex valued in general. As mentioned in 
\cite{diplom}, any such third rank tensor admits a ``4-vector 
decomposition''
\beq\label{4v}
  \Sigma_{\a\b\c}=
   Q_\a g_{\b\c}
  +S_\b g_{\a\c}
  +g_{\a\b}U_\c
  -\frac{1}{12}\eta_{\a\b\c\d}V^\d
  +\Upsilon_{\a\b\c}\;,
\eeq
where the four vectors $Q_\a$, $S_\b$, $U_\c$, $V_\d$ and the ``tensor rest'' 
$\Upsilon_{\a\b\c}$ are defined in the appendix, see (\ref{ap4v2}). This 
tensor rest%
\footnote{For an explicit example of $\Upsilon_{\a\b\c}$ see (\ref{ap4v5}). 
} 
satisfies $\Upsilon^\a{}_{\!\a\c}=\Upsilon^\a{}_{\!\c\a}
=\Upsilon_\c{}^\a{}_{\!\a}=\Upsilon_{[\a\b\c]}=0$. 
The volume element $\eta_{\a\b\c\d}$ was defined in (\ref{volume}). 
With (\ref{sigma}) and (\ref{4v}) the connection can be decomposed as%
\footnote{Note that this decomposition is not irreducible in the 
sense of \cite{mcc92}, but is introduced for computational convenience.} 
\beq\label{decomposition}
  \Con\a\m\b = \Chr\a\m\b + Q^\a g_{\m\b} + S_\m\d^\a{}_{\!\b}
             + \d^\a{}_{\!\m}U_\b - \frac{1}{12}\eta^\a{}_{\!\m\b\c}V^\d
             + \Upsilon^\a{}_{\!\m\b}\;.
\eeq
Using (\ref{sigma}) only, the field equation for $\Con{a}{\mu}{b}$ 
following from (\ref{lagrangian}) reads 
\beq\label{eqcon0} 
  0 = \frac{\d}{\d\Con{a}{\m}{b}}
      \left({\cal L}_m+{\cal L}_G+{\cal L}_Y\right)\cdot 
  \d^\c{}_{\!\m}\,e^{a\a}\,e_b{}^{\!\b}
  \cdot\frac{k}{g}\qquad\Leftrightarrow
\eeq
\beq\label{eqcon1} 
  0 = 
  -\frac{1}{4}i\planck^2\,\PSI\c^\c\c^\b\c^\a\psi 
  -\frac{1}{2}
   \left[ \Sigma^{\b\e}{}_{\!\e}g^{\a\c}
         +\Sigma^\e{}_{\!\e}{}^\a g^{\c\b}
         -\Sigma^{\b\a\c}-\Sigma^{\c\b\a} \right]
  -l^2 g^{\a\b}\,\nabla^\ast_{\!\n}Y^{\n\c} \;,\quad
\eeq
where $\nabla^\ast_{\!\n}$ denotes the covariant differentiation with respect 
to $\Chr\a\m\b$. The bracket $[ \ldots ]$ on the left-hand side pertains to 
the Lagrangian ${\cal L}_G$ and can be expressed with the 4-vector 
decomposition (\ref{4v}) as follows: 
\baq\label{eq4v} 
  -\frac{1}{2}\big[\ldots\big] &=& 
  -\frac{1}{2}\big[ (Q^\a+3U^\a)g^{\b\c}+(U^\b+3Q^\b)g^{\a\c}
                   -(Q^\c+U^\c)g^{\a\b} \nonumber\\ 
  & &\phantom{-\frac{1}{2}\big[ } 
               +\frac{1}{6}\eta^{\c\b\a\d}V_\d 
               -(\Upsilon^{\b\a\c}+\Upsilon^{\c\b\a}) \big]\;. 
\eaq 
Note that in (\ref{eq4v}) there is no term containing derivatives.%
\footnote{This pleasant feature is due to the decomposition (\ref{sigma}): 
In calculating $\d{\cal L}_G / \d\Con{a}\m{b}$, we encounter the expression 
$\partial_\n(\d{\cal L}_G / \partial_\n\d\Con{a}\m{b})$ according to 
(\ref{euler}). As can be seen from the structure of the curvature 
(\ref{ricci}), this expression contains merely partial derivatives of 
various tetrad components. But these cancel exactly with the Levi--Civita 
parts contained in $\partial {\cal L}_G / \partial\Con{a}\m{b}$. For 
details we refer to \cite{diplom}.} 
We now define the vector current $j^\a:=\PSI\c^\a\psi$ and the axial current 
$j^{{\sssty 5}\,\d} := \PSI\c^5\c^\d\psi$. Inserting (\ref{eq4v}) into 
(\ref{eqcon1}) and contracting (\ref{eqcon1}) successively with $g_{\b\c}$, 
$g_{\a\c}$, $g_{\a\b}$ and $1/6 \cdot \eta_{\c\b\a\d}$ we obtain in this 
order 
\eqlabel{eqcon2}
\begin{eqroman} 
\baq 
  0 &=& -i\planck^2\cdot j^\a - 3Q^\a - 6U^\a - l^2\nabla^\ast_{\!\n}Y^{\n\a}
\label{eqcon2a}\\
  0 &=& \frac{1}{2}i\planck^2\cdot j^\b - 6Q^\b 
       -3U^\b - l^2\nabla^\ast_{\!\n}Y^{\n\b}
\label{eqcon2b}\\
  0 &=& -i\planck^2\cdot j^\c - 4l^2\;\nabla^\ast_{\!\n}Y^{\n\c}
\label{eqcon2c}\\
  0 &=& -\frac{1}{4}\planck^2\cdot j^{\sssty 5}{}_{\!\d} + \frac{1}{12}V_\d
\label{eqcon2d}  
\eaq%
\end{eqroman}%
One can easily derive 
\beq\label{quvj} 
  -Q^\a=U^\a=-i\planck^2 / 4 \cdot j^\a \;,\qquad 
  V_\d=3\planck^2 j^{\sssty 5}_\d \;. 
\eeq 
Inserting this result and (\ref{eqcon2c}) back into 
(\ref{eqcon1}) we obtain $\Upsilon_{\a\b\c}=0 $. Since the Levi--Civita 
connection is compatible with the metric, it follows from (\ref{metricity}) 
that its anholonomic components fulfill $ \Chr{a}{\mu}{a}=0 $. Using this 
fact and the equations (\ref{sigma}) and (\ref{quvj}), it follows 
\beq\label{cons} 
 \Con{a}{\mu}{a}=\Sigma^a{}_{\!\mu a}=Q_\mu+4S_\mu+U_\mu=4S_\mu 
\eeq 
and thus from (\ref{riccid}) 
\beq\label{ymm} 
 Y_{\mu\nu}=4S_{\mu\nu}:=4(\partial_\mu S_\nu-\partial_\nu S_\mu)\;.  
\eeq 
Inserting (\ref{quvj}) and $\Upsilon_{\a\b\c}=0$ into 
(\ref{decomposition}) we obtain 
\baq
  \Con{a}{\m}{b} &=&
  {\widehat{\Gamma}}^a{}_{\!\m b}
  +\d^a{}_{\!b}\cdot S_\m\;,\qquad\mbox{where}\label{eqcon3}
\\
  {\widehat{\Gamma}}^a{}_{\!\m b} &:=&
  \Chr{a}{\m}{b}+\frac{1}{4}\planck^2
  \left( i\cdot j^a e_{b\m}-i\cdot e^a{}_{\!\m}j_b
        -\eta^a{}_{\!\m bd} j^{{\sssty 5}\,d} \right)\label{eqcon4}
\eaq
and, from (\ref{eqcon2c}), 
\beq\label{eqcon5}   
    16i (l^2 / \planck^2)\:\nabla^\ast_{\!\nu}S^{\nu\c} = j^\c\;.
\eeq 
The last equation implies the current conservation
\beq\label{eqcon6}
  \nabla^\ast_{\!\c}j^\c = 16i (l^2 / \planck^2)\:
  \nabla^\ast_{\!\c}\nabla^\ast_{\!\n}S^{\n\c} = 0\;.
\eeq 
Note that so far we have not used the complex extension of the connection 
explicitly in the calculations. But from (\ref{eqcon3}), (\ref{eqcon4}) and  
(\ref{eqcon5}) it follows that 
the connection must be complex valued. In other words, these equations can 
not be solved using a real connection only. This is exactly the reason why we 
have chosen a complex rather than a real linear connection as our field 
variable. It is obvious that these complex contributions are solely due 
to the presence of Dirac fields, since, if Dirac fields were not present, all 
of the above field equations could be considered real valued: In this case, 
we would obtain instead of (\ref{eqcon3}) and (\ref{eqcon5}) 
\eqlabel{eqcon7} 
\begin{eqroman} 
\baq
  \Con{a}{\m}{b} &=& \Chr{a}\m{b}+\d^a{}_{\!b}\cdot S_\m\;, 
\label{eqcon7a}\\ 
  0 &=& 16i (l^2 / \planck^2)\:\nabla^\ast_{\!\nu}S^{\nu\c} \;. 
\label{eqcon7b} 
\eaq%
\end{eqroman}%
Thus, without Dirac spinors, complex geometric structures become superfluous. 

We observe, that the connection part ${\widehat{\Gamma}}^a{}_{\!\m b}$ 
satisfies 
\beq\label{eqcon8} 
  {\widehat{\Gamma}}_{a\m b} = -{\widehat{\Gamma}}_{b\m a} 
\eeq 
and therefore defines a {\em Lorentzian} connection, that is, a connection 
compatible with the metric, cf.\ (\ref{metricity}).

             \subsection{Dirac equations}

The Lagrangian (\ref{lagrangian}) immediately yields 
$0=\d{\cal L} / \d\PSI =\partial{{\cal L}_m} / \partial\PSI$ or, 
equivalently, 
\beq\label{dirac1} 
  i\c^\m\nabla_{\!\m}\psi-\frac{mc}{\hbar}\psi=0\;, 
\eeq 
wherein $\nabla_{\!\m}$ is given by the extended covariant spinor derivative 
(\ref{esd}). The equation (\ref{dirac1}) can be reexpressed with the help of 
(\ref{esd}) and (\ref{4v}) as follows: 
\beq\label{dirac2} 
  i\c^\m(\nabla^\ast_{\!\m}-S_\m)\psi-\frac{mc}{\hbar}\psi
  +\left(\frac{3}{2}iU_\m+\frac{1}{8}V_\m\c^5\right)\c^\m\psi=0\;,
\eeq 
where the symbol $\nabla^\ast_{\!\mu}$ is used for the covariant
derivative with respect to $\Chr\a\m\b$ as well as for the corresponding 
spinor derivative 
\beq\label{usualsd}
  \nabla^\ast_{\!\mu}\psi:=\partial_\mu\psi
  -\frac{1}{4}\chr{a}{\mu}{b}\gamma^b\gamma^a\psi\;.
\eeq 
Upon inserting the field equation for the connection (\ref{quvj}), replacing  
the vectors $U_\m$ and $V_\m$ by $U_\m = -i\planck^2 / 4\,j_\m$ and 
$V_\m = 3\planck^2 j^{\sssty 5}_\m$, respectively, the Dirac equation 
becomes 
\beq\label{dirac}
   i\c^\m(\nabla^\ast_{\!\m}-S_\m)\psi - \frac{mc}{\hbar}\psi 
   +\frac{3}{8}\planck^2 (j_\m+j^{\sssty 5}{}_{\!\m}\c^5)\c^\m\psi = 0\;. 
\eeq
The spinor equation for $\PSI$ is more difficult to compute \cite{diplom}. 
The result is
\beq\label{adjdirac}
   i(\nabla^\ast_{\!\mu}+S_\m)\PSI\cdot\c^\m+\frac{mc}{\hbar}\PSI
   -\frac{3}{8}\planck^2 \PSI (j_\m+j^{\sssty 5}{}_{\!\mu}\c^5)\c^\mu = 0
\eeq
with $\nabla^\ast_{\!\mu}\PSI=\overline{\nabla^\ast_{\!\mu}\psi}$. Contrary 
to the Dirac equation of the Einstein--Cartan theory (\ref{ecdirac}), the  
nonlinear terms in (\ref{dirac}) and (\ref{adjdirac}) vanish due to the 
identity
\beq\label{pauli}
  (j_\mu+j^{\sssty 5}{}_{\!\m}\c^5)\c^\m\psi = 
  (\PSI\c_\m\psi+\PSI\c^5\c_\m\psi\,\c^5)\c^\m\psi = 0\;,
\eeq
see \cite{diplom}. Note that (\ref{pauli}) is more stringent than the 
well-known Pauli relation 
$j^\m j_\m+j^{{\sssty 5}\m} j^{\sssty 5}{}_{\!\m}=0$, which can be 
obtained from (\ref{pauli}) by contracting from left by $\PSI$.  

Since (\ref{adjdirac}) is the spinor equation for the adjoint spinor $\PSI$, 
it must agree with the adjoint of the first equation (\ref{dirac}). This 
immediately implies that $S_\mu$ is purely imaginary,
\beq\label{imaginary}
  \mbox{Re}(S_\mu)=0\;.
\eeq

           \subsection{Field equation for the tetrad}

The Lagrangian (\ref{lagrangian}) contains no derivatives of $e_a{}^{\!\mu}$.
With the help of (\ref{tetrad}) we get
\baq
    0 &=& \frac{\d{\cal L}}{\d e_c{}^{\!\a}}e_{c\b}
        = \frac{\partial}{\partial e_c{}^{\!\a}} 
          \left({\cal L}_m+{\cal L}_G+{\cal L}_Y\right)\cdot e_{c\b}
\nonumber\\
     &=&  \left[-{\cal L}_m\,g_{\a\b}\,
                +g\,i\hbar c\PSI\c_\a\nabla_{\!\b}\psi \right]
         -\frac{1}{2k}\left[-gR\,g_{\a\b}
                            +gR_\a{}^\m{}_{\!\b\m}
                            +gR^\m{}_{\!\a\m\b} \right]
  \nonumber\\ \label{eqtet1} 
     & &\phantom{\big[} + 
        \frac{1}{4k}l^2\left[-gY_{\m\n}Y^{\m\n}g_{\a\b}
                             +4gY_{\m\a}Y^\m{}_{\!\b}\right]\;.
\eaq
In order to elaborate the physical content of this equation, we insert 
the field equations for the connection (\ref{eqcon3}), (\ref{eqcon4}), 
(\ref{eqcon5}) and the Dirac equations (\ref{dirac}) and (\ref{adjdirac}) 
together with (\ref{pauli}) and obtain the following equation, where the 
brackets $[ \ldots ]$ correspond to respective brackets in (\ref{eqtet1}) 
\baq 
  0&=& \phantom{-g}
       g\big[ i\hbar c\PSI\c_\a(\nabla^\ast_{\!\b}-S_\b)\psi 
             -\frac{1}{8}\hbar c\planck^2
              (j_\a j_\b+j^{\sssty 5}_\a j^{\sssty 5}_\b) \big] 
\nonumber\\ 
   & &-\frac{g}{2k}
        \big[ 2G^\ast_{\a\b}+i\hbar ck 
             \big( \PSI\c_\a\nabla^\ast_{\!\b}\psi
                  +\nabla^\ast_{\!\b}\PSI\cdot\c_\a\psi
                  -\frac{1}{2}\nabla^{\ast\c}(\PSI\c_{[\a}\c_\b\c_{\c]}\psi) 
             \big) 
\nonumber\\ 
  & & \phantom{-\frac{g}{2k}\big[} 
             -\frac{1}{4}i\hbar ck\planck^2 
              (j_\a j_\b+j^{\sssty 5}_\a j^{\sssty 5}_\b) \big] 
\nonumber\\ \label{eqtet2} 
  & &+\frac{g}{4k}l^2
       \big[ 64S_{\a\c}S_\b{}^\c
            -16S_{\m\n}S^{\m\n}g_{\a\b} \big]\;. 
\eaq 
In the second bracket, the Einstein-tensor 
\beq\label{einsteintensor} 
  G^\ast_{\a\b} = R^\ast_{\a\b}-\frac{1}{2}R^\ast g_{\a\b} 
\eeq 
is built from the Ricci tensor $R^\ast_{\a\b}$ and the 
Ricci scalar $R^\ast$ of the Levi--Civita connection. Note that the 
current-current term $(j_\a j_\b+j^{\sssty 5}_\a j^{\sssty 5}_\b)$ in the 
first bracket comes from the spinor derivative, whereas the corresponding 
expression in the second bracket is the result of the variation of the 
curvature scalar. Both current-current terms cancel each other and we can 
reexpress (\ref{eqtet2}) as follows 
\eqlabel{eqtet} 
\begin{eqroman} 
\baq 
  T^G_{\a\b} &=& T^m_{\a\b}+T^S_{\a\b}\;,
  \quad\mbox{where} 
\label{eqteta}\\
  T^G_{\a\b} &:=& \frac{1}{k}
  \left( R^\ast_{\a\b}-\frac{1}{2}R^\ast g_{\a\b}\right)\;; 
\label{eqtetb}\\
  T^m_{\a\b} &:=& \frac{i\hbar c}{2}
  \left[ \PSI\c_\a(\nabla^\ast_{\!\b}-S_\b)\psi
        -(\nabla^\ast_{\!\b}+S_\b)\PSI\cdot\c_\a\psi
        +\frac{1}{2}\nabla^{\ast\c}
         (\PSI\c_{[\a}\c_\b\c_{\c]}\psi)\right]\;;\qquad 
\label{eqtetc}\\
  T^S_{\a\b} &:=& \frac{16}{k}l^2
    \left[ S_{\a\c}S_\b{}^\c
          -\frac{1}{4}S_{\m\n}S^{\m\n}g_{\a\b}\right]\;.
\label{eqtetd} 
\eaq%
\end{eqroman}%
Since $T^G_{\a\b}$ and $T^S_{\a\b}$ are symmetric in $\a$ and $\b$, 
$T^m_{\a\b}$ has this property too, due to (\ref{eqteta}). Indeed, a 
lengthy calculation \cite{diplom} gives 
\beq
  T^m_{\a\b}=\frac{i\hbar c}{4}
  \left[ \PSI\c_\a(\nabla^\ast_{\!\b}-S_\b)\psi
        -(\nabla^\ast_{\!\b}+S_\b)\PSI\cdot\c_\a\psi
        +(\a\leftrightarrow\b) \right]\;.
\eeq 
Since $T^G_{\a\b}$ is proportional to the Einstein--tensor, we obtain the 
conservation law 
\beq
  0 = \nabla^\ast_{\!\a}(T^G){}^{\a\b}
    = \nabla^\ast_{\!\a}(T^m){}^{\a\b}
     +\nabla^\ast_{\!\a}(T^S){}^{\a\b}\;.
\eeq

%
                   \section{Physical interpretation}
%

  \subsection{Formal aspects of gravity and electromagnetism}

The field equations (\ref{dirac}), (\ref{adjdirac}), (\ref{eqcon5}) 
and (\ref{eqtet}), 
\eqlabel{field} 
\begin{eqroman} 
\beq 
  i\c^\m(\nabla^\ast_{\!\m}-S_\m)\psi-\frac{mc}{\hbar}\psi = 0 \;; 
\label{fielda} 
\eeq 
\beq 
  i(\nabla^\ast_{\!\m}+S_\m)\PSI\cdot\c^\m+\frac{mc}{\hbar}\PSI = 0 \;; 
\label{fieldb} 
\eeq 
\beq 
  16i (l^2 / \planck^2)\,\nabla^\ast_{\!\n}S^{\n\c} = j^\c \;; 
\label{fieldc} 
\eeq 
\beq 
  T^G_{\a\b}=T^m_{\a\b}+T^S_{\a\b} \;, 
\label{fieldd} 
\eeq 
\end{eqroman}%
resemble the well-known structures of the Einstein--Maxwell theory, 
provided that the vector $S_\m$ is identified with the electromagnetic 
potential $A_\m$. But the factor of proportionality between $S_\m$ and $A_\m$ 
remains undetermined because the charge of the Dirac particle is not fixed. 
However, the gauge property (\ref{gauge}) below shows that this charge has to 
be negative, cf.\ \cite{itz80}. We therefore identify $\psi$ with electron 
carrying the negative elementary charge $-e$, and we make the following 
identification 
\beq\label{s-a}
  S_\mu = \frac{ie}{\hbar c} A_\mu \;. 
\eeq 
Since here the vector $S_\m$ is purely imaginary, this identification is 
in accordance with (\ref{imaginary}). With (\ref{s-a}) the equation 
(\ref{fielda}) describes the Dirac equation for an electron in a 
curved spacetime, (\ref{fieldb}) being its adjoint. In order to accomodate 
(\ref{fieldc}) exactly to the inhomogeneous Maxwell equation in the curved 
spacetime of general relativity (cf.\ \cite{mtw}) 
\beq\label{maxwell} 
  -e j^\c = \nabla^\ast_{\!\n}F^{\n\c} \;, 
\eeq 
we adjust the length scale $l$ 
\baq\label{l}
  j^\c = 16i(l^2/\planck^2)\,\nabla^\ast_{\!\n}S^{\n\c} 
       = 16i(l^2/\planck^2)\,\frac{ie}{\hbar c}\,
         \nabla^\ast_{\!\n}F^{\n\c} 
  \stackrel{!}{=}
         \frac{1}{-e}\,\nabla^\ast_{\!\n}F^{\n\c}
  & \Leftrightarrow\qquad\quad 
\nonumber\\
  l^2 = \frac{1}{64\pi}\planck^2\frac{\hbar c}{e^2 / 4\pi}
      = \frac{1}{64\pi\,\a}\,\planck^2 & \Rightarrow 
  l \approx 0.83 \,\planck\;, 
\eaq 
where $\a$ is the fine structure constant and where we have employed 
Heaviside--Lorentz units, cf.\ \cite{jac65}. Inserting this result into 
(\ref{eqtetd}) the last equation (\ref{fieldd}) becomes the energy-momentum 
equation of general relativity including the energy-momentum tensors of 
gravity $T^G_{\a\b}$ (\ref{eqtetb}), of electron $T^m_{\a\b}$ (\ref{eqtetc}), 
and of the electromagnetic field $T^S_{\a\b}$ (\ref{eqtetd}). Moreover, if 
(\ref{s-a}), (\ref{l}), and the field equations (\ref{eqcon3}), 
(\ref{eqcon4}), (\ref{eqcon5}) for the connection are inserted back into the 
Lagrangian (\ref{lagrangian}), this Lagrangian becomes the familiar 
Einstein--Maxwell Lagrangian 
\beq\label{lagrangian2} 
  {\cal L} = g\cdot \hbar c
  \left[ i\PSI\c^\m(\nabla^\ast_{\!\m}
        -\frac{ie}{\hbar c}A_\m)\psi
        -\frac{mc}{\hbar}\PSI\psi \right]
  -\frac{g}{2k}\,R^\ast
  -\frac{g}{4}\,F_{\m\n}F^{\m\n}\;.
\eeq

       \subsection{Geometric interpretation of electromagnetism}
       \label{geometricint} 

We have shown that {\em formally}, the field equations of our theory 
completely agree with those of Einstein--Maxwell theory. But there are 
important differences concerning the physical understanding of the 
electromagnetism: Contrary to the ordinary theory {\em our interpretation 
of electromagnetism is geometric.}  To explain this view, we briefly discuss 
the fibre bundle structure expounded in the next chapter and give rigorous 
geometric meanings to the field equation (\ref{eqcon3}) for the connection 
and to the identification (\ref{s-a}).

We first remark that the whole connection (\ref{eqcon3}) is not a Lorentzian 
connection, since it is not compatible with the metric, that is, its 
covariant derivative of the metric does not vanish: 
\baq
  \nabla_{\!\m}g_{\a\b} &=& 
  \partial_\m g_{\a\b} 
  -({\widehat{\Gamma}}^\e{}_{\!\m\a}+\d^\e{}_{\!\a}S_\m)g_{\e\b} 
  -({\widehat{\Gamma}}^\e{}_{\!\m\b}+\d^\e{}_{\!\b}S_\m)g_{\a\e}    
  \qquad\Leftrightarrow 
\nonumber\\ 
  \nabla_{\!\m}g_{\a\b} &=& -2S_\m\cdot g_{\a\b} \;\neq\; 0\;,  
\label{nonmetricity} 
\eaq 
where we have used the fact that ${\widehat{\Gamma}}^a{}_{\!\m b}$ in 
(\ref{eqcon4}) defines a Lorentzian connection compatible with the metric, 
see (\ref{eqcon8}). From the equation (\ref{nonmetricity}) we conclude that 
the vector $S_\m$, which was identified with the 
electromagnetic potential via (\ref{s-a}), is a non-metricity vector, 
cf.\ \cite{mcc92}. Thus we can say that electromagnetism is related to 
non-metricity rather than to torsion. But this hasty conclusion must be 
regarded with caution, since the fibre bundle geometry of our theory 
interprets $S_\m$ as something completely different, and, in this 
setting, $S_\m$ is really a true \uone\ potential and nothing else, as the 
next chapter and also the discussions below will show. 

The fact that the resultant connection in (\ref{eqcon3}) is not compatible 
with the metric means that our theory can not be immersed into a 
Riemann--Cartan geometry, where the connection is assumed to be metric 
compatible from the outset, but that it requires the full \glfourc\ 
structure. A similar statement also holds for the unified field theories 
mentioned in the introduction, since in these theories the geometry requires 
the full \glfourr\ structure of the real frame bundle \pfm.  
As mentioned thereby, one of the problems with unified field 
theories is the lack of an unique geometric prescription of how to separate 
the resultant connection (\ref{mckellar}). 

In principle, it not difficult to provide such a decomposition prescription, 
which will be developed in detail in the next chapter. Here we shall explain 
the main idea of this decomposition principle: The starting point is the 
well-known fact, that a general linear connection is represented by a 
connection 1-form $\o$ (see \ref{connections}) on the tangent frame bundle 
$F(M)$ of the spacetime manifold $M$. Suppose now that $\o$ can definitely 
be mapped (to be more precise, be pulled back) 
to a connection 1-form defined on a certain {\em fibre-product 
bundle} (see \ref{productbundles}) built of the special Lorentz 
bundle \lplusm\ (\ref{lorentz}) and some yet unknown \uone\ bundle 
$\uone (M)$. As explained in \ref{productbundles} this is a principal bundle 
with structure group $\lplus\times\uone$ and will be denoted simply by 
$(\lplus\times\uone)(M)$. Since this fibre-product bundle is built 
canonically from both bundles \lplusm\ and $\uone(M)$, it is now possible to 
decompose $\o$ uniquely into a Lorentzian connection 1-form on \lplusm\ and 
a \uone\ potential on $\uone(M)$ and represent $\o$ as the sum of these two 
connection 1-forms (see {\bf Proposition 2} in 
\ref{connections} for the proof of this general feature of fibre-product 
bundles). Since a Lorentzian connection 1-form on \lplusm\ defines a 
connection compatible with the metric (cf.\ \ref{covsd}), this decomposition 
of $\o$ would provide the desired separation prescription of the connection 
(\ref{mckellar}) obtained by McKellar \cite{mck79}. 

To make this idea of the pull-back more concrete and to employ it to our 
complex resultant connection (\ref{eqcon3}), let us look at the extended 
spinor derivative (\ref{esd}) and explain its geometric foundation. 
Before doing so we first consider the usual spinor derivative 
(\ref{spinorderivative}):  Any metric connection 1-form, to be more precise, 
any connection 1-form, which defines a Lorentzian connection compatible with 
the metric, $\o_m$ (with or without torsion) is defined on the Lorentz bundle 
\lplusm, which --- provided that the spacetime $M$ admits a spin structure, 
cf.\ \ref{spingeometry} --- is endowed with a ``spin structure'' 
$\mbox{Spin}(M)\rightarrow\!\!\!\! \rightarrow\lplusm$. This is a twofold 
covering bundle map and induces a $\complex^4$-spinor bundle, on which 
spinors with their spin 1/2 representation are defined properly. With this 
spin structure, $\o_m$ can be pulled back to $\mbox{Spin}(M)$ and yields a 
spin connection, which in turn defines the spinor derivative 
(\ref{spinorderivative}) (for details see \ref{covsd}). On the other hand, 
a complex linear connection $\o$, as in our theory, is defined on the whole 
complex frame bundle $F_c(M)$ built from all tangent frames of 
$\complex\otimes TM$. Since there is no comparable twofold mapping onto 
$F_c(M)$, $\o$ does not yield a spin connection directly. Therefore, it must 
be pulled back to an ``intermediate bundle'', for which an appropriate spin 
structure exists. Such a bundle is given by $(\clplus\times\uone)(M)$, which 
is the complex analogue of $(\lplus\times\uone)(M)$ mentioned above and is 
built from the complexified orthonormal frame bundle \clplusm\ and a trivial 
\uone\ bundle $M\times\uone$. The fact that $\o$ can indeed be pulled back to 
this fibre-product, which in itself is not a natural subbundle of the frame 
bundle, is not as trivial as it might look at first sight, see 
\ref{espinconnection} for details. Once $\o$ is pulled back onto this 
intermediate bundle, a complexified spin structure 
$\cspinm\rightarrow\!\!\!\!\rightarrow\clplusm$ can be employed to pull it 
back further to $(\cspin\times\uone)(M)$, which then gives rise to the 
extended spinor derivative (\ref{esd}). 

According to this geometric background, the resulting connection 
$\Con{a}{\m}{b}$ (\ref{eqcon3}) can be decomposed uniquely into its 
Lorentzian  connection compatible with the metric 
${\widehat{\Gamma}}^a{}_{\!\m b}$ on the complex Lorentz bundle \clplusm\ 
and a true \uone\ potential $S_\m$($=\frac{1}{4}\Con{a}{\m}{a}$) on 
$M\times\uone$, see (\ref{extended5a}), (\ref{extended6a}) and 
(\ref{gtrans11}). It is now clear that the identification 
$S_\m=\frac{ie}{\hbar c}A_\m$ in (\ref{s-a}) is not only a formal one, but 
is a true geometric identity on the trivial \uone\ bundle $M\times\uone$. 
We can therefore interpret electromagnetism 
geometrically by choosing $S_\m$ to be the true potential rather than 
$A_\m$ itself, and describing the electromagnetic interaction through the 
field equations in (\ref{field}) together with the definite value of $l$ 
in (\ref{l}) only, thereby completely disregarding (\ref{s-a}). This 
geometrical point of view respects the way in which the \uone\ potential 
$S_\mu$ together with the ``gravitational'' Lorentzian connection 
(\ref{eqcon4}) originated from a single spacetime connection. 

Let us stress here that the above discussion of the underlying fibre 
bundle structure is only a sketch of the more detailed and careful treatment 
expounded in chapter 3. It should be also noted that this rigorous 
mathematical treatment is needed for the completion of our theory. 

Contrary to other unified field theories mentioned in the introducing 
chapter, where the whole connection (\ref{mckellar}) is supposed to unify 
gravity, represented by the Christoffel symbol, and electromagnetism, 
represented by the torsion trace $T_\m$, in our theory we see that the 
non-metric part $S_\m$ must be {\em detached} from the whole connection on 
the frame bundle and must be pulled back to the trivial \uone\ bundle in 
order to obtain the electromagnetic potential. This decomposition principle 
is in accordance with the well-known theorem that it is impossible to combine 
spacetime and internal symmetry in any but a trivial way \cite{ora65}. 
We can say, however, that it is not necessary to include the electromagnetic 
potential into the spacetime as something foreign or, as has been done by 
Infeld and van der Waerden \cite{inf33}, only on the spin connection level, 
but that electromagnetic phenomena can be viewed as phenomena 
originating from the intrinsic geometry of spacetime. 

Note that the length scale $l$ (\ref{l}) determining the electromagnetic 
field strength is very close to the Planck length $\planck$, which is 
the characteristic length of quantum gravity. This supports the point of view 
that gravity and electromagnetism have the same geometrical origin.

           \subsection{Torsion and electromagnetism}
           \label{torsionandem} 

We now want to make some clarifying remarks on the identification 
$T_\m\sim A_\m$ (\ref{t-a}), which has been proposed by many authors so far 
\cite{bor76a,mof77,kun79,mck79,fer82,jak85,ham89}. None of them has 
considered the geometry behind this formal identification.%
\footnote{One exception is \cite{fer82}, in which however a \uone\ gauge 
theory differing from the usual setting was derived. A charged particle is 
represented by a complexified density on $\bigwedge^4 TM$.} 

According to the geometric background briefly outlined in the previous 
subsection, the true geometric interpretation of electromagnetism is given 
by\\  
\parbox{1cm}{\mbox{ }}\hfill 
\parbox{3cm}{ 
\[ 
  S_\m := \frac{ie}{\hbar c}A_\mu\;. 
\] 
} 
\hfill\parbox{1cm}{(\ref{s-a})}\\ 
Strictly speaking, $S_\m = \frac{1}{4}\Gamma^a{}_{\!\mu a} = 
\frac{ie}{\hbar c}A_\m$ is not a \uone\ potential on the principal bundle 
$M\times\uone$, but a 1-form defined on the spacetime manifold $M$ itself, 
which has been obtained by pulling back the \uone\ potential $\o_c$ on 
$M\times\uone$ onto $M$ via a special \uone\ cross section, namely 
the trivial cross section $\hat{1}$ defined in (\ref{localsections3}) on p.\ 
\pageref{localsections3}, which prescribes to each point on $M$ the 
constant value $1\in\uone$: 
\[ 
  \hat{1}\o_c = S_\m dx^\m = \frac{ie}{\hbar c}A_\m dx^\m \;, 
\] 
see (\ref{extended6a}), where we have omitted the superfluous matrix 
indices $\d^a{}_{\! b}$. If, instead, another \uone\ cross section is 
used for the pull-back, then it will result in an \uone\ gauge 
transformation of (\ref{s-a}). To be more presice, if the cross section reads 
$\exp(\l)\hat{1}$, which is a \uone-valued function assigning to $p\in M$ 
the value $\l(p)\in\uone$, this cross section will result in the following 
gauge transformation (see (\ref{gtrans4}))  
\beq\label{gauge} 
  e_a{}^{\!\m} \mapsto e_a{}^{\!\m}\;,\quad
  S_\m \mapsto S_\m + \partial_\m\l\;,\quad 
  \psi \mapsto \exp(\l)\psi\;. 
\eeq 
Since this transformation takes place only on the \uone\ bundle and on the 
associated spinor bundle, the tetrad fields and the Lorentzian connection 
(\ref{eqcon4}) as cross sections and as connection 1-form on \clplusm\ remain 
invariant, cf.\ 3.4. 

Now, the identification (\ref{s-a}) can be inserted into the expression of 
the whole connection (\ref{eqcon3}), and its torsion trace can be computed,
\baq 
  T_\m &=& \Con\a\m\a-\Con\a\a\m 
      \;=\;\Con{a}\m{a}-\Con{a}\a{b}e_a{}^{\!\a}e^b{}_{\!\m} 
\nonumber\\ 
       &=& \big[{\widehat{\Gamma}}^a{}_{\m a}+\d^a{}_{\!a}S_\m\big] 
          -\big[{\widehat{\Gamma}}^a{}_{\a b}+\d^a{}_{\!a}S_\a\big] 
           e_a{}^{\!\a}e^b{}_{\!\m} 
      \;=\;\big[4S_\m\big] 
          -\big[-\frac{3}{4}i\planck^2j_\m+S_\m\big] 
\nonumber\\ 
       &=& 3\frac{ie}{\hbar c}A_\m+\frac{3}{4}i\planck^2\,j_\m\;.
\label{t-aj}
\eaq 
This shows that the simple ansatz $T_\m\sim A_\m$ is no more valid in our 
theory, if matter is present. However, since the above equation (\ref{t-aj}) 
contains both the torsion trace $T_\m$ and the potential $A_\m$, they still 
seem to be related to each other. But in 
contrast to (\ref{s-a}), the torsion components in (\ref{t-aj}) are 
derived from the coordinate connection components (\ref{coordinatecon}), 
which are obtained by pulling back the general linear connection $\o$ 
from the frame bundle to $M$ via the cross section given by a coordinate 
reference frame ($\partial / \partial x^\mu$). Thus, there is no possibility 
of an \uone\ gauge transformation in (\ref{t-aj}), see (\ref{gtrans15}). 

If we employ the cross section $\exp(\l)\hatone$ instead of $\hatone$, so 
that the gauge transformation (\ref{gauge}) takes place, then the 
formal application of the formula (\ref{t-aj}) to $S_\m + \partial_\m\l$ 
instead of $S_\m$ would result in 
\beq\label{t-aj2} 
  T_\m = 3\big(\frac{ie}{\hbar c}A_\m+\partial_\m\l\big) 
        +\frac{3}{4}i\planck^2\,j_\m \;, 
\eeq 
which is not equal to (\ref{t-aj}). This implies that, to obtain 
(\ref{t-aj}) from (\ref{s-a}), the special \uone\ gauge $\hat{1}$ 
implicitly chosen in (\ref{s-a}) must be held fixed. Since (\ref{t-aj}) 
is valid in this gauge only, the relation $T_\mu\sim A_\mu$ is merely a 
formal remnant of the true \uone\ identity (\ref{s-a}). 

It is important to note that, in accordance with the decomposition principle 
explained in \ref{geometricint},  the parallel displacements of (uncharged) 
vectors and tensors on the spacetime are to be performed with the resultant 
Lorentzian connection ${\widehat{\Gamma}}^a{}_{\!\m b}$ (\ref{eqcon4}) only 
and not with the full connection $\Con{a}\m{b}$. Now, it is easy to see that 
the torsion trace ${\widehat{T}}_\m$ of this Lorentzian connection 
${\widehat{\Gamma}}^a{}_{\!\m b}$ does not contain the electromagnetic vector 
potential, but only the vector current of the Dirac field, 
\beq\label{t-j} 
  {\widehat{T}}_\m = \frac{3}{4}i\planck^2\,j_\m \;. 
\eeq 
Thus, once the resultant connection $\Con{a}\m{b}$ (\ref{eqcon3}) is 
decomposed in its Lorentzian connection part and the \uone\ potential part, 
there is not even a formal relation between the torsion trace and the 
electromagnetic potential. So, torsion and electromagnetism seem to be two 
completely different physical quantities, at least in the end. But in order 
to motivate our theory, especially the extended spinor derivative (\ref{esd}),
the earlier unified field theories based on the simple ansatz 
$T_\m \sim A_\m$ (\ref{t-a}) must be considered seriously. We may say that 
this ansatz is to be viewed as a first formal hint that electromagnetic 
phenomena originate from the intrisic spacetime geometry.

%
           \section{Extension of the theory}
           \label{theextension} 
%

So far we have considered only an electron. In order to include 
other, differently charged particles we observe that in the case of a 
Lorentzian connection with $\con{a}{\m}{b} = -\con{b}{\m}{a}$ the 
spinor derivative (\ref{spinorderivative}) can be written in many ways 
\eqlabel{xsd} 
\begin{eqroman} 
\baq\label{xsda} 
   \nabla_{\!\mu}\psi &=& 
   \partial_\mu\psi-\frac{1}{4}\Gamma_{\!a\mu b}\gamma^b\gamma^a\psi
  =\left( \partial_\mu-\frac{1}{4}\Gamma_{\!a\mu b}\sigma^{ba}
          -\frac{1}{4}\Gamma^a{}_{\!\mu a}  \right) \psi\;;
\\ \label{xsdb}  
  \nabla_{\!\mu}\psi &=&  
   \partial_\mu\psi+\frac{1}{4}\Gamma_{\!a\mu b}\gamma^a\gamma^b\psi
  =\left( \partial_\mu-\frac{1}{4}\Gamma_{\!a\mu b}\sigma^{ba}
          +\frac{1}{4}\Gamma^a{}_{\!\mu a}  \right) \psi\;;
\\ \label{xsdc}
\nabla_{\!\mu}\psi &=&  
   \partial_\mu\psi-\frac{1}{4}\Gamma_{\!a\mu b}\sigma^{ba}\psi\;;
\\ \label{xsdd} 
  \nabla_{\!\mu}\psi &=&  
  \partial_\mu\psi-\frac{1}{4}\Gamma_{\!a\mu b}\sigma^{ba}\psi 
               +\frac{\varepsilon}{4}\Gamma^a{}_{\!\mu a}\psi \;,
  \quad \vare \in \real \;. 
\eaq 
\end{eqroman}%
All these four expressions are equivalent to each other due to the 
metricity condition. 

But if we now insert our complex connection, these four 
spinor derivatives become different and correspond to derivatives of 
Dirac spinors with charges $-e$, $+e$, $0$, and, more generally,  
$\vare e$, where $\vare \in \real$, respectively. The last 
case is necessary if fractionally charged particles are considered. 
Otherwise, the first three cases suffice. Under the \uone\ gauge 
transformation (\ref{gauge}) the Dirac spinors belonging to each of the 
above spinor derivatives (\ref{xsda}) to (\ref{xsdd}) transform as 
\eqlabel{xgauge} 
\begin{eqroman} 
\baq
  S_\m \mapsto S_\m+\partial_\m\l \;&\; 
  \psi \mapsto \phantom{\vare}\exp(+\l)\psi 
\label{xgaugea}\\
  S_\m \mapsto S_\m+\partial_\m\l \;&\; 
  \psi \mapsto \phantom{\vare}\exp(-\l)\psi 
\label{xgaugeb}\\
  S_\m \mapsto S_\m+\partial_\m\l \;&\; 
  \psi \mapsto \phantom{\exp(-\vare\l)}\psi 
\label{xgaugec}\\
  S_\m \mapsto S_\m+\partial_\m\l \;&\; 
  \psi \mapsto \exp(-\vare\l)\psi 
\label{xgauged} 
\eaq 
\end{eqroman}%
see (\ref{gtrans5}) and (\ref{gtrans6}). To confirm the extension principle 
in (\ref{xsd}), according to which we can 
incorporate Dirac spinors with arbitrary charges into our theory, we will 
consider a many-particle system in chapter 4.

                \chapter{Fibre Bundle Geometry}
                \label{fibrebundlegeometry}  

In this chapter we shall fully clarify the underlying fibre bundle structure 
of the unified field theory discussed in the foregoing chapter. 
Although the general setting of the fibre bundle background has been already 
explained in the appendix of the diploma thesis \cite{diplom}, there are 
still several important aspects of the fibre geometry which deserve a more 
detailed consideration.

The salient feature of the bundle structure explained below is that it 
provides an unique prescription how to construct a spinor derivative out 
of a given general complex linear connection. 
The whole construction of the fibre bundle geometry was in fact 
motivated by the problem of giving the formally extended 
spinor derivative (\ref{esd}) a rigorous mathematical foundation. 
Without the consideration of Dirac spinors, there would be no 
clear guideline for the construction of a fibre geometrical background. 
For example, it is easy to see that in the field equations (\ref{quvj}) 
to (\ref{eqcon5}) all complex contributions will vanish if no Dirac spinor 
is present, and the connection will simply be given by the 
real solution of McKellar (\ref{mckellar}), cf.\ (\ref{eqcon7a}). 
Also, in this case, the inhomogeneous Maxwell equation (\ref{eqcon5}), 
which follows from (\ref{eqcon2c}), would not 
contain the imaginary unit $i$, see (\ref{eqcon7b}). Thus, as has been said 
in the discussion following (\ref{eqcon7}), a complex geometrical structure 
would become unnecessary, and this would lead to the conclusion of the 
important non-metricity vector $S_\m$ being real valued, making the 
identification (\ref{s-a}) incorrect. In this case, the question would 
arise how to make a real valued vector a true \uone\ potential.%
\footnote{A solution to this problem has 
been suggested by Jakubiec and Kijowski in \cite{jak85}. Instead of the 
real valued torsion trace in (\ref{mckellar}) these authors considered 
the trace $\Con\a\m\a$ of the whole linear connection as the 
electromagnetic potential. Since 
$\Con\a\m\a$ is not a vector but a connection on the bundle of scalar 
densities $\bigwedge^4TM$, it was necessary to introduce the notion of 
scalar densities of ``complex weight'' in order to make contact with the 
commonly used \uone\ gauge theory of electromagnetism, see for details 
\cite{jak85}. In other words, they had to introduce the complex structure 
from the outside. This artificial feature of the geometry is one of the 
main drawbacks of their theory.} 

Note that the consideration of Dirac fields enforces a geometrical 
decomposition of the whole linear connection (\ref{eqcon3}), since otherwise 
the extended spinor derivative (\ref{esd}) remains only a formal definition, 
and, even worse, could be in contradiction to the well-known twofold 
representation structure of Dirac fields: A general linear connection, 
as used in (\ref{esd}), can not be ``lifted'' to a spin connection in 
contrast to a Lorentzian connection defined on an orthonormal 
(or Lorentz) frame bundle. The algebraic reason for this fact is that, 
contrary to the Lorentz group, the structure group \glfourc\ of the 
complex frame bundle $F_c(M)$ does not possess a twofold covering map, 
which accounts for the spin 1/2 nature of spinors. Therefore it is not   
sufficient to decompose the linear connection only formally, but the 
whole fibre bundle geometry of such a decomposition must be clarified. 

In the first section we discuss some fundamental aspects of the general 
fibre bundle geometry. This does not mean a mere recapitulation of 
well-known facts, which can be found in textbooks like \cite{kob63}, 
\cite{gre72}, or \cite{nak90}, but the discussion comprises several 
special topics in great detail, which are essential for the construction 
of the special fibre geometry of our theory, see also \cite{bos93,bos94}. 
The topics are the following:  
principal bundles, bundle mappings, product bundles, associated 
vector bundles, local cross sections, gauge transformations, and connections 
and their covariant derivatives. 

In the second section the spin geometry is expounded in some detail. The 
material was gathered from various textbooks 
\cite{bau81,ber91,ble81,ben87,har90,law89} (see also \cite{due89}) 
but it also contains own computations. 
Especially, the notion of a complex spin geometry can be found only 
en passant in a few textbooks \cite{ber91,har90,str64}. Although most of 
the structures of the complex spin geometry is merely an exact complexified 
copy of the real spin geometry, some care is needed because 
of various possible representations of the complex spin group and of the 
imbedding of the real structure into the complex one. The complex spin  
geometry is needed as a central device in the next section, where we 
develop the fibre bundle geometry of our theory.

The third section contains every detail of the fibre bundle structure 
needed to complete our geometrical theory of gravity and electromagnetism. 
The strategy of the construction is as follows: To obtain the 
``intermediate bundles'' (cf.\ \ref{geometricint}) between the complex 
frame bundle and a yet unknown spin bundle we first concentrate only on the 
corresponding structure groups of the principal fibre bundles to be 
determined. We construct a special diagram of Lie group homomorphisms, which 
then can easily be translated to a corresponding diagram of bundle mappings. 
We then use this bundle diagram to map a general complex linear connection 
1-form on the frame bundle $F_c(M)$ onto an unique spin connection. To see 
how this spin connection gives rise to the extended spinor derivative 
(\ref{esd}) we employ the concept of local cross sections to obtain local 
expressions of the various connections and their covariant derivatives. In 
doing so we will notice that the resultant non-Lorentzian connection in 
equation (\ref{eqcon3}) can be decomposed unambiguously. Keeping in mind 
the main goal of the construction of the special fibre bundle background, 
namely the unique prescription of building a covariant spinor derivative 
out of a general complex linear connection, will help to survey these 
technicalities. 

Finally, in the forth section, the \uone\ gauge transformation in this 
geometric setting is explained and compared with the naive gauging of the 
torsion vector.

         \section{Some aspects of differential geometry}

\subsection{Principal bundles}\label{principalbundles}  

Let $M$ be a differentiable manifold, which in this work denotes the real 
4-dimensional spacetime manifold, although, of course, the following 
considerations are valid for an arbitrary manifold. Let $G$ be a Lie 
group. A {\em principal fibre bundle} over $M$ with group $G$ consists of a 
manifold $G(M)$ with the following conditions \cite{kob63}:\\ 
\begin{enumerate} 
\item The right action, denoted by $G(M)\times G \rightarrow G(M)$, 
  $(u,\Lambda) \mapsto u\Lambda$, is free. That is, if $u\Lambda = u$ for 
  some $u \in G(M)$ , then $\Lambda$ is already the trivial element 
  $\Lambda = \One \in G$. 
\item Let $\sim$ be the equivalence relation on $G(M)$ defined by $u 
  \sim v \Leftrightarrow u = v\Lambda$ for some (and hence exactly one) 
  $\Lambda \in G$. Then the quotient space $G(M) /_\sim$ is precisely 
  $M$. If $\pi_G$ denotes the differentiable canonical projection 
  \beq\label{pribu1} 
    \pi_G: G(M) \rightarrow G(M) /_\sim = M\;, 
  \eeq 
  then each equivalence class 
  corresponds to exactly one {\em fibre} $\pi_G^{-1}(p)$, $p\in M$, which 
  is diffeomorphic to $G$ itself, $\pi_G^{-1}(p) \cong G$. 
\item Furthermore, every point $p$ has an open neighbourhood $\calu$ 
  such that $\pi_G^{-1}(\calu)$ is {\em isomorphic} with $\calu\times G$ 
  (local triviality). This means that there exists a diffeomorphism 
  \beq\label{pribu2}  
  \eqalign{ 
    \Psi = \pi_G\times\phi: \pi_G^{-1}(\calu) &\longrightarrow 
    \phantom{(\pi} \calu\times G  
  \\ 
    \phantom{\Psi=\pi_G\times\phi:\pi_G} 
    u &\longmapsto \left(\pi\left(u\right), \phi\left(u\right)\right)\;, 
  } 
  \eeq 
  such that 
  $\phi(u\Lambda) = \phi(u)\cdot\Lambda$, where $\Lambda\in G$ and the dot  
  on the left-hand side denotes the group multiplication in $G$. 
\end{enumerate}
We call $G(M)$ the {\em total space}, $M$ the {\em base space}, $G$ 
the {\em structure group}, and $\pi_G$ the {\em (bundle) projection\/}. 
If no confusion is to be expected, we will denote the principal bundle 
simply by \pgm. Exceptions are, for example, the bundle of linear frames 
\pfm\ and the complex frame bundle $F_c(M)$, whose structure groups are 
\glfourr\ and \glfourc, respectively.

\subsection{Bundle mappings}\label{bundlemappings} 

Bundle mappings will be used in \ref{fibrebackground} to pull back linear 
connection 1-form from the frame bundle $F_c(M)$ to intermediate bundles 
(see \ref{geometricint}) and, finally, to an extended spin principal bundle. 
In this way we obtain a special spin connection 1-form, which defines 
the extended spinor derivative (\ref{esd}). 

Let \pgm\ and \phm\ denote two principal bundles over the same base 
manifold \mm. A {\em bundle homomorphism} is a pair $(f,f_o)$, where $f$ is 
a mapping between the total spaces $f:\pgm\rightarrow\phm$ and $f_o$ is a 
Lie group homomorphism $f_o: G\rightarrow H$, where $f$ and $f_o$ must 
satisfy $f(u\Lambda) = f(u)f_o(\Lambda)$ for all $u\in\pgm$ and $\Lambda 
\in G$. Here the product $f(u)f_o(\Lambda)$ denotes the right action of $H$ 
on \phm. This implies that each fibre $\pi_G^{-1}(p)$ of \pgm\ is mapped into 
a fibre of \phm. Therefore, a bundle homomorphism $(f,f_o)$ defines a mapping 
$f_M$ on the base manifold \mm\ by $f_M: M\rightarrow M$, $p\mapsto \pi_H
(f(u))$, where $u$ is an arbitrary element of the fibre $\pi_G^{-1}(p)$. 

In this work, we will consider such bundle homomorphisms $(f,f_o)$, 
which induce the identity mapping $f_M \equiv 1$ on \mm. Often we denote 
$(f,f_o)$ 
simply by $f$ and call it the {\em bundle mapping\/}. This means that the 
following diagram is commutative:\\ 
\refstepcounter{equation}\label{bundlemap} 
\unitlength1cm
\begin{picture}(15,6.5)

\put(2.5,2.5){\makebox(3,0.5)[b]{\pgm}}
\put(2.5,4.5){\makebox(3,0.5)[b]{$\pgm\times G$}}
\put(8.5,2.5){\makebox(3,0.5)[b]{\phm}}
\put(8.5,4.5){\makebox(3,0.5)[b]{$\phm\times H$}}
\put(6,0.5){\makebox(2,0.75)[b]{$M$}}
\put(6,2.9){\makebox(2,0.5)[b]{$f$}}
\put(6,4.9){\makebox(2,0.5)[b]{$f\times f_o$}}
\put(4.7,1.2){\makebox(0.5,0.5)[b]{$\pi_G$}}
\put(4.1,3.5){\makebox(0.5,0.5)[b]{$R$}}
\put(8.8,1.2){\makebox(0.5,0.5)[b]{$\pi_H$}}
\put(10.1,3.5){\makebox(0.5,0.5)[b]{$R$}}
\put(13.9,2.5){\makebox(1,0.5)[r]{(\ref{bundlemap})}}
\put(5.8,4.65){\vector(1,0){2.8}}
\put(5.2,2.65){\vector(1,0){3.9}}
\put(4,4.2){\vector(0,-1){1.3}}
\put(10,4.2){\vector(0,-1){1.3}}
\put(4.1,2.3){\vector(2,-1){2.8}}
\put(9.9,2.3){\vector(-2,-1){2.8}}
\end{picture}
Here $R$ denotes the right group actions. 

If, in particular, $f:\pgm\rightarrow\phm$ is an imbedding and 
$f_o: G\rightarrow H$ a Lie group  monomorphism, then $f$ is called a {\em 
bundle imbedding}. Since $f$ is a topological imbedding, we can identify 
\pgm\ with its image $f(\pgm)$ and transfer the principal bundle structure 
of \pgm\ to $f(\pgm)$. This makes $f(\pgm)$ itself a principal bundle, 
which is contained in \phm. We call $f(\pgm)$ or \pgm\ a {\em (reduced) 
subbundle} of \phm\ and $f$ a {\em bundle reduction}. 

For example, the special Lorentz bundle \lplusm\ (p.\ \pageref{lorentz}) is 
a subbundle of the frame bundle \pfm, where $f$ is simply the canonical 
inclusion of \lplusm\ in \pfm. Also, the frame bundle \pfm\ is a natural 
subbundle contained in the complex frame bundle $F_c(M)$.

\subsection{Product bundles}\label{productbundles} 

The notion of product bundles is needed for the construction of the 
``intermediate bundle'', whose very product structure will lead to a natural 
decomposition of the linear connection (\ref{eqcon3}) into its Lorentzian 
part and its \uone\ part in section \ref{fibrebackground}. 

Let again \pgm\ and \phm\ be two principal bundles over \mm. Then their 
(topological) product $\pgm\times\phm$ is naturally a principal bundle over 
the base manifold $M\times M$ with structure group $G\times H$. The fibre 
over a base point $(p,q)\in M\times M$ is given by $\pi_G^{-1}(p)\times 
\pi_H^{-1}(q)$. Now, if we consider not the whole base space $M\times M$, 
but only the diagonal space $\Delta:=\{(p,p)\in M\times M\}$, then the 
totality of its fibres, $(\pi_G\times\pi_H)^{-1}(\Delta)$, is easily seen 
to be a principal bundle again.%
\footnote{This construction is of course valid in a more general setting: 
If \pkm\ is a principal bundle over $M$ and $N\subset M$ is a submanifold 
of $M$, then $\pi_K^{-1}(N)$ is a principal bundle over N with the same 
structure group $ K $. }  
We identify the diagonal $\Delta$ with \mm\ itself and denote this $G\times 
H$ principal bundle by $(G\times H)(M)$, 
\beq\label{pb1} 
  (\pi_G\times\pi_H)^{-1}(\Delta) =: (G\times H)(M)\;.  
\eeq 
Note that in $(G\times H)(M)$ only the fibres are ``multiplied''. We call 
this bundle the {\em (fibre) product bundle}. 

An element of $(G\times H)(M)$ is given by $(u,v) \in \pgm\times\phm$ with 
the diagonality condition $\pi_G(u) = \pi_H(v)$. If we denote the canonical 
projections of the total spaces by 
\eqlabel{pb2} 
\begin{eqroman} 
\baq 
  p:& (G\times H)(M) \rightarrow \pgm & \qquad\mbox{and}\\ 
  q:& (G\times H)(M) \rightarrow \phm & , 
\eaq 
\end{eqroman}%
and the corresponding canonical projections of the Lie groups $G\times H
\rightarrow G$ and $G\times H\rightarrow H$ by $p_o$ and $q_o$, respectively,
then, $(p,p_o)$ and $(q,q_o)$ define canonical bundle mappings of the fibre 
product bundle $(G\times H)(M)$ onto its building blocks. But it is important 
to note that these building blocks \pgm\ and \phm\ are {\em not} canonical 
subbundles of $(G\times H)(M)$ in general, that is, it is not always possible 
to imbed \pgm\ or \phm\ naturally into $(G\times H)(M)$.

\subsection{Associated vector bundles}\label{avb} 

Let \pgm\ be a principal bundle and $V$ a vector space (real or complex), 
upon which the structure group $G$ acts on the left by a representation 
$\rho$:\\  
\parbox{1cm}{ }\hfill 
\parbox{12cm}{
  \[\renewcommand{\arraystretch}{1.4}
    \arraycolsep0.2mm
    \begin{array}{c@{\;\;}c@{\;\;}cl}
        G \times V 
      & \longrightarrow 
      & V 
      &
      \\
        (\Lambda,\,v) 
      & \longmapsto
      & \rho(\Lambda)v 
      & ,\;\;\;
    \end{array}
  \]}
\hfill
\parbox{1cm}{\baq\label{avb1}\eaq}\\
where $\rho(\Lambda)$ is an element of the general linear group of $V$. 
Consider now $\pgm\times V$ and introduce an equivalence relation through 
\beq\label{avb2} 
  \pgm\times V \ni (u,v) \sim 
  \left( u\Lambda,\,\rho(\Lambda^{-1}) v\right) \,,\quad \Lambda\in G\;, 
\eeq 
and denote the resulting quotient space by \pvm, 
\beq\label{avb3} 
  \pvm := \pgm\times V\big/_\sim = \pgm\times_\rho V\;. 
\eeq
An element $\phi$ of \pvm\ is thus an equivalence class, which will be 
denoted by 
\beq\label{avb4} 
  \phi = [u,\,v] = [u\Lambda,\,\rho(\Lambda^{-1})v] \in \pvm\;. 
\eeq 
The manifold \pvm\ has a canonical projection mapping $\pi_V$ defined 
through the principal bundle projection $\pi_G$, 
\baq 
  \pi_V : \pvm &\longrightarrow & \;\; M \nonumber\\
  {[u,\,v]}    &\longmapsto       & \pi_G(u)\;,\label{avb5} 
\eaq 
where the definition is independent of the choice of representative 
$u$. Each point $p\in M$ has an open neighbourhood \calu\ 
such that $\pi_V^{-1}(\calu)$ is diffeomorphic to $\calu\times V$. The 
diffeomorphism $\Psi_V$ can be constructed using the local 
trivialization of the principal bundle \pgm. With the help of the 
isomorphism $\Psi = \pi_G\times \phi$ defined in 
\ref{principalbundles} we define 
\baq 
  \Psi_V : \pi_V^{-1}(\calu) &\stackrel{\cong}{\longrightarrow}&  
  \qquad\calu\times V
\nonumber\\ 
  {[u,\,v]} &\longmapsto & \left(\pi_G\left(u\right),\, 
  \rho\left(\phi\left(u\right)\right)v \right)\;, 
\label{avb6} 
\eaq 
which is easily seen to be independent of the choice of representative 
for the equivalence class.%
\footnote{If we start with another representative of the same equivalence 
class $[u\Lambda,\,\rho(\Lambda^{-1})v]$ instead, we 
obtain the same result $\left(\pi_G\left(u\Lambda\right),\,\rho\left(\phi 
\left(u\Lambda\right)\right)\rho\!\left(\Lambda^{-1}\right)v\right) = 
\left(\pi_G\left(u\right),\,\rho\left(\phi\left(u\right)\right)v \right)$.} 
Thus \pvm\ is a fibre bundle with fibre $V$, which is called the {\em vector 
bundle associated with the principal bundle \pgm.}

\subsection{Local cross sections} 

Cross sections provide a link between the abstract concept of connection 
1-forms defined on a principal bundle and the more familiar notion of 
connection components on the spacetime manifold. These components lead to a 
convenient representation of the covariant derivative.

A local cross section $\s$ in a principal bundle \pgm\ is a mapping from an 
open subset $\calu \subset M$ of the base manifold \mm\ to \pgm, which 
respects the fibre structure. That is, for each $p\in \calu$ the image 
$\s(p)$ lies in the fibre above $p$, $\pi_G\left(\s\left(p\right)\right) 
= p$. 

Given such a local cross section $\s$, it is possible to trivialize \pgm\ 
on \calu\ by defining the following diffeomorphism 
\baq 
  \Psi_\s : \pi_G^{-1}({\cal U}) &\stackrel{\cong}{\longrightarrow}& 
  \quad\calu\times G
\nonumber\\ 
  u\quad &\longmapsto & [\pi_G(u),\,\phi_\s(u)]\;, 
\label{lcs1} 
\eaq 
where $\phi_\s(u) \in G$ is uniquely determined by the definition 
\beq\label{lcs2} 
  u =: \s\left(\pi_G\left(u\right)\right)\phi_\s\left(u\right)\;. 
\eeq 
Since principal bundles are in general not globally trivial, cross sections 
are normally defined only locally and can not be extended to a global cross 
section on \mm. 

A cross section $v$ in an associated vector bundle \pvm\ is defined 
analogously by demanding $v(p) \in \pi_V^{-1}(p)$. Contrary to the case of 
principal bundles, any vector bundle admits global cross sections,%
\footnote{For example, a special cross section is given by the zero cross 
section, which prescribes to each $p\in M$ the zero vector in its fibre
$\pi_V^{-1}(p)$.} 
and these are called {\em vector fields\/}. With the help of a local cross 
section $\s$ over \calu\ in the corresponding principal bundle \pgm, a 
vector field in \pvm\ can be represented locally by a $V$-valued function 
$v_\s$ on \calu\ as follows (see (\ref{avb4})) 
\baq 
  v|_{\cal U} : \calu & \longrightarrow & \qquad\pvm|_{\cal U} 
\nonumber\\ 
  p\; &\longmapsto & v(p) = [\s(p),\,v_\s(p)]\;. 
\label{lcs3} 
\eaq 
As an example, let \pvm\ be the tangent bundle $TM$ associated to the frame 
bundle \pfm. For the cross section $\s$ of \pfm, we take a local coordinate 
frame $(\partial_\m)$. Now, if $v$ is a tangent vector field on $TM$, then 
it can be represented by a $\real^4$-valued function $v^\m$, $\m = 0,1,2,3$, 
so that 
\beq 
  v(p) = [(\left.\partial_\m\right|_p),\,v^\m(p)] \;. 
\eeq 
This is of course nothing but a sophisticated way to express $v$ in its 
coordinate components via $v = v^\m\partial_\m$. The representation 
(\ref{lcs3}) will be used for the definition of the covariant derivative and 
also for the local description of a Dirac spinor field $\psi$ below.

\subsection{Gauge transformation}\label{gaugetransformations} 

A gauge transformation (see e.g.\ \cite{ble81}) on a principal bundle \pgm\ 
is a special bundle mapping $f:\pgm\rightarrow\pgm$, which is a 
diffeomorphism and induces the identity mapping $f_M=1$ on \mm, see 
\ref{bundlemappings}. Since $f$ is a diffeomorphism, its corresponding Lie 
group homomorphism $f_o$ is in fact an isomorphism. 

Consider now a local cross section $\s$ over $\calu\subset M$. For each 
$p \in\calu$, the gauge transformation $f$ acts on $\s$ via $f: \s(p) 
\mapsto f\left(\s\left(p\right)\right)$. Since both elements $\s(p)$ 
and $f\left(\s\left(p\right)\right)$ lie in the same fibre $\pi_G^{-1} 
(p)$, there exists an unique local $G$-valued function $\Lambda : \calu
\rightarrow G$, such that 
\beq\label{gt1} 
  f\left(\s\left(p\right)\right) = \s(p)\Lambda(p),\quad
  p\in\calu\;. 
\eeq 
Note that $f\left(\s\left(p\right)\right)$ defines another local cross 
section on \calu, which we denote by $\t(p)$. In the foregoing subsection we 
have represented a vector field $v$ on \pvm\ by a $V$-valued function $v_\s$ 
on \calu, using the cross section $\s$ on \pgm. Analogously, we may define 
another $V$-valued function $v_\t$ corresponding to $\t$ via (\ref{lcs3}). 
With (\ref{gt1}) and (\ref{avb4}), we can relate both functions $v_\s$ and 
$v_\t$ 
\baq 
  v(p) &=& [\s(p),\,v_\s(p)]\;=\;[\s(p)\Lambda(p),\,\rho\left(\Lambda 
            \left(p\right)^{-1}\right)v_\s(p)] 
\nonumber\\ 
       &=& [\t(p),\,\rho\left(\Lambda\left(p\right){}^{-1}\right)v_\s(p)] 
            \qquad\Rightarrow 
\nonumber\\
  v_\t(p) &=& \rho\left(\Lambda\left(p\right){}^{-1} \right) v_\s(p)\;. 
\label{gt2} 
\eaq

\subsection{Connections}\label{connections} 

There are three ways of defining a connection on a principal bundle \pgm: 
The first way is to define it as a special assignment of a subspace $Q_u$ of 
tangent space $T_uG(M)$ to each point $u \in \pgm$ \cite{kob63}. Another 
way to define a connection is to determine its so-called connection 1-form 
on the principal bundle \pgm\ \cite{kob63}. Besides these well-known 
definitions, there is yet another interesting definition (which in turn 
leads to two other definitions of a connection) based on the notion of the 
tangential group equivariance \cite{bos94}. In our work, we shall employ 
the second definition. 

In this subsection we denote the right action of $G$ on a principal bundle 
\pgm\ by $R$, $u\Lambda =: R_\Lambda(u)$. Let $\fff{g}$ be the Lie 
algebra of $G$ and $A$ be an arbitrary element of $\fff{g}$. Let 
$\exp(tA)$ be the exponential mapping of $A$, which defines a path on $G$. 
{\em The fundamental vector field} $A^+$ on \pgm\ is defined as follows: 
If $f$ is a function on \pgm, then the action of $A^+$ on $f$ is determined 
by $A^+(f) := d/dt|_{t=0} f\circ R_{\exp(tA)}$. Evaluated at a point 
$u\in\pgm$ this definition means $(A^+(f))(u) = d/dt_{t=0}f(u\exp(tA))$. 
Since the right action $R$ acts only in the fibres of \pgm, but not 
between different fibres, $A^+$ is a vector field tangent to the fibres 
$\pi_G^{-1}(u)$ of \pgm. 

A connection 1-form $\o$ on a principal bundle \pgm\ is a 1-form on \pgm\ 
with values in the Lie algebra $\fff{g}$ of $G$, which satisfies the 
following conditions:\\[1mm] 
\eqlabel{concon} 
\begin{eqroman}
\mbox{{ }}\hspace{4mm}1.  $\o(A^+) = A$ for each $A\in\fff{g}$.
\hfill\parbox{1cm}{\baq\label{concona}\eaq}\\[1mm] 
\mbox{{ }}\hspace{7mm}2.  $R^\ast_\Lambda\o = \Lambda^{-1}\o\Lambda$ for each 
$\Lambda\in G$.\hfill\parbox{1cm}{\baq\label{conconb}\eaq}\\[1mm] 
\end{eqroman}%
Here $R^\ast_\Lambda\o$ is the pull-back of $\o$ by the right action 
$R_\Lambda$. Explicitly, $(R^\ast_\Lambda\o)_u(X)=\o_{u\Lambda}
(R_{\Lambda\,\ast}X)=\Lambda^{-1}\o_u(X)\Lambda$ for a tangent 
vector $X$ at the point $u \in G(M)$. $R_{\Lambda\,\ast}X$ is the 
push-forward of the vector $X$ and is defined through its action on a 
function $f$ as follows: $(R_{\Lambda\,\ast}X)_{u\Lambda}(f):=
X\left(f\left(u\Lambda\right)\right)$. 

For the sake of simplicity, from now on we consider only matrix Lie groups, 
so that in (\ref{conconb}) the adjoint mapping $\Lambda^{-1}\o\Lambda$ can 
be read simply as a matrix multiplication. 

In the third section the following theorem (see \cite{kob63}) plays a 
crucial role:\\[2mm] 
{\bf Proposition 1.} Let $H$ be a Lie subgroup of another Lie group $G$ and
let \fff{h} and \fff{g} be the corresponding Lie algebras, where \fff{h} 
is a Lie subalgebra of \fff{g}. Let \phm\ and \pgm\ be principal bundles 
with structure groups given by $H$ and $G$, respectively, and suppose that 
\phm\ is a subbundle of \pgm. 

If there exists a vector subspace \fff{m} of \fff{g}, such that \fff{g} 
can be written as a direct sum (as a vector space) $\fff{g} = \fff{h} 
\oplus \fff{m}$, and if $\Lambda\fff{m}\Lambda^{-1} \subset \fff{m}$ for 
all $\Lambda \in H$, then from every connection $\o$ on \pgm\ we can 
build a connection $\o'$ on \phm\ by restring $\o$ to \phm\ and taking 
its \fff{h}-component.\\[2mm] 
{\it Proof\/}. (See \cite{kob63}.) To verify the first condition 
(\ref{concona}) for the connection let $A$ be an element of \fff{h}. Then 
$A$ is also an element of \fff{g} and thus $\o'(A^+) = \o(A^+) = A$. 

Now let $\Lambda$ be an element of $H$. To verify the second condition 
(\ref{conconb}) we study $R_\Lambda\o'$ at a point $u \in \phm$. Let $X$ 
be a tangent vector at $T_u\phm$ and denote the \fff{h}-component of $\o$ 
by $\varpi$. Then we have 
\baqn 
  R_\Lambda^\ast\o'_u(X) + R_\Lambda^\ast\varpi_u(X) &=& 
  (R_\Lambda^\ast\o)_u(X)  
\\   
  &=& \Lambda^{-1}\o_u(X)\Lambda 
\\
  &=& \Lambda^{-1}(\o'_u(X)+\varpi_u(X))\Lambda 
\\ 
  &=& \Lambda^{-1}\o'_u(X)\Lambda+\Lambda^{-1}\varpi_u(X)\Lambda\;. 
\eaqn 
If we now take the \fff{h}-components only, then we obtain the desired 
result $R_\Lambda^\ast\o' = \Lambda^{-1}\o'\Lambda$ due to the 
assumption $\Lambda^{-1}\fff{m}\Lambda \subset \fff{m}$ in the last line.
$\square$ \\[2mm] 

The following proposition (see \cite{kob63}) provides the central device for 
the desired decomposition of our linear connection (\ref{eqcon3}):\\[2mm] 
{\bf Proposition 2.} Let $(G\times H)(M)$ be the fibre product bundle 
built from \pgm\ and \phm. Let $\o$ be a connection 1-form on 
$(G\times H)(M)$. Then there are unique connection 1-forms $\o_G$ on \pgm\ 
and $\o_H$ on \phm\ such that $\o=p^\ast\o_G + q^\ast\o_H$.\\[2mm] 
Here $p$ and $q$ denote the canonical bundle mappings defined in 
\ref{productbundles}.\\[2mm] 
{\it Proof\/}. Using the trivial projection mappings $p_o : G\times H 
\rightarrow G$ and $q_o : G\times H \rightarrow H$ we can decompose $\o$ 
algebraically as $\o = p_o\o + q_o\o$, where we have used the same letters 
to denote the  Lie group homomorphisms and their Lie algebra homomorphisms. 
To define the connection $\o_G$ on \pgm, let $u$ 
be a point in \pgm\ and $X$ a tangent vector at $u$. Now, take any point 
$v \in\phm$ over the same base point as $u$, that is $\pi_G(u) = \pi_H(v)$. 
Then, $(u,v) \in (G\times H)(M)$ and clearly $X \in T_uG(M) \subset 
T_uG(M)\oplus T_vH(M) = T_{(u,v)}(G\times H)(M)$. We then define 
\[ 
  {\o_G}_u(X) := p_o\o_{(u,v)}(X)\;. 
\] 
To show that this definition does not depend on the particular choice of 
$v$, let $v' = v\Lambda$ be another point in \phm\ over the same base 
point, where $\Lambda \in H$. Since $X$ is a vector in the tangent 
bundle of \pgm, it is not affected by the push-forward of the right 
action $R_\Lambda$ of the other group $H$, which can be written 
on the whole product bundle as $R_{(\Ones,\Lambda)}$. Thus $X = 
{R_{(\Ones,\Lambda)}}_\ast X$, and therefore 
\baqn  
  p_o\o_{(u,v')}(X) &=& p_o\o_{(u,v\Lambda)} 
  \left({R_{(\Ones,\Lambda)}}_\ast X\right) 
 \;=\; 
  p_o\big[(\One,\Lambda)^{-1}\cdot\o_{(u,v)}(X)\cdot (\One,\Lambda)\big] 
\\ 
  &=& p_o\big[ \One p_o\o_{(u,v)}(X)\One 
              +\Lambda^{-1}q_o\o_{(u,v)}(X)\Lambda\big] 
\\ 
  &=& p_o^2\o_{(u,v)}(X) \;=\; p_o\o_{(u,v)}(X)\;.
\eaqn 
Using similar techniques it is easy to verify the two conditions for the 
connection given above. The connection $\o_H$ is defined similarly. It is 
obvious, that $\o = p^\ast\o_G + q^\ast\o_H$. $\square$ \\[2mm] 
The following proposition, which is slightly more general than 
Proposition 6.2.\ on page 81 in \cite{kob63}, is needed for the definition 
of the spin connection.\\[2mm] 
{\bf Proposition 3.} Let $f: \pgm \rightarrow \phm$ be a bundle homomorphism 
such that its Lie group homomorphism $f_o$ induces a Lie algebra 
isomorphism, which we denote by the same letter $f_o$. For every connection 
$\o$ on \phm, there is an unique connection $\o'$ on \pgm\ such that 
$f^\ast\o = f_o\o'$.\\[2mm] 
{\it Proof\/}. Simply define $\o' := f_o^{-1}(f^\ast\o)$, where $f_o^{-1}$ 
is the inverse Lie algebra isomorphism. We first prove the condition 
(\ref{concona}) for connection 1-forms. Let \fff{g} denote the Lie algebra of 
$G$ and let $A\in \fff{g}$. We evaluate $\o'(A^+)$ at a point 
$u \in \pgm$ and obtain: 
\beq\label{help1} 
  \o'_u(A^+) = f_o^{-1}(f^\ast\o)_u(A^+) 
             = f_o^{-1}\o_{f(u)}\left(f_\ast A^+(u)\right)\;. 
\eeq 
Since the bundle homomorphism $f$ is not necessarily a diffeomorphism, it 
is not possible to push-forward the whole fundamental vector field $A^+$ 
from \pgm\ to \phm, but only pointwise. To see what $f_\ast A^+(u)$ means, 
we evaluate this vector at $f(u) \in \phm$ on a smooth function $k$ on 
\phm. Remembering the definition of the bundle homomorphism (which in this 
proof does not necessarily induce the identity mapping $f_M = 1$ on the base 
space) in \ref{bundlemappings} and the definition of the fundamental 
vector fields we obtain 
\baqn 
  f_\ast(A^+(u))_{f(u)}(k) &=& A^+(u)(k\circ f) = 
  \left.\frac{d}{dt}\right|_{t=0}k(f(u\,\exp(tA))\,) 
\\ 
  &=& \left.\frac{d}{dt}\right|_{t=0}k(f(u)\,\exp(t f_o(A))\,) = 
  f_o(A)^+{}_{\! f(u)}(k) \quad \Rightarrow 
\\ 
  f_\ast(A^+(u)) &=& f_o(A)^+{}_{\! f(u)}\;. 
\eaqn 
Using (\ref{concona}) for $\o$, we therefore obtain in (\ref{help1}) the 
desired result 
\[ 
  \o'_u(A^+) = f_o^{-1}(f^\ast\o)_u(A^+) 
             = f_o^{-1}\o_{f(u)}(f_o(A)^+{}_{\! f(u)}) 
  = f_o^{-1}(f_o(A)) = A\;. 
\] 
To verify the second condition (\ref{conconb}), let $\Lambda \in G$ and $X$ a 
tangent vector at $u \in \pgm$. Because of the commutative diagram 
(\ref{bundlemap}) (to be more precise, only the rectangle part of it)  we 
have $f\circ R_\Lambda = R_{f_o(\Lambda)}\circ f$, which is important in the 
following computation 
\baqn 
  \left(R^\ast_{\Lambda}(f_o^{-1}f^\ast\o)\right){}_u(X) &=& 
  f_o^{-1}\o_{f(u\Lambda)}\left((f\circ R_\Lambda)_\ast X\right) 
\\ 
  &=& f_o^{-1}\o_{f(u)f_o(\Lambda)} 
  \left((R_{f_o(\Lambda)}\circ f)_\ast X\right) 
\\ 
  &=& f_o^{-1}\big(R^\ast_{f_o(\Lambda)}\o\big)_{f(u)}(f_\ast X) 
\\ 
  &=& f_o^{-1}\big(f_o(\Lambda^{-1})\o_{f(u)}(f_\ast X)f_o(\Lambda)
  \big) 
\\ 
  &=& \Lambda^{-1}\cdot f_o^{-1}(f^\ast\o)_u(X)\cdot\Lambda\;. 
\eaqn 
This completes the proof. $\square$ \\[2mm] 
Note the important fact that the Lie group homomorphism $f_o$ 
needs not to be an isomorphism, but only its concommitant Lie algebra 
mapping. Thus the group homomorphism can be a twofold mapping, which  
is the case for the universal covering map of the Lorentz group by 
its spin group \sltwoc, see below at \ref{realspingeometry}.

\subsection{Covariant derivatives} 

Given a connection $\o$ on a principal bundle \pgm, we now construct its 
covariant derivative on the associated vector bundle \pvm. For our 
purposes it is convenient to define it by using a local cross section. A 
fairly detailed account of this topic can be found for example in 
\cite{nak90}. In the following, proofs are omitted in order 
to keep the presentation lucid. 

Let $v$ be a vector field in \pvm. From (\ref{lcs3}) we know that $v$ 
can be represented by a $V$-valued function $v_\s$, when a local cross 
section $\s$ is given on \calu\ in \pgm, 
\[ 
  v = [\s,\,v_\s]\;. 
\] 
Let $X$ be a tangent vector at $p \in \calu \subset M$. We then define 
the {\em covariant derivative $\nabla_{\! X}v$ 
of $v$ at $p$ in the direction $X$} as follows 
\beq\label{covariantderivatives} 
  \nabla_{\! X}v := \big[ \s(p),\, X(v_\s(p)) 
                   +\rho\left(\s^\ast\o(X)\right)v_\s(p) \big]\;, 
\eeq 
where $\s^\ast\o$ is the pull-back of the connection 1-form by $\s$, and 
the same symbol $\rho$ is used to denote the Lie algebra homomorphism 
defined by the representation $\rho$ of the structure group $G$ into 
$V$. Thus, $\rho(\s^\ast\o)$ is a 1-form defined on \calu\ with values in 
the Lie algebra of the general linear group of $V$. 

It can be shown that this definition of the covariant derivative is 
equivalent to the other, more common, definition which is directly 
built on the notion of parallel displacements of vector fields, see e.g.\ 
\cite{nak90}. Here we are not going to prove this equivalence, since the 
proof is very technical, but we show that the definition 
(\ref{covariantderivatives}) is independent of the special choice of the 
local cross section $\s$. 

Let $\t$ be another local cross section. Since for each $p \in \calu$ 
the values $\s(p)$ and $\t(p)$ lie in the same fibre over $p$, we can  
find a $G$-valued function $\Lambda$ such that 
\beq\label{help2} 
  \t(p) = \s(p)\Lambda(p)\;. 
\eeq 
Before pulling back $\o$ by $\t$, we evaluate the push-forward $\t_\ast 
X$ of the tangent vector $X$ at the base point $p \in M$. In (\ref{help2}) 
let $\t_0$, $\s_0$, and $\Lambda_0$ denote the values of the corresponding 
fields at this fixed point $p$. Using the Leibniz rule we can calculate the 
action of $\t_\ast X$ on a function $f$ on \pgm\ as follows 
\baq 
  (\t_\ast X)_{\t(p)}f &=& X_p(f(\t)) \;=\; X_p(f(\s\Lambda)) 
\nonumber\\ 
  &=& X_pf(\s\Lambda_0) + X_pf(\s_0\Lambda) 
\nonumber\\ 
  &=& X_pf(R_{\Lambda_0}\circ\s) + X_pf(\t_0\Lambda_0^{-1}\Lambda) \,. 
\label{help3}  
\eaq 
In the last line the first term can be simply expressed as 
\beq\label{help4}  
  X_pf(R_{\Lambda_0}\circ\s) =
    \left(\s_\ast X\right){}_{\s_0}(f\circ R_{\Lambda_0}) 
  = \left({R_{\Lambda_0}}_\ast(\s_\ast X)\right){}_{\t_0}f\;. 
\eeq 
Note that this equation would be incorrect if $\Lambda_0$ was not constant. 
To bring the second term in (\ref{help3}) into a more familiar form, let us 
introduce a curve $\c(t)$ in \mm, which runs through $p=\c(0)$ and 
whose tangent vector $\frac{d}{dt}|_{t=0}\c$ is precisely the vector $X$. 
Then this curve defines a curve in the structure group $G$ via 
$\Lambda(\c(t))$, whose derivative $\frac{d}{dt}|_{t=0}\Lambda(\c(t))$ 
is equal to $X(\Lambda)$. We obtain 
\baq 
  X_pf(\t_0\Lambda_0^{-1}\Lambda) &=& 
  \left.\frac{d}{dt}\right|_{t=0} 
  f(\t_0\Lambda(\c(0))^{-1}\Lambda(\c(t))\,) 
\nonumber\\ 
  &=& \left.\frac{d}{dt}\right|_{t=0} 
      f(\t_0\exp(t\Lambda(\c(0))^{-1}X(\Lambda))\,) 
\label{help5}\\ 
  &=& \big(\Lambda^{-1}X(\Lambda)\big){}^+_{\t_0}f\;, 
\label{help6} 
\eaq 
where we note that in (\ref{help5}) the curve $\Lambda(\c(0))^{-1} 
\Lambda(\c(t))$ runs through $\One \in G$ and has, at this point, the same 
tangent as $\;\exp(t\Lambda(\c(0))^{-1}\frac{d}{dt}|_{t=0} 
\Lambda(\c(t))\,)\;$, which defines a fundamental vector field as in 
(\ref{help6}). With the help of (\ref{help3}) to (\ref{help6}) we obtain 
\beq\label{help7} 
  \t_\ast X = {R_{\Lambda}}_\ast(\s_\ast X) 
             +\big(\Lambda^{-1}X(\Lambda)\big){}^+\;, 
\eeq 
where we have omited the point of evaluation. With this result we can 
relate $\t^\ast\o$ and $\s^\ast\o$ immediately as follows, using the  
characteristic conditions (\ref{concon}) for the connection 1-form $\o$ 
\baq 
  \t^\ast\o(X) &=& \o(\t_\ast X) 
\nonumber\\ 
  &=& \o({R_{\Lambda}}_\ast(\s_\ast X)) + \o((\Lambda^{-1}X(\Lambda))^+) 
\nonumber\\ 
  &=& \Lambda^{-1}\o(\s_\ast X)\Lambda + \Lambda^{-1}X(\Lambda) 
\nonumber\\ 
  &=& \Lambda^{-1}\s^\ast\o(X)\Lambda + \Lambda^{-1}X(\Lambda)\;. 
\label{connectiongauge} 
\eaq 
This formula displays the gauge transformation law of the connection 1-form. 
To see that the covariant derivative defined in (\ref{covariantderivatives}) 
is independent of the choice of the cross section, we use (\ref{gt2}) to 
obtain first 
\baq 
  X(v_\t) &=& X(\rho(\Lambda^{-1})v_\s) 
\nonumber\\ 
  &=& \rho(X(\Lambda^{-1}))v_\s + \rho(\Lambda^ {-1})X(v_\s) 
\nonumber\\ 
  &=& \rho(\Lambda^{-1})\big(\rho(\Lambda X(\Lambda^{-1}))+X(v_\s)\big) 
\nonumber\\ 
  &=& \rho(\Lambda^{-1})\big(-\rho(\Lambda^{-1}X(\Lambda))+X(v_\s)\big)\;. 
\label{help8} 
\eaq 
Finally, if we use the cross section $\t$ in the definition 
(\ref{covariantderivatives}), then from (\ref{connectiongauge}), 
(\ref{help8}), and (\ref{gt2}) we obtain the desired invariant result: 
\baq 
  \nabla_{\! X}v &=& \big[\t(p)\;,\; X(v_\t(p)) 
                   + \rho((\t^\ast\o)(X))v_\t(p)\big] 
\nonumber\\ 
  &=& \big[\s(p)\Lambda(p)\;,\; \rho(\Lambda(p)^{-1}) 
  \big(X(v_\s(p))+\rho(\s^\ast\o(X))v_\s(p)\big)\big] 
\nonumber\\ 
  &=& \big[\s(p)\;,\; X(v_\s(p))+\rho(\s^\ast\o(X))v_\s(p)\big]\;. 
\label{invariance} 
\eaq

             \section{Spin geometry}\label{spingeometry} 

In this section we will discuss the spin structure over a spacetime 
manifold \mm\ with a Lorentzian (pseudo-Riemannian) metric $g_{\m\n}$. 
The spin structure consists of a so-called spin bundle over \mm\ and a 
twofold bundle map from this spin bundle to the special Lorentz bundle 
introduced on p.\ \pageref{lorentz}, and is necessary in order to introduce 
Dirac spinors on a curved manifold. Contrary to this ``real spin structure'', 
its complexified version, which we call ``complex spin structure'', is less 
well-known but is needed for our theory, since, roughly speaking, the linear 
connection determining the covariant derivative of the Dirac spinors is 
complex in our theory. This complexification is, as we will see, rather 
straightforward if written in local representations using tensor components%
\footnote{For example, Ashtekar's formalism of general relativity 
\cite{ash91} employs a complexified theory of general relativity. Although  
spinor fields are considered in this complex geometry, the notion of a 
complex spin structure is absent, because, working with 
local representations, a recourse to the complex spin structure is not 
necessary, since all computations go through by simply allowing the real 
variables to be complex valued.} 
but is not so trivial if the underlying global geometry is taken into 
account.

\subsection{Spin structure of Minkowski spacetime} \label{spinminkowski} 

Let us first consider Minkowski spacetime and its familiar spin structure, 
which will be generalized to the case of an arbitrary curved spacetime in 
the next subsection. 

The special orthochronous Lorentz group \lplus\ was defined in 
(\ref{lorentz}) as 
\beq\label{lorentz2} 
  \lplus = \big\{\Lambda\in\mbox{Mat}(4,\real)| 
           \Lambda^T\eta\Lambda=\eta,\;\det\Lambda=1,\;
           \Lambda^0{}_{\! 0}\ge 1\big\} \;. 
\eeq 
The spin group ``Spin'' of \lplus\ is, by definition, the universal 
covering space of \lplus, which is simply given by $\mbox{Spin}=\sltwoc$. 
The covering map is a twofold Lie group homomorphism, denoted by 
\beq\label{spinmap1} 
  \xi_o: \mbox{Spin}\cong\sltwoc \longrightarrow \lplus\;. 
\eeq 
To make this map $\xi_o$ more explicit, let $x^a$ be the cartesian 
components of a point in flat Minkowski spacetime. We define the following
vector space isomorphism from the Minkowski space to the vector space of 
hermitian $2\times 2$ matrices, 
\beq\label{minkowski} 
  \sim: x=(x^a) \longmapsto \widetilde{x} =  
  \left( \begin{array}{cc} x^0+x^3  & x^1-ix^2 
                        \\ x^1+ix^2 & x^0-x^3 \end{array} \right)\;. 
\eeq 
Then the action of the Lorentz matrix $\xi_o(A)$, where $A\in\sltwoc$, is 
defined by 
\beq\label{spinmap2} 
  \widetilde{\xi_o(A)(x)} := A(\widetilde{x}) A^\dagger\;. 
\eeq 
This homomorphism $\xi_o$ is twofold, since, obviously, both $A$ and $-A$ 
result in the same Lorentz map. 

By definition, a covering map is locally a diffeomorphism. Especially, the 
spin map $\xi_o$ (\ref{spinmap1}) induces an isomorphism between the Lie 
algebras of \sltwoc\ and \lplus, which we shall now state explicitly. 

Let $D_0{}^{\! a}$, $a = 1,2,3$, denote the infinitesimal generators of the 
Lorentz boosts in the three coordinate directions $x^a$. Let $D_a
{}^{\! b}$, $a,b = 1,2,3$ and $a\neq b$, be the infinitesimal generators 
of ordinary rotations of space, whose rotation axes are given by  
$\pm x^c$, where $c$  and the rotation direction $\pm$ are determined by 
demanding that the triplet $(a,b,c)$ should be a positive (negative) 
permutation of $(1,2,3)$; in this way, there are, up to sign, of course only 
three generators of rotations.%
\footnote{For example $D_0{}^{\!1}=\frac{1}{2}\left(\begin{array}{cccc}
0&1&0&0\\1&0&0&0\\0&0&0&0\\0&0&0&0\end{array}\right)$ and $D_3{}^{\!2}=
\frac{1}{2}\left(\begin{array}{cccc}0&0&0&0\\0&0&0&0\\0&0&0&1\\0&0&-1&0
\end{array}\right)$.} 
Altogether, there are 6 independent generators of the Lie algebra \fff{l} 
(\ref{lorentz}) of the Lorentz group. With the help of the canonical 
generators $E_a{}^{\! b}$ of the whole matrix group $\mbox{Mat}(4,\real)$, 
whose entries are 1 in the $a$-th row of the $b$-th column and 0 otherwise, 
that is, 
\beq\label{gl4rgenerators} 
  (E_a{}^{\!b})^c{}_{\!d}=\delta^c{}_{\!a}\,\delta^b{}_{\!d} \;, 
\eeq 
we can write all Lorentz generators systematically as 
\beq\label{generators} 
  D_a{}^{\!b} = 
  \frac{1}{2}\left( E_a{}^{\!b}-\eta_{ac}E_d{}^{\!c}\eta^{db} \right) \;. 
\eeq 
Here, $a$ and $b$ are allowed to take any value 0, 1, 2, 3, but only those  
combinations satisfying $a\neq b$ yield non-zero results. 

On the other hand, the Lie algebra of \sltwoc, denoted by \liesltwoc, 
has the 6 generators (to be more precise, generators of the real algebra) 
given by the three Pauli matrices and their imaginary multiples, 
\beq\label{generators2} 
  \s^1,\;\s^2,\;\s^3,\;i\s^1,\;i\s^2,\;i\s^3\;, 
\eeq 
where the Pauli matrices are given by 
\beq\label{paulimatrices} 
    \sigma^1
   =\left( \begin{array}{cc} 0 & 1 \\ 1 & 0 \end{array} \right)
  , \;\;
    \sigma^2
   =\left( \begin{array}{cc} 0 & -i \\ i & 0 \end{array} \right)
  , \;\;
    \sigma^3
   =\left( \begin{array}{cc} 1 & 0 \\ 0 & -1 \end{array} \right)   
  \;.
\eeq 

The Lie algebra isomorphism induced by $\xi_o$, which we will denote by the 
same letter, is determined through the following relations between the 
generators of the Lie algebras \liesltwoc\ and \fff{l}: 
\beq\label{lieisomorphism}  
\eqalign{ 
  \s^a &\longmapsto 4D_0{}^{\! a},\;\; a=1,2,3 \;, 
\\ 
  i\s^c &\longmapsto 4D_a{}^{\! b},\;\; 
  (a,b,c)=(1,2,3),\,(2,3,1),\,(3,1,2)\;. 
} 
\eeq 
This result can be directly deduced from the definition (\ref{spinmap2}) of 
the spin map $\xi_o$. 

In Minkowski spacetime, Dirac spinors $\psi$ are vector fields with 
values in $\complex^4$, which, however, do not obey the ordinary 
transformation law of vectors. If, for example, the Minkowski spacetime is 
rotated by a Lorentz matrix $\Lambda$, then an ordinary vector 
$X^a$ at a spacetime point $x^a$ is transformed into $\Lambda^a{}_{\! b}
X^b$ at $\Lambda^a{}_{\! b} x^b$. On the other hand, a Dirac spinor $\psi$ 
is transformed according to the following law: Let $A$ be one of the 
two elements in \sltwoc, which is mapped by $\xi_o$ onto $\Lambda$. Then 
$\psi$ transforms as 
\beq\label{spinmap3} 
  \psi \longmapsto \zeta(A)\psi := 
  \left( \begin{array}{cc} A & 0 \\ 0 & (A^\dagger)^{-1} \end{array} \right)
                   \psi \;, 
\eeq 
where 
\beq\label{spinrep1} 
\eqalign{  
  \zeta: \sltwoc &\longrightarrow \quad\glfourc 
\\ 
  \qquad\quad A &\longmapsto {\left( \begin{array}{cc}
                     A & 0 \\ 0 & (A^\dagger)^{-1} \end{array} \right)}\;, 
} 
\eeq
is called a {\em spin representation}. The peculiar feature of the 
transformation law (\ref{spinmap3}) is, that, if the Minkowski spacetime 
is rotated gradually from $0^\circ$ up to $360^\circ$, so that vectors 
retain their original values again, a Dirac spinor $\psi$ will be 
transformed into $-\psi$. This characteristic spin 1/2 behaviour of spinors 
is due to the twofold spin homomorphism $\xi_o$. 

In the following, we need an explicit expression of the Lie algebra 
homomorphism $\zeta\circ\xi_o^{-1}$ from the Lie algebra \fff{l} of the 
Lorentz group into the Lie algebra \lieglfourc\ of \glfourc. This is most 
easily displayed by using the generators of the Lorentz algebra defined 
above, where we only give the result and omit the derivation: 
\beq\label{liehomomorphism} 
   \zeta\circ\xi^{-1}_o:\fff{l}\longrightarrow\fff{gl}(4,\complex)\,,\quad
   D_a{}^{\!b}\longmapsto -\frac{1}{4}\c^b\c_a \;,
\eeq 
where $a \neq b$ is to be understood. Note that the factor $1/4$ arises from 
the inverting of the factor 4 in (\ref{lieisomorphism}).

\subsection{Real spin geometry}\label{realspingeometry}

Let $(M,\,g_{\m\n})$ be the spacetime manifold with a Lorentzian 
(pseudo-Riemannian) metric. A {\em spin structure} is a copy of all the 
structures discussed so far for a flat Minkowski spacetime to the case 
of a non-flat spacetime \mm. Now the spin homomorphism $\xi_o$ is 
replaced by a bundle map $\xi$ and the Dirac spinors become cross sections 
of a spinor bundle \psm. However, these structures can only be defined if 
some global topological conditions are met by the manifold \mm. For the 
sake of simplicity, we will assume that this is the case.%
\footnote{For a detailed discussion of the topological conditions we refer 
to \cite{bau81,law89,ger68,ger70}.} 

Let \lplusm\ be the special Lorentz bundle introduced on p.\ 
\pageref{lorentz}. It is a \lplus\ principal bundle and it is built from 
certain orthonormal tangent frames in $TM$. A spin bundle \spinm\ is a 
principal bundle with structure group $\mbox{Spin}=\sltwoc$, which is 
defined together with the bundle map $\xi:\spinm\rightarrow\lplusm$ via 
the following commutative diagram (cf.\ \ref{bundlemappings}):\\ 
\eqlabel{realspinmap}
\unitlength1cm
\begin{picture}(15,6.5)

\put(2.5,2.5){\makebox(3,0.5)[b]{$\mbox{Spin}(M)$}}
\put(2.5,4.5){\makebox(3,0.5)[b]{$\mbox{Spin}(M)\times\mbox{Spin}$}}
\put(8.5,2.5){\makebox(3,0.5)[b]{\lplusm}}
\put(8.5,4.5){\makebox(3,0.5)[b]{$\lplusm\times\lplus$}}
\put(6,0.5){\makebox(2,0.75)[b]{$M$}}
\put(6,2.9){\makebox(2,0.5)[b]{$\xi$}}
\put(6,4.9){\makebox(2,0.5)[b]{$\xi\times\xi_o$}}
\put(4.7,1.2){\makebox(0.5,0.5)[b]{$\pi$}}
\put(4.1,3.5){\makebox(0.5,0.5)[b]{$R$}}
\put(8.8,1.2){\makebox(0.5,0.5)[b]{$\pi$}}
\put(10.1,3.5){\makebox(0.5,0.5)[b]{$R$}}
\put(14,2.5){\makebox(1,0.5)[r]{(\ref{realspinmap})}}
\put(5.8,4.65){\vector(1,0){2.8}}
\put(5.2,2.65){\vector(1,0){3.9}}
\put(4,4.2){\vector(0,-1){1.3}}
\put(10,4.2){\vector(0,-1){1.3}}
\put(4.1,2.3){\vector(2,-1){2.8}}
\put(9.9,2.3){\vector(-2,-1){2.8}}
\end{picture}
\stepcounter{equation}\\
We call this bundle mapping $\xi$ a {\em spin structure}.%
\footnote{Note that there may be more than one spin structure for a 
given Lorentz bundle, see \cite{bau81}.} 
According to (\ref{realspinmap}), the bundle map $\xi$ is compatible with 
the spin map $\xi_o$ (\ref{spinmap1}), that is, 
\beq\label{realspinmap2} 
  \xi(uA) = \xi(u)\xi_o(A),\quad u\in\spinm,\quad A\in\sltwoc\;. 
\eeq 
Since the spin structure $\xi$ can be replaced by the spin map $\xi_o$ in 
each fibre of \spinm, $\xi$ is also a twofold covering map and therefore 
surjective. 

Once such a spin structure (\ref{realspinmap}) is given, we can define the 
{\em spinor bundle} \psm, which is the $\complex^4$ vector bundle associated 
to \spinm, the representation of the structure group $\mbox{Spin}$ precisely 
being the spin representation $\zeta$ of (\ref{spinrep1}). Thus, 
\beq\label{spinorbundle} 
  \psm = \spinm \times_\zeta\complex^4\;, 
\eeq 
see (\ref{avb3}). Cross sections into this associated vector bundle \psm\ 
are called {\em Dirac spinors\/}, see e.g.\ \cite{ble81}. 

We shall now define the covariant spinor derivative built from a Lorentzian  
connection compatible with the metric $g_{\m\n}$, cf.\ (\ref{metricity}). 
According to (\ref{covariantderivatives}) a covariant spinor derivative can 
be constructed from a connection on the spin bundle \spinm, 
which is called a {\em spin connection}.%
\footnote{We remark that the converse statement is not true: A covariant 
derivative on an associated vector bundle can be defined without the 
notion of connection 1-forms, and there might exist a covariant 
derivative, which can not be derived from a connection 1-form on the
principal bundle via (\ref{covariantderivatives}), see \cite{gre72}.} 
So the only task is to obtain a spin connection from a Lorentzian connection. 
But from {\bf Proposition 3} it follows that this is indeed possible: Since 
the Lie algebra homomorphism $\xi_o$ is actually an isomorphism, see 
(\ref{lieisomorphism}), the spin structure $\xi$ in (\ref{realspinmap}) 
yields a spin connection $\o_s$ starting from {\em any} metric connection 
1-form $\o_m$ defined on the Lorentz bundle \lplusm, 
\beq\label{realspinconnection} 
  \o_s = \xi_o^{-1}(\xi^\ast\o_m) \;. 
\eeq 

\subsection{Covariant spinor derivative}\label{covsd} 

In the following we shall study two questions: First, how does a metric 
connection 1-form $\o_m$ define a Lorentzian connection $\Con{a}\m{b}$ on 
the spacetime manifold \mm? Secondly, how does this connection yield the 
covariant spinor derivative (\ref{spinorderivative})? 

According to our definition (\ref{covariantderivatives}) of the covariant 
derivative, we first of all need a local cross section $\s$ in the Lorentz 
bundle \lplusm, that is, an orthonormal tetrad field $(e_a{}^{\!\m})$. 
Then, since the spin structure is a surjective mapping, there exists a local 
cross section \hats\ in the spin bundle \spinm, such that 
\beq\label{spinsection} 
  \xi(\hats) = \s \;. 
\eeq 
In fact there are exactly two such cross sections in \spinm, namely 
\hats\ and $\hats(-\One)$. The metric connection 1-form $\o_m$ on \lplus\ 
can be pulled back to the base manifold \mm\ via $\s$ yielding a 
matrix-valued 1-form, whose components are defined as
\beq\label{connection1} 
  \Con{a}\m{b}dx^\m := (\s^\ast\o_m)^a{}_{\! b} \;.
\eeq 
Since these components belong to a matrix of the Lie algebra \fff{l} 
of the Lorentz group, they satisfy, according to (\ref{lorentz}), 
\beq\label{lorentzcondition} 
  \Con{c}\m{b}\eta_{ca}+\eta_{bc}\Con{c}\m{a} = 
  \con{a}\m{b}         +         \con{b}\m{a} = 0 \;,  
\eeq 
which is precisely the metricity condition (\ref{metricity}). Instead of 
giving the matrix components of the connection like in (\ref{connection1}) 
we can of course give the full \fff{l}-valued connection 1-form on \mm\ by 
using the canonical generators $E^a{}_{\! b}$, see (\ref{gl4rgenerators}), 
and the Lorentz generators (\ref{generators}) as follows 
\beq\label{connectionfull} 
\eqalign{ 
    \sigma^\ast\omega_m
  &= 
    \Con{a}{\mu}{b}\,dx^\mu\cdot E_a{}^{\!b}
   =\frac{1}{2}(\Con{a}{\mu}{b}-\Gamma_{b\mu}{}^{\!a})\cdot E_a{}^{\!b}\,dx^\mu
  \nonumber\\
  &= 
    \Con{a}{\mu}{b}\cdot\frac{1}{2}
    \left( E_a{}^{\!b}-\eta_{ac}E_d{}^{\!c}\eta^{db} \right)\,dx^\mu
  \nonumber\\
  &= 
    \Con{a}{\mu}{b}\cdot D_a{}^{\!b}\,dx^\mu \;.
} 
\eeq 
This expression will be used below to derive the covariant spinor derivative. 

To define the covariant derivative of a vector field $v = v^\m \partial_\m$ 
in the tangent bundle $TM$, we represent it by a $\real^4$-valued function 
$v^a$, see (\ref{lcs3}), 
\beq\label{lcs4}             
  v = [\s,\,v^a] = [(e_a{}^{\!\n}\partial_\n),\,v^a] \;, 
\eeq 
so that $v^a$ are merely the anholonomic tetrad components of the vector 
field $v$. With the definition (\ref{covariantderivatives}), the covariant 
derivative of $v$ now reads 
\beq\label{connection2} 
  \nabla_{\!\m}v = [(e_a{}^{\!\n}\partial_\n)\;,\;
                    \partial_\m v^a+\Con{a}\m{b}v^b] \;, 
\eeq 
which is commonly and more loosely written as
\beq\label{connetion3} 
  \nabla_{\!\m}v^a = \partial_\m v^a+\Con{a}\m{b}v^b \;. 
\eeq 
This is exactly the usual covariant derivative of a vector field in 
orthonormal, anholonomic components, showing together with 
(\ref{lorentzcondition}) that the metric connection 1-form $\o_m$ indeed 
defines a Lorentzian connection (in anholonomic components) via 
(\ref{connection1}). 

We now define the covariant spinor derivative with the help of this 
metric connection $\o_m$. In close analogy to the case of the vector 
field $v$ before, we first represent a Dirac field $\psi$ by a 
$\complex^4$-valued function $\psi_{\hat{\s}}$ via the cross section \hats\ 
(\ref{spinsection}) 
\beq\label{lcs5} 
  \psi = [\hats,\,\psi_{\hat{\s}}] \;. 
\eeq 
We then employ the spin connection (\ref{realspinconnection}) to define the 
covariant spinor derivative 
\beq\label{sd1} 
  \nabla_{\!\mu}\psi = 
    \left[ \hats\;,\; \partial_\mu\psi_{\hat{\s}}
          +\zeta(\hat{\s}^\ast\o_s)(\partial_\m)\psi_{\hat{\s}}
    \right]\;, 
\eeq 
where 
\beq\label{sd2} 
\eqalign{ 
  \zeta(\hat{\s}^\ast\o_s)(\partial_\m)\psi_{\hat{\s}} &= 
  \zeta((\hats^\ast\xi^{-1}_o(\xi^\ast\o_m))(\partial_\m))\psi_{\hat{\s}} 
\\
  &= \zeta\circ\xi_o^{-1}(\xi(\hats)^\ast\o_m(\partial_\m))\psi_{\hat{\s}} 
\\
  &= \zeta\circ\xi_o^{-1}(\s^\ast\o_m(\partial_\m))\psi_{\hat{\s}} 
\\ 
  &= \zeta\circ\xi_o^{-1}(\Con{a}\m{b}D_a{}^{\!b})\psi_{\hat{\s}} 
\\ 
  &= -\frac{1}{4}\c^b\c_a\Con{a}\m{b}\psi_{\hat{\s}} \;. 
} 
\eeq 
(This is derived with the help of (\ref{connectionfull}) and 
(\ref{liehomomorphism}).) The result may now be written as 
\beq\label{sd3} 
  \nabla_{\!\m}\psi_{\hat{\s}} = \partial_\mu\psi_{\hat{\s}} 
                    -\frac{1}{4}\Con{a}\m{b}\c^b\c_a\,\psi_{\hat{\s}}\;.
\eeq
Usually, the subscript \hats, denoting the special cross section used to 
represent $\psi$ as a $\complex^4$-valued function, is skipped. We see 
that this covariant derivative is precisely the one given in 
(\ref{spinorderivative}). 

In summary, we have exploited the spin structure (\ref{realspinmap}) to 
obtain a covariant spinor derivative out of an arbitrary metric connection 
1-form $\o_m$. Note that we have never spoken of the Levi--Civita connection. 
In fact, the metric connection may have non-vanishing torsion.

     \subsection{Complex spin geometry}\label{complexspingeometry} 

We now introduce the notion of complex spin geometry. This complex 
extension is necessary in order to accomodate the real spin structure to the 
complex tangent bundle geometry used in our theory.

\subsubsection{Algebraic preliminaries} 

It is well-known that the full (real) Lorentz group $L$, 
\beq\label{lorentzfull} 
  L := \big\{\Lambda\in\glfourr |\, \Lambda^t\eta\Lambda = \eta\big\} \;, 
\eeq 
consists of four 
topological components characterized by the sign of the determinant and 
the sign of the component $\Lambda^0{}_{\!0}$. The complex Lorentz group 
\cl\ is defined analogously 
\beq\label{lorentzcomplexfull} 
  \complex L := \big\{\Lambda\in\glfourc |\,\Lambda^t\eta\Lambda = \eta\big\} 
  \;. 
\eeq 
Contrary to $L$ however, \cl\ consists of only two components, because those 
components of $L$, which are separated by the sign of $\Lambda^0{}_{\!0}$ 
are now connected by a path over complex Lorentz matrices. The two components 
of \cl\ are characterized by the sign of the determinant only, see for a 
detailed discussion \cite{str64}. The special complex Lorentz group \clplus\ 
is the component containing $\One$, 
\beq\label{lorentzcomplex} 
  \clplus := \big\{\Lambda\in\glfourc | 
  \Lambda^t\eta\Lambda = \eta\,,\;\det(\Lambda)=1\big\} \;. 
\eeq 
Contrary to the real case, where \lplus\ is of course not isomorphic to 
$\mbox{SO}(4)$, the special complex Lorentz group \clplus\ is isomorphic to 
the complex special orthogonal group $\complex\mbox{SO}(4)$, which is 
defined analogously to the real case, $\complex\mbox{SO}(4) = \{ 
\Lambda' \in \glfourc |\,{\Lambda'}^T\Lambda' = \One\}$. The isomorphism 
is given by $\Lambda' = W^{-1}\Lambda W$, where $W =\mbox{diag}(i,1,1,1)$ 
is simply the Wick-rotation. 

Since \clplus\ is a complex 6-dimensional Lie group, it has twice as much 
real dimensions as the real Lorentz group \lplus. Correspondingly, the spin 
group of \clplus\ has also 12 real dimensions and is given by \cite{str64} 
\beq\label{complexspingroup} 
  \cspin := \sltwoc\times\sltwoc \;. 
\eeq 
The twofold spin map will be denoted by the same letter $\xi_o$ as in the 
real  case (\ref{spinmap1}), 
\beq\label{cspinmap1} 
  \xi_o:\cspin \cong \sltwoc\times\sltwoc \longrightarrow \clplus \;, 
\eeq 
and is now defined by%
\footnote{Our convention differs from that used in \cite{str64}.} 
(compare (\ref{spinmap2})) 
\beq\label{cspinmap2} 
  \widetilde{\xi_o((A,B))(x)} := A(\widetilde{x})B^\dagger \;, 
\eeq
where $(A,B) \in\sltwoc\times\sltwoc$. That (\ref{cspinmap2}) indeed defines 
a complex Lorentz matrix can be seen as follows: The Lorentz group is 
characterized by the transformation invariance of the metric measure, 
which is given by (see (\ref{minkowski})) 
\beq\label{metricmeasure} 
  x^T\eta x = \det(\widetilde{x})\;. 
\eeq 
This yields the desired invariance property%
\footnote{To prove that $\xi_o((A,B))$ has positive determinant, as is 
required by the definition of \clplus, let us notice $\xi_o((\Ones,\Ones)) = 
\Ones$. This means that there is one point in \cspin\, which is mapped into 
\clplus. Since \cspin\ is definitely connected (because \sltwoc\ is 
connected), $\xi_o$ maps the whole domain group \cspin\ into \clplus, 
proving the assertion. That $\xi_o$ is actually surjective 
can be seen using familiar topological arguments.} 
\beq\label{help10} 
  (\xi_o((A,B))(x))^T\eta(\xi_o((A,B))(x)) 
  = \det(A(\widetilde{x})B^\dagger) = 1\cdot\det(\widetilde{x})\cdot 1 \;.  
\eeq 

Instead of the 6 Lorentz generators in (\ref{generators}) there are now 
12 generators of \clplus\ given by 
\beq\label{complexgenerators} 
  D_a{}^{\!b} \quad\mbox{and}\quad iD_a{}^{\!b} \;. 
\eeq 
There are also 12 generators of the complex spin group \cspin, and these are 
mapped onto the Lorentz generators by the Lie algebra isomorphism $\xi_o$ 
as follows: 
\beq\label{complexlieisomorphism} 
\eqalign{ 
  (\s^a, \s^a) &\longmapsto 4D_0{}^{\! a},\;\; a=1,2,3 \;, 
\\ 
  (i\s^\c, i\s^c) &\longmapsto 4D_a{}^{\! b},\;\; 
  (a,b,c)=(1,2,3),\,(2,3,1),\,(3,1,2)\;, 
\\ 
  (\s^c,-\s^c) &\longmapsto -4iD_a{}^{\! b},\;\; 
  (a,b,c)=(1,2,3),\,(2,3,1),\,(3,1,2)\;, 
\\ 
  (i\s^a,-i\s^a) &\longmapsto 4iD_0{}^{\! a},\;\; a=1,2,3 \;. 
} 
\eeq 
The complex spin representation $\zeta$ of \cspin\ into \glfourc\ is defined 
as 
\beq\label{cspinrep1} 
\eqalign{  
  \zeta: \sltwoc\times\sltwoc &\longrightarrow \quad\glfourc 
\\ 
  \qquad\qquad (A,B) &\longmapsto {\left( \begin{array}{cc}
                     A & 0 \\ 0 & (B^\dagger)^{-1} \end{array} \right)}\;, 
} 
\eeq
where we have used the same letter $\zeta$ as in the real spin 
representation (\ref{spinrep1}). According to this complex representation, 
Dirac spinors transform as 
\beq\label{cspinmap3} 
  \psi \longmapsto \left( \begin{array}{cc}
                     A & 0 \\ 0 & (B^\dagger)^{-1} \end{array} \right)
                   \psi \;. 
\eeq 
If we now look at both the complexified Lie group homomorphisms $\xi_o$ and 
the complex spin representation $\zeta$, then these two maps are constructed 
in such a pleasant way that the resultant Lie algebra homomorphism 
$\zeta\xi_o^{-1}$ from the complex Lorentz Lie algebra 
$\complex\otimes\fff{l}$ into \lieglfourc\ is the same 
as the real homomorphism (\ref{liehomomorphism}), that is, 
\beq\label{cliehomo} 
  \zeta\circ\xi_o^{-1}: \complex\otimes\fff{l} \longrightarrow 
  \lieglfourc\,,\quad D_a{}^{\!b} \longmapsto -\frac{1}{4}\c^b\c_a \;. 
\eeq 
So, especially, 
\beq\label{cliehomo2} 
  iD_a{}^{\!b} \longmapsto -i\frac{1}{4}\c^b\c_a \;. 
\eeq

\subsubsection{Bundle analogue} 

Precisely as in the real spin geometry, the complex spin geometry is an 
exact translation of the complex spin algebra into the framework of fibre 
bundles. Instead of the real Lorentz bundle \lplusm, we now have its 
complexified version, namely a complex Lorentz bundle \clplusm, which not 
only contains real orthonormal tangent frames of $TM$, but also complex 
orthonormal tangent frames of \ctm. The structure group is \clplus. 

The {\em complex spin structure} will be denoted by the same letter $\xi$ as 
in the real case and consists of a complex spin bundle \cspinm\ with 
structure group \cspin\ together with a twofold covering bundle mapping 
$\xi$ defined by the following commutative diagram, analogous to 
(\ref{realspinmap}):\\ 
\eqlabel{complexspinmap}
\unitlength1cm
\begin{picture}(15,6.5)

\put(2.5,2.5){\makebox(3,0.5)[b]{\cspinm}}
\put(2.5,4.5){\makebox(3,0.5)[b]{$\cspinm\times\cspin$}}
\put(8.5,2.5){\makebox(3,0.5)[b]{\clplusm}}
\put(8.5,4.5){\makebox(3,0.5)[b]{$\clplusm\times\clplus$}}
\put(6,0.5){\makebox(2,0.75)[b]{$M$}}
\put(6,2.9){\makebox(2,0.5)[b]{$\xi$}}
\put(6,4.9){\makebox(2,0.5)[b]{$\xi\times\xi_o$}}
\put(4.7,1.2){\makebox(0.5,0.5)[b]{$\pi$}}
\put(4.1,3.5){\makebox(0.5,0.5)[b]{$R$}}
\put(8.8,1.2){\makebox(0.5,0.5)[b]{$\pi$}}
\put(10.1,3.5){\makebox(0.5,0.5)[b]{$R$}}
\put(14,2.5){\makebox(1,0.5)[r]{(\ref{complexspinmap})}}
\put(5.8,4.65){\vector(1,0){2.7}}
\put(5.2,2.65){\vector(1,0){3.8}}
\put(4,4.2){\vector(0,-1){1.3}}
\put(10,4.2){\vector(0,-1){1.3}}
\put(4.1,2.3){\vector(2,-1){2.8}}
\put(9.9,2.3){\vector(-2,-1){2.8}}
\end{picture} 
\stepcounter{equation}\\ 
Exactly as in (\ref{spinorbundle}) the {\em complex spinor bundle}, which 
we denote by the same symbol \psm, is defined by 
\beq\label{cspinorbundle} 
  \psm = \cspinm\times_\zeta\complex^4 \;, 
\eeq 
where $\zeta$ is the complex spin representation of (\ref{cspinrep1}). 

Proceeding as on p.\ \pageref{spinorbundle}, any complex metric connection 
1-form $\o_m$ on \clplusm\ defines an unique complex spin connection 
$\o_s$ via (\ref{realspinconnection}). Using a complex tetrad field as 
local cross section into \clplusm, this complex spin connection defines 
precisely the same covariant spinor derivative as in (\ref{sd3}), since 
the Lie algebra homomorphism $\zeta\circ\xi_o^{-1}$ in (\ref{cliehomo}) 
has exactly the same structure as in (\ref{liehomomorphism}). Because of  
this formal resemblance of the real and the complex spin geometry, we may 
speak of a ``natural'' extension of the real spin geometry to the complex 
case.%
\footnote{I could not find any textbook, where the complex 
extension of the spin geometry is discussed in such a great detail as here.} 

      \section{Fibre bundle background} 
      \label{fibrebackground} 

\subsection{Group structure} 

As it was outlined in the introduction to this chapter, we first construct a 
diagram of Lie group homomorphisms, which will then be copied into the 
framework of bundle mappings. Consider now the following diagram:\\ 
\eqlabel{groupstructure} 
\unitlength1cm
\begin{picture}(15,3.5)

\put( 0.7,2.2){\makebox(3.0,0.5)[b]{$\cspin\times\complex^\times$}}
\put( 6.0,2.2){\makebox(2.0,0.5)[b]{$\clplus\times\complex^\times$}}
\put( 9.5,2.2){\makebox(1.0,0.5)[b]{$\fff{G}$}}
\put(12.0,2.2){\makebox(1.5,0.5)[b]{\glfourc}}
\put( 4.0,2.5){\makebox(1.5,0.5)[b]{$\xi_o\! \times\!\mbox{id}$}}
\put( 8.1,2.5){\makebox(1.5,0.5)[b]{$\theta_o$}}
\put(10.4,2.5){\makebox(1.5,0.5)[b]{$j_o$}}
\put( 3.5,2.3){\vector( 1, 0){2.3}}
\put( 8.0,2.3){\vector( 1, 0){1.6}}
\put(10.5,2.3){\vector( 1, 0){1.4}}
\put( 1.7,0.5){\makebox(1.0,0.5)[b]{\glfourc}}
\put( 1.2,1.3){\makebox(1.0,0.5)[b]{$\zeta_c$}}
\put( 2.2,2.0){\vector( 0,-1){1.0}}
\put(14.0,1.5){\makebox(1.0,0.5)[r]{(\ref{groupstructure})}}
\end{picture}
\stepcounter{equation} 
In the following, we shall explain the details of this diagram: 
First, the complex spin group \cspin\ and the complex Lorentz group 
\clplus, together with the spin mapping $\xi_o$, were defined in the 
foregoing section. The group of invertible elements of $\complex$ is the 
abelian multiplicative group of non-zero complex numbers and is isomorphic 
to \glonec. It was denoted in the above diagram by 
\beq\label{cx}  
  \ctimes := \complex\setminus\{0\} \cong \glonec \;. 
\eeq 
If \ctimes\ is restricted to unit elements, one gets of course \uone, which 
will become the electromagnetic gauge group later on. The reason, why 
\ctimes\ instead of \uone\ is considered here, is that a general 
complex linear connection might posses a trace part $\Con{a}\m{a}$, which 
is not purely imaginary as in the field equation (\ref{eqcon3}), 
and thus is not an \uone\ potential, but a \ctimes\ potential, see below 
(\ref{extended6}). Note that we must explain the geometry of our extended 
spinor derivative in (\ref{esd}) {\em before} we take into account the field 
equations, since otherwise, the spinor derivative (\ref{esd}) and the 
Lagrangian ${\cal L}_m$ (\ref{lagrangian}) based upon this derivative are 
not defined mathematically. 

The representation $\zeta_c$ in (\ref{groupstructure}) of the product 
group $\cspin\times\ctimes$ into \glfourc\ will be called the {\em extended 
spin representation} and will be needed below to construct the spinor 
bundle on the basis of the spin bundle. The representation $\zeta_c$ is 
defined in the following way, 
\beq\label{espinrep} 
\eqalign{ 
  \zeta_c : \cspin\times\ctimes &\longrightarrow \quad\glfourc
\\
  \qquad\left(\,(A,B)\,,\,c\right) &\longmapsto 
  \zeta((A,B))\!\cdot\! c^{-1}\;.
} 
\eeq 
Here we have written the complex spin group as $\sltwoc\times\sltwoc$, 
see (\ref{complexspingroup}). The choice $c^{-1}$ for the representation of 
\ctimes\ is necessary in (\ref{espinrep}) in order to obtain the spinor 
derivative (\ref{esd}) and corresponds to the negative charge of the 
spinor. Other possible representations $c^\vare$, $\vare \in \real$, 
correspond to spinors with electric charge $\vare e$, see below, 
(\ref{gtrans5}). In subsesction \ref{extended}, we will need the Lie algebra 
homomorphism of (\ref{espinrep}), which is simply given by 
\beq\label{espinrep2} 
  \zeta_c((\Lambda,\Lambda'),\,\l)) = \zeta((\Lambda,\Lambda'))-\l\One 
  \;,\quad
  ((\Lambda,\Lambda'),\l) \in 
  (\liesltwoc\!\times\!\liesltwoc)\times\complex \;. 
\eeq 

It remains to explain $\theta_o$, $\fff{G}$, and $j_o$ in the diagram 
(\ref{groupstructure}). The Lie group homomorphism $\theta_o$ is defined by 
the following rule: 
\beq\label{theta} 
\eqalign{ 
  \theta_o: \clplus\times\ctimes &\longrightarrow \glfourc 
\\
  \qquad\quad (\Lambda \, ,\, c) &\longmapsto \quad \Lambda c \;.
} 
\eeq 
The Lie group $\fff{G}$ is the image of $\theta_o$, 
\beq\label{groupg} 
  \fff{G} := \theta_o(\clplus\times\ctimes) 
           = \big\{\Lambda c |\,\Lambda\in\clplus,\,c\in\ctimes\big\} \;, 
\eeq 
and $j_o$ denotes the canonical inclusion of this group $\fff{G}$ into 
the full \glfourc. Thus, by the definition of $\fff{G}$, $\theta_o$ in the 
diagram (\ref{groupstructure}) is a surjective map. Moreover, it induces a 
Lie algebra isomorphism: The Lie algebra of $\clplus\times\ctimes$ is clearly 
the cartesian product $\complex\otimes\fff{l}\times\complex$, where 
\fff{l} is the Lie algebra of \lplus\ defined in (\ref{lorentz}).  
Let $(A,\l)$ be an arbitrary element of this Lie algebra. Then it is 
mapped by $\theta_o$ (to be more precise, by its differential at the 
unit element $(\One,1)$) to 
\baqn 
  \theta_o((A,l)) &=& {\theta_o}_\ast(A,\l) 
  \;=\; 
  \left.\frac{d}{dt}\right|_{t=0}\theta_o(\exp(t(A,\l))\,) 
\\ &=& 
  \left.\frac{d}{dt}\right|_{t=0}\theta_o(\exp(tA,t\l)\,) 
  \;=\;  
  \left.\frac{d}{dt}\right|_{t=0}\theta_o(\,(\exp(tA),\exp(t\l))\,) 
\\ &=& 
  \left.\frac{d}{dt}\right|_{t=0}\big[\exp(tA)\cdot\exp(t\l)\big]  
\\ &=& 
  \left.\frac{d}{dt}\right|_{t=0}\big[\exp(tA)\cdot 1+\One\cdot\exp(t\l)\big]
\\ &=&
  A + \l\One \;. 
\eaqn 
Since the elements of the Lorentz Lie algebra $\complex\otimes\fff{l}$ 
do not contain any diagonal 
elements but only off-diagonal ones, the sum in the last line is direct.%
\footnote{If $A+\l\Ones = \l'\Ones$, then $A = 0$, and if $A+\l\Ones = A'$, 
then $\l = 0$.} 
Therefore, the Lie algebra of $\fff{G}$, denoted henceforth by $\fff{g}$, is 
the direct sum 
\beq\label{algebrag} 
  \fff{g} = \complex\otimes\fff{l} \oplus \complex\One \;, 
\eeq 
and $\theta_o$ is obviously an isomorphism between the two Lie algebras, 
\beq\label{theta1} 
  \theta_o : \complex\otimes\fff{l}\times\complex 
               \stackrel{\cong}{\longrightarrow} 
             \complex\otimes\fff{l}\oplus\complex\One \;. 
\eeq 
This simple but subtle isomorphism property of $\theta_o$ will become crucial 
for the construction of the extended spin connection, see (\ref{omegalc}). 
We remark that, commonly, the Lie algebra of a product group such as  
$\complex\otimes\fff{l}\times\complex$ is already identified with 
$\complex\otimes\fff{l}\oplus\complex\One$. Thus, when the Lie algebra 
isomorphism $\theta_o$ is considered without its underlying Lie group mapping 
$\theta_o$, (\ref{theta1}) rather becomes a tautology. \eqlabel{tautology} 

We further remark that the Lie group homomorphism $\theta_o$ is a twofold 
map.%
\footnote{Due to the property of the Lorentz matrices we get 
from $\Lambda c=\Lambda' c'$ first the equality $c^2\eta=(\Lambda c)^T\eta
(\Lambda c)=(\Lambda' c')^T\eta(\Lambda' c')=c'^2 \eta\;\Leftrightarrow\;
c'=\pm c$. This yields $\Lambda =\pm \Lambda'$ and thus $(\Lambda,c)=
(\pm\Lambda',\pm c')$.} 

\subsection{Bundle structure}\label{bundle} 

Having explained the basic group structure (\ref{groupstructure}), we now 
construct its exact translation to the framework of fibre bundles. Thereby 
the Lie groups become the structure groups of principal bundles, and the 
Lie group homomorphisms become the accompanying group homomorphisms of 
bundle mappings (cf.\ \ref{bundlemappings}). 

The main fibre bundle structure of our theory can be summarized in the 
following ``copy-diagram'' of (\ref{groupstructure})\\ 
\eqlabel{bundlestructure} 
\unitlength1cm
\begin{picture}(15,3.5)

\put( 0.7,2.2){\makebox(3.0,0.5)[b]{$(\cspin\!\times\!\ctimes)(M)$}}
\put( 6.0,2.2){\makebox(2.0,0.5)[b]{$(\clplus\!\times\!\ctimes)(M)$}}
\put(10.0,2.2){\makebox(1.0,0.5)[b]{$\fff{G}(M)$}}
\put(12.5,2.2){\makebox(1.5,0.5)[b]{$F_c(M)$}}

\put( 4.0,2.5){\makebox(1.5,0.5)[b]{$\xi\!\times\!\mbox{id}$}}
\put( 8.5,2.5){\makebox(1.5,0.5)[b]{$\theta$}}
\put(11.0,2.5){\makebox(1.5,0.5)[b]{$j$}}

\put( 4.0,2.3){\vector( 1, 0){1.5}}
\put( 8.5,2.3){\vector( 1, 0){1.3}}
\put(11.2,2.3){\vector( 1, 0){1.3}}

\put( 1.7,0.5){\makebox(1.0,0.5)[b]{$S_c(M)$}}
\put( 1.2,1.3){\makebox(1.0,0.5)[b]{($\ast$)}}
\put( 2.2,2.0){\vector( 0,-1){1.0}}

\put(14.0,1.5){\makebox(1.0,0.5)[r]{(\ref{bundlestructure})}}
\end{picture}
\stepcounter{equation}\\
Let us first explain the various fibre bundles in this diagram: First, 
define the following trivial \ctimes\ principal bundle \cxm\ by 
\beq\label{bundle1} 
  \cxm := M \times \ctimes \;. 
\eeq 
Then $(\cspin\!\times\!\ctimes)(M)$ and $(\clplus\!\times\!\ctimes)(M)$ 
are fibre product bundles of \cspinm\ and \cxm, and of \clplus\ and \cxm, 
respectively, see \ref{productbundles}. The bundle on the left-hand side of 
(\ref{bundlestructure}), $F_c(M)$, is the complex frame bundle defined on 
p.\ \pageref{con}. The fibre bundle $\fff{G}(M)$ is a special subset of this 
complex frame bundle containing only tangent frames of the special form 
$(c\cdot e_a{}^{\!\m}\partial_\m)$. Here $(e_a{}^{\!\m}\partial_\m)$ is a 
complex orthonormal frame of \clplusm\ and $c$ is a non-zero complex number, 
thus, 
\beq\label{bundleg} 
  \fff{G}(M) := \big\{(c\cdot e_a{}^{\!\m}\partial_\m)\; | \; 
  (e_a{}^{\!\m}\partial_\m)\in\clplusm \, , \;\; c\in\ctimes \big\} \;. 
\eeq 
Then, $\fff{G}(M)$ is obviously a $\fff{G}$ principal bundle, where the right 
action of the group $\fff{G}$ is given by 
\beq\label{bundleg2} 
  (c\cdot e_a{}^{\!\m}\partial_\m)(\Lambda c') := 
  (c'c\cdot e_b{}^{\!\m}\Lambda^b{}_{\!a}\partial_\m) \;. 
\eeq 
It is easy to show that this action is free and that the other axioms for 
the principal bundles in \ref{principalbundles} are fulfilled. 

The fibre bundle $S_c(M)$ at the bottom of (\ref{bundlestructure}) is not 
a principal bundle, but is the $\complex^4$ vector bundle 
\beq\label{bundle2} 
  S_c(M) := (\cspin\!\times\!\ctimes)(M)\times_{\zeta_c}\complex^4 
\eeq 
associated to the product bundle $(\cspin\!\times\!\ctimes)(M)$ via 
the extended spin representation $\zeta_o$ in (\ref{espinrep}), see 
\ref{avb}. We call $S_c(M)$ the {\em extended spinor bundle\/}. 

We shall now explain the bundle mappings between the principal bundles of 
(\ref{bundlestructure}). Remembering that $\xi$ in (\ref{bundlestructure}) 
denotes the complex spin structure as defined in (\ref{complexspinmap}), 
the bundle mapping $\xi\times \mbox{id}$ is simply defined as follows: An 
element $(u,v)$ of $(\cspin\!\times\!\ctimes)(M)$ is mapped to $(\xi(u),v)$ 
in $(\clplus\!\times\!\ctimes)(M)$. Because of the trivial relation 
\beq\label{bundle3} 
  (\xi\!\times\!\mbox{id})(u\Lambda,vc) = (\xi(u)\xi_o(\Lambda),vc) 
  = \big((\xi\!\times\!\mbox{id})(u,v)\big) 
    ((\xi_o\!\times\!\mbox{id})(\Lambda,c)\,) \;, 
\eeq 
where $(\Lambda,c) \in \cspin\times\ctimes$, we see that 
$(\xi\!\times\!\mbox{id}\, , \, \xi_o\!\times\!\mbox{id})$ is a bundle 
mapping as explained in \ref{bundlemappings}. 

To explain the bundle map $\theta$, we denote an element of $(\clplus\! 
\times\!\ctimes)(M)$ by $(e_a{}^{\!\m}\partial_\m,c)$, where $c$ is the 
\ctimes-component of the respective element in \cxm\ over the same base point 
as the orthonormal frame $(e_a{}^{\!\m}\partial_\m)$. Note that such a 
simplified notation is possible here because \cxm\ is a trivial bundle. 
Then, $\theta$ can be defined as follows 
\beq\label{bundle4} 
\eqalign{ 
  \theta: (\clplus\!\times\!\ctimes)(M) &\longrightarrow \; G(M) 
\\
  \qquad\quad\left((e_a{}^{\!\mu}\partial_\mu)\,,\:c \,\right) &\longmapsto 
  (c\,e_a{}^{\!\mu}\partial_\mu) \;. 
} 
\eeq 
It is straightforward to show that $(\theta,\theta_o)$ (cf.\ (\ref{theta})) 
indeed defines a bundle mapping. Furthermore it is important at this point 
to note that the above construction of $\theta$ necessitates a trivial 
structure of the principal bundle \cxm, since otherwise there would be no 
well-defined multiplication of a tangent frame with a complex number. Since 
\cxm, when restricted to its \uone\ subbundle, will constitute the 
electromagnetic \uone\ bundle, see (\ref{gtrans11}), we may say that 
the electromagnetic \uone\ bundle is {\em necessarily trivial} in our theory. 
                                                             
Finally, the bundle map $j$ in (\ref{bundlestructure}) is simply the 
canonical inclusion of $\fff{G}(M)$ into the frame bundle $F_c(M)$. 

Let us briefly discuss the main feature of the bundle diagram 
(\ref{bundlestructure}): Our aim is to construct a covariant spinor 
derivative out of an arbitrary complex linear connection $\o$ defined on the 
complex frame bundle $F_c(M)$. As outlined in \ref{geometricint} on p.\ 
\pageref{geometricint}, this can be done by pulling $\o$ back onto an 
``intermediate bundle'', which possesses a spin structure. In the following 
subsections, this will be realized with the help of the above diagram: We 
first pull $\o$ back via $j$ onto $\fff{G}(M)$, then via $\theta$ onto the 
product bundle $(\clplus\times\ctimes)(M)$, which possesses the spin 
structure $(\cspin\times\ctimes)(M)$. Finally, if $\o$ is further pulled 
back to this spin bundle via $\xi\times\mbox{id}$, then it will define an 
unique covariant spinor derivative on the spinor bundle $S_c(M)$. The 
principal bundle $\fff{G}(M)$ located between the product bundle 
$(\clplus\times\ctimes)(M)$ and the frame bundle $F_c(M)$ is introduced in 
the diagram in order to make the pull-back procedure especially simple.

\subsection{Extended spin connection}\label{espinconnection} 

As has been said before, the fibre bundle diagram (\ref{bundlestructure}) 
will enable us to construct an unique spin connection on 
$(\cspin\!\times\!\ctimes)(M)$ starting from an arbitrary complex linear 
connection of the spacetime manifold by pulling back its connection 1-form on 
$F_c(M)$ along the horizontal line of the diagram from the right to the left. 

To see that this procedure really works, let $\o$ be an arbitrary connection 
1-form on the complex frame bundle $F_c(M)$. The first step is to construct a 
connection on the bundle $\fff{G}(M)$. Since $\fff{G}(M)$ is a subbundle of 
$F_c(M)$, we may apply {\bf Proposition 1} of \ref{connections}. For the 
application, it is necessary to find a vector subspace \fff{m} of 
\lieglfourc, such that \lieglfourc\ is the direct sum of \fff{m} and the Lie 
algebra \fff{g} of \fff{G} (\ref{algebrag}) having the additional property 
stated in that proposition. Define the vector subspace \fff{m} by 
\beq\label{subspace1} 
  \fff{m}:= \big\{A\in\lieglfourc\, 
  |\: A^T\eta-\eta A=0\;,\;\;\mbox{Tr}(A)=0\big\} \;. 
\eeq 
It is straightforward to show that this \fff{m} is indeed a 
$\complex$-vector subspace of \lieglfourc. Note that \fff{m} is {\em not} 
a Lie subalgebra. Then, with the definition (\ref{algebrag}) of \fff{g}, 
\beq\label{subspace2}
  \lieglfourc = \fff{g}\oplus\fff{m} 
                         = \complex\!\otimes\!\fff{l}\oplus
                           \complex\One \oplus \fff{m} \;. 
\eeq 
To prove this assertion, we explicitly give the components of an element 
of \lieglfourc\ according to this decomposition,\\ 
\parbox{13.8cm}{
  \[\renewcommand{\arraystretch}{1.5}
    \arraycolsep0.2mm
    \begin{array}{ccccccc}
      \lieglfourc\;\; &=& \complex\otimes\fff{l}      &\oplus&
      \complex\One         &\oplus& \fff{m} \\
               A              &=\;& \frac{1}{2}(A-\eta A^T\eta) &\;+\;&
      \frac{1}{4}\mbox{Tr}A
      \!\cdot\!\One              &\;+\;& \frac{1}{2}(A+\eta A^T\eta
                                  -\frac{1}{2}\mbox{Tr}A\!\cdot\!\One)
    \end{array}
  \] }
\hfill
\parbox{1cm}{\baq\label{subspace3}\eaq}\\ 
It is easy to show that the components given in (\ref{subspace3}) indeed 
fulfill the required algebraic properties. In order to employ {\bf 
Proposition 1}, we must prove $(\Lambda c)\fff{m}(\Lambda c)^{-1} \subset 
\fff{m}$ for all $\Lambda c \in \fff{G}$. Using $\Lambda^T\eta\Lambda=\eta$, 
we have for an arbitrary element $A$ of $\fff{m}$ 
\baqn 
    \left( \Lambda c\,A\,(\Lambda c)^{-1}\right)^T\eta
  & \!\!\!=\!\!\! & 
    \Lambda^{-1\,T}A^T\Lambda^T\eta \;= \Lambda^{-1\,T}A^T\eta\Lambda^{-1} 
  \\ 
  & \!\!\!=\!\!\! &
    -\Lambda^{-1\,T}\eta A\Lambda^{-1} = -\eta\Lambda A\Lambda^{-1}
  = -\eta\left( \Lambda c\,A\,(\Lambda c)^{-1}\right)  \;,
  \\
    \mbox{Tr}\left(\Lambda c\,A\,(\Lambda c)^{-1}\right)
  & \!\!\!=\!\!\! &
    \mbox{Tr}A = 0 \;.
\eaqn
With the help of {\bf Proposition 1}, we now obtain a connection 1-form on 
$\fff{G}(M)$ by restricting $\o$ to $\fff{G}(M)$ and taking its 
\fff{g}-component. This connection will be denoted by $\o_G$, 
\beq\label{omegag} 
  \o_G := \fff{g}\mbox{-component of}\quad \o|_{G(M)} \;. 
\eeq 

As the next step, we construct a connection on the product bundle 
$(\clplus\!\times\!\ctimes)(M)$. Since the Lie algebra homomorphism of 
$\theta_o$ is actually an isomorphism, see (\ref{theta1}), we may apply 
{\bf Proposition 3} of \ref{connections} to the bundle map $\theta$ and 
obtain the following connection 1-form on $(\clplus\!\times\!\ctimes)(M)$ 
\beq\label{omegalc} 
  \o_{lc} := \theta_o^{-1}\theta^\ast\o_G \;. 
\eeq 
Since $(\clplus\!\times\!\ctimes)(M)$ is a fibre product bundle, we can 
decompose this connection $\o_{lc}$ by using {\bf Proposition 2} of 
\ref{connections} as 
\beq\label{decompo1} 
  \o_{lc} = p^\ast\o_l + q^\ast\o_c \;, 
\eeq                  
where $p$ and $q$ are the canonical projections from 
$(\clplus\!\times\!\ctimes)(M)$ to \clplusm\ and \cxm, respectively, and 
$\o_l$ and $\o_c$ denote the connections on \clplus\ and \cxm\ constructed 
canonically from $\o_{lc}$, see the proof of {\bf Proposition 2}. Thus, 
$\o_l$ is a complex Lorentz connection on \clplus, whose algebraic components 
are given by the restriction of $\o_{lc}$ to its $\complex\otimes\fff{l}$%
-component, and $\o_c$ is a \ctimes\ potential on \cxm, whose algebraic 
component is the \ctimes-component of $\o_{lc}$.%
\footnote{In the diploma thesis \cite{diplom} these two connections $\o_l$ 
and $\o_c$ were assumed to be some {\em restrictions} of the full connection 
on $(\clplus\!\times\!\ctimes)(M)$ onto its ``compontents'' \clplusm\ and 
\cxm, see p.\ 67, above the formula (A.40). But this is not the correct 
way to express these two connections, since \clplusm\ and \cxm\ might not be 
contained in $(\clplus\!\times\!\ctimes)(M)$ as natural subbundles, that is 
as ``components''. The problem is here, that there is no natural inclusion 
mapping from \cxm\ into $(\clplus\!\times\!\ctimes)(M)$ if \clplusm\ is not 
a trivial bundle. Nevertheless, the formula (A.41) in the diploma thesis is 
formally correct, and can be used to decompose a linear connection into its 
metric part and its non-metric vector part.} 

We can now construct the {\em extended spin connection} on 
$(\cspin\!\times\!\ctimes)(M)$ from $\o_{lc}$ by using again {\bf 
Proposition 3}, since the Lie algebra mapping $\xi_o\times\mbox{id}$ 
(\ref{bundlestructure}) is an isomorphism, cf.\ 
(\ref{complexlieisomorphism}). If we denote this spin connection by 
$\o_{sc}$, then 
\baq 
  \o_{sc} &=& (\xi_o\times\mbox{id})^{-1}(\xi\times\mbox{id})^\ast\o_{lc}
\label{omegasc1}\\ 
  &=& \xi_o^{-1}\xi^\ast p^\ast\o_l + q^\ast\o_c \;. 
\label{omegasc2} 
\eaq

\subsection{Local cross sections}\label{localsections} 

In order to obtain the components of the connection on the base manifold \mm\ 
from the various connection 1-forms introduced in the foregoing subsection we 
shall consider local cross sections of the principal bundles in the 
diagram (\ref{bundlestructure}). 

Let \calu\ be an open subset of \mm, on which all the principal bundles 
considered so far admit local cross sections. Let 
\beq\label{localsections1} 
  \s = (e_a{}^{\!\m}\partial_\m) 
\eeq 
be a local cross section of the complex Lorentz bundle \clplusm. Thus, $\s$ 
is a complex orthonormal tetrad field. Although we could restrict our 
considerations only to the case of real tetrad fields as in chapter 2, we 
shall allow here for arbitrary complex tetrad fields, because we want to 
study the full {\em mathematical structure} of the bundle geometry without 
bothering about physics. As remarked on p.\ \pageref{con}, complex tetrad 
fields are also allowed in our theory, if one does not consider the physical 
role of the tetrad fields themselves. 

As in the case of real spin geometry, there exists a local cross 
section $\hat{\s}$ of the complex spin bundle \cspinm, such that the 
spin mapping $\xi$ maps it onto $\s$, cf.\ (\ref{spinsection}), 
\beq\label{localsections2} 
  \s = \xi(\hats) \;. 
\eeq 

Since we want to consider cross sections of the product bundles 
$(\cspin\!\times\!\ctimes)(M)$ and $(\clplus\!\times\!\ctimes)(M)$ in 
the diagram (\ref{bundlestructure}), we need a cross section of the 
principal bundle \cxm, which is merely a \ctimes-valued function, because 
\cxm\ is a trivial bundle. At the moment, we choose a special function, 
denoted by $\hat{1}$, whose values equal constantly $1 \in \ctimes$, 
\beq\label{localsections3} 
  \hat{1} : \calu \longmapsto \ctimes,\quad p \longmapsto 1 \;. 
\eeq 
Later on, we will consider arbitrary functions and elaborate the gauge 
transformations connected with the change from \hatone\ to these functions. 

Now, $(\s,\hatone)$ and $(\hats,\hatone)$ are clearly cross sections of 
the product bundles $(\clplus\!\times\!\ctimes)(M)$ and 
$(\cspin\!\times\!\ctimes)(M)$, respectively. Remembering the definition 
of the bundle map $\theta$ in (\ref{bundle4}), we obtain the following 
commutative diagram of various cross sections constructed so far:\\ 
\eqlabel{sectionstructure} 
\unitlength1cm
\begin{picture}(15,4.6)

\put( 0.4,3.2){\makebox(3.0,0.5)[b]{$(\cspin\!\times\!\ctimes)(M)$}}
\put( 5.7,3.2){\makebox(2.0,0.5)[b]{$(\clplus\!\times\!\ctimes)(M)$}}
\put( 9.7,3.2){\makebox(1.0,0.5)[b]{$\fff{G}(M)$}}
\put(12.2,3.2){\makebox(1.5,0.5)[b]{$F_c(M)$}}

\put( 3.7,3.5){\makebox(1.5,0.5)[b]{$\xi\!\times\!\mbox{id}$}}
\put( 8.2,3.5){\makebox(1.5,0.5)[b]{$\theta$}}
\put(10.8,3.5){\makebox(1.5,0.5)[b]{$j$}}

\put( 3.7,3.3){\vector( 1, 0){1.5}}
\put( 8.3,3.3){\vector( 1, 0){1.3}}
\put(10.9,3.3){\vector( 1, 0){1.3}}

\put( 1.4,0.5){\makebox(1.0,0.5)[b]{$\cal U$}}
\put( 6.2,0.5){\makebox(1.0,0.5)[b]{$\cal U$}}
\put( 9.7,0.5){\makebox(1.0,0.5)[b]{$\cal U$}}
\put(12.5,0.5){\makebox(1.0,0.5)[b]{$\cal U$}}

\put( 0.8,1.8){\makebox(1.2,0.5)[b]{$(\hats,\hatone)$}}
\put( 5.3,1.8){\makebox(1.2,0.5)[b]{$(\s,\hatone)$}}
\put( 9.0,1.8){\makebox(1.0,0.5)[b]{$1\cdot\s$}}
\put(11.8,1.8){\makebox(1.0,0.5)[b]{$1\cdot\s$}}

\put( 1.9,1.0){\vector( 0, 1){2.0}}
\put( 6.7,1.0){\vector( 0, 1){2.0}}
\put(10.2,1.0){\vector( 0, 1){2.0}}
\put(13.0,1.0){\vector( 0, 1){2.0}}

\put( 2.4,0.6){\vector( 1, 0){3.8}}
\put( 7.2,0.6){\vector( 1, 0){2.5}}
\put(10.7,0.6){\vector( 1, 0){1.8}}

\put( 4.2,0.7){\makebox(0.5,0.5)[b]{=}}
\put( 8.3,0.7){\makebox(0.5,0.5)[b]{=}}
\put(11.5,0.7){\makebox(0.5,0.5)[b]{=}}

\put(14.0,1.3){\makebox(1.0,0.5)[r]{(\ref{sectionstructure})}}
\end{picture}
\stepcounter{equation}\\

\subsection{Extended spinor derivative}\label{extended} 
\subsubsection{Connections on the base space} 

Let $\o$ be a connection 1-form on the complex frame bundle $F_c(M)$ and let 
\beq\label{extended1} 
  \Con{a}\m{b}dx^\m := \big((1\cdot\s)^\ast\o\big)^a{}_{\!b} 
\eeq 
be the \lieglfourc-components of the pulled back connetion on the base space 
\mm. They are the anholonomic tetrad components of the general complex 
linear connection as introduced in (\ref{con}). The superfluous factor 1 in 
front of $\s$ is inserted here as well as in the diagram 
(\ref{sectionstructure}) 
in view of the \uone\ gauge transformation considered in the next section. 

Now consider the connection $\o_G$ on $\fff{G}(M)$ defined in (\ref{omegag}). 
If this connection is pulled back by the same local cross section $1\cdot\s$ 
to $M$, then the resultant connection on the spacetime \mm\ will not be the 
same as in (\ref{extended1}), but it will have only its \fff{g}-components. 
Thus, although the diagram (\ref{sectionstructure}) of the various cross 
sections is perfectly commutative, this property is lost when considering 
the connections, because the ``mappings'' between them, cf.\ equations 
(\ref{omegag}) to (\ref{omegasc2}), do not include only the mappings between 
the underlying topological spaces, but also various Lie algebra 
homomorphisms. Using the explicit decomposition (\ref{subspace3}), we take 
the \fff{g}-components of (\ref{extended1}) to obtain 
\beq\label{extended2} 
  \big((1\cdot\s)^\ast\o_G\big)^a{}_{\!b} =
  \frac{1}{2}(\Con{a}\m{b}-\Gamma_{b\m}{}^a)dx^\m \,+\, 
  \frac{1}{4}\Con{c}\m{c}\d^a{}_{\!b}dx^\m \;. 
\eeq 
Next, the pull-back of $\o_{lc}$ (\ref{omegalc}) via the cross section 
$(\s,\hatone)$ in (\ref{sectionstructure}) results in the same expression, 
\beq\label{extended3} 
\eqalign{ 
  \big((\s,\hatone)^\ast\o_{lc}\big)^a{}_{\!b} &=  
  \big((\s,\hatone)^\ast\theta_o^{-1}\theta^\ast\o_G\big)^a{}_{\!b}  
\\ 
  &=  
  \big(\theta_o^{-1}(\s\cdot\hatone)^\ast\o_G\big)^a{}_{\!b} 
\\ 
  &=  
  \frac{1}{2}(\Con{a}\m{b}-\Gamma_{b\m}{}^a)dx^\m \,+\, 
  \frac{1}{4}\Con{c}\m{c}\d^a{}_{\!b}dx^\m \;, 
} 
\eeq 
where we exploited the commutative rectangle at the centre of the diagram 
(\ref{sectionstructure}). Note that here the Lie algebra of 
$\clplus\times\ctimes$ has been trivially identified 
with $\complex\otimes\fff{l}\oplus\complex\One$, see the remark on p.\ 
\pageref{tautology}. If we do not make such an identification, then the 
correct, but somewhat pedantic, expression reads 
\beq\label{extended4} 
  \big((\s,\hatone)^\ast\o_{lc}\big)^a{}_{\!b}  
  = 
  \big(\frac{1}{2}(\Con{a}\m{b}-\Gamma_{b\m}{}^a)dx^\m \;,\; 
  \frac{1}{4}\Con{c}\m{c}\d^a{}_{\!b}dx^\m\big) \;. 
\eeq 
Of course, care must be taken when (\ref{extended2}) and (\ref{extended3}) 
are compared, since they belong to connection 1-forms on different principal 
bundles: Whereas in (\ref{extended2}) the plus-sign denotes merely an 
addition of different {\em algebraic} components, the plus-sign in 
(\ref{extended3}) means the sum of two {\em geometrically} different 
connections, namely of (see (\ref{decompo1})) 
\baq 
  \big(\s^\ast\o_l\big)^a{}_{\!b} &=& 
  \frac{1}{2}(\Con{a}\m{b}-\Gamma_{b\m}{}^a)dx^\m \quad\mbox{and} 
\label{extended5}\\ 
  \big(\hatone^\ast\o_c\big)^a{}_{\!b} &=& 
  \frac{1}{4}\Con{c}\m{c}\d^a{}_{\!b}dx^\m \;. 
\label{extended6}
\eaq                
As we have said below (\ref{decompo1}), $\o_l$ is a complex Lorentzian 
connection on \clplusm, and $\o_c$ is a \ctimes\ potential on \cxm. 

In a similar fashion, using the left commutative rectangle of 
(\ref{sectionstructure}) and the decomposition (\ref{omegasc2}), we obtain 
the extended spin connection on the base space \mm, 
\baq 
  (\hats,\hatone)^\ast\o_{sc} &=& 
  (\hats,\hatone)^\ast(\xi_o^{-1}\xi^\ast p^\ast\o_l + q^\ast\o_c) 
\nonumber\\ 
  &=& \xi_o^{-1}\s^\ast\o_l \;+\; \hatone^\ast\o_c \;. 
\label{extended7} 
\eaq 
We now employ the extended spin representation $\zeta_c$ (\ref{espinrep}), 
its Lie algebra homomorphism (\ref{espinrep2}), and equation (\ref{cliehomo}) 
to obtain 
\baq 
  \zeta_c\big((\hats,\hatone)^\ast\o_{sc}\big) &=& 
  (\zeta\circ\xi_o^{-1})\s^\ast\o_l \;-\; \hatone^\ast\o_c 
\label{extended8}\\ 
  &=& 
  -\frac{1}{4}\c^b\c_a\cdot\frac{1}{2}(\Con{a}\m{b}-\Gamma_{b\m}{}^a)dx^\m 
  \;-\;  
  \One\cdot\frac{1}{4}\Con{c}\m{c}dx^\m 
\label{extended0}\\ 
  &=& 
  -\frac{1}{4}\con{a}\m{b}\c^b\c^a dx^\m \;. 
\label{extended9} 
\eaq 

\subsubsection{The extended covariant spinor derivative}

We are now able to construct the extended spinor derivative (\ref{esd}) on 
p.\ \pageref{esd} by following analogous steps as in (\ref{lcs5}) to 
(\ref{sd3}) for the construction fo the ordinary spinor derivative 
(\ref{spinorderivative}). 

In the bundle diagram (\ref{bundlestructure}), Dirac spinors $\psi$ are 
vector fields on the spinor bundle $S_c(M)$, which we represent as 
\beq\label{extended10} 
  \psi = [(\hats,\hatone)\;,\;\psi_{(\hat{\s},\hat{1})}] \;, 
\eeq 
where $\psi_{(\hat{\s},\hat{1})}$ is a $\complex^4$-valued function on 
\calu. With the help of (\ref{extended9}), the extended covariant spinor 
derivative reads 
\beq\label{extended11} 
\eqalign{ 
  \nabla_{\!\m}\psi &= \big[(\hats,\hatone)\;,\;
  \partial_\m\psi_{(\hat{\s},\hat{1})} + 
  \zeta_c((\hats,\hatone)^\ast\o_{sc}(\partial_\m)) 
  \psi_{(\hat{\s},\hat{1})} \big] 
\\ 
  &= \big[(\hats,\hatone)\;,\;
  \partial_\m\psi_{(\hat{\s},\hat{1})} - 
  \frac{1}{4}\con{a}\m{b}\c^b\c^a\psi_{(\hat{\s},\hat{1})} \big] \;, 
} 
\eeq 
which may be written simply as 
\beq\label{extended12} 
  \nabla_{\!\m}\psi = 
  \partial_\m\psi - \frac{1}{4}\con{a}\m{b}\c^b\c^a\psi \;. 
\eeq 
This is precisely our extended covariant spinor derivative (\ref{esd}).

\subsubsection{Decomposition principle}\label{decompositionprinciple} 

We shall now turn our attention to the mathematical structure of the 
connection (\ref{eqcon3}), which is the result of the field equation 
(\ref{eqcon0}): 
\beq 
  \Con{a}\m{b} = {\widehat{\Gamma}}^a{}_{\!\m b} + \d^a{}_{\!b}\cdot S_\m \;, 
\eeq 
where ${\widehat{\Gamma}}^a{}_{\!\m b}$ is a complex Lorentzian connection 
compatible with the metric, ${\widehat{\Gamma}}_{a\m b} = 
-{\widehat{\Gamma}}_{b\m a}$. If we insert this connection (\ref{eqcon3}) 
into the above formulae (\ref{extended5}) and (\ref{extended6}), then 
\baq 
  \big(\s^\ast\o_l\big)^a{}_{\!b} &=& 
  {\widehat{\Gamma}}^a{}_{\!\m b}dx^\m \;, 
\label{extended5a}\\
  \big(\hatone^\ast\o_c\big)^a{}_{\!b} &=& 
  \d^a{}_{\!b}\cdot S_\m dx^\m \;.
\label{extended6a}
\eaq 
Thus, we can uniquely decompose the resultant connection in (\ref{eqcon3}) 
in accordance with (\ref{decompo1}), that is, as the sum of a complex 
Lorentzian connection on \clplusm\ and a \ctimes\ potential on \cxm. In so 
doing, we of course interpret the connection (\ref{eqcon3}) as a connection 
resulting from the product bundle $(\clplus\!\times\!\ctimes)(M)$ and not 
from the frame bundle $F_c(M)$. This point of view can only be taken after 
the field equations for the connection have been considered, but not before, 
since an arbitrary linear connection does not possess the special  
structure of (\ref{eqcon3}). 
  
We now discuss the extended spinor derivative (\ref{esd}), 
\beq 
  \nabla_{\!\m}\psi = \partial_\m\psi-\frac{1}{4}\con{a}\m{b}\s^{ba}\psi 
                     -\frac{1}{4}\Con{c}\m{c}\psi \;. 
\eeq 
We see that this extended spinor derivative is already decomposed 
{\em formally} into two contributions $-\frac{1}{4}\con{a}\m{b}\s^{ba}$ and 
$-\frac{1}{4}\Con{c}\m{c}$. But now regarding the equations (\ref{extended7}) 
to (\ref{extended9}) it is clear that this decomposition is based on a 
geometric foundation: The extended spin connection $\o_{sc}$ is indeed the 
sum of two different connections, namely the ``ordinary'' complex spin 
connection $\o_s$ defined by $\o_s = \xi_o^{-1}\xi^\ast p^\ast\o_l$ in 
equation (\ref{omegasc2}), cf.\ \ref{complexspingeometry}, and a 
\ctimes\ potential $\o_c$ on \cxm, see (\ref{omegasc2}). Whereas $\o_s$ 
gives rise to the covariant derivative characterized by 
\beq\label{extended8b} 
  \zeta_c\big(\hats^\ast\o_s\big) 
  = \zeta_c\big(\hats^\ast(\xi_o^{-1}\xi^\ast p^\ast\o_l)\big) 
  = (\zeta_c\xi_o^{-1})\hats^\ast\o_l 
  = -\frac{1}{4}\con{a}\m{b}\s^{ab}dx^\m \;, 
\eeq 
the \ctimes\ potential $\o_c$ leads to 
\beq\label{extended9b} 
  -\hatone^\ast\o_c = -\frac{1}{4}\Con{c}\m{c}dx^\m \;, 
\eeq 
cf.\ (\ref{extended7}) to (\ref{extended9}). This decomposition of the 
spinor derivative is, contrary to the decomposition of the linear 
connection as considered in (\ref{extended5a}) and (\ref{extended6a}), 
valid already before the field equations have been taken into account. 
This property of the extended spinor derivative (\ref{esd}) is indeed 
necessary for the construction of the basic matter Lagrangian ${\cal L}_m$ 
(\ref{lagrangian}), as was said in the discussion following (\ref{cx}). 

Note that the factor $\frac{1}{4}$ in front of the trace $\Con{c}\m{c}$ 
in (\ref{extended0}) has its real origin in the algebraic decomposition 
(\ref{subspace3}), whereas in (\ref{esd}), this factor seems to be caused 
by the overall factor $1/4$ of the usual covariant spinor derivative 
(\ref{spinorderivative}).

\section{Electromagnetic gauge transformation} 

Let us now study the \ctimes\ gauge transformation, which, if restricted to 
\uone, will become the electromagnetic phase transformation. Let $\l$ be a 
$\complex$-valued function on \calu\ and let 
\beq\label{gtrans1} 
  \hat{\l} := \hatone\cdot\exp(\l) 
\eeq 
be an arbitrary \ctimes-valued function on \calu, viewed as a cross section 
of \cxm. Then, according to the gauge transformation law 
(\ref{connectiongauge}), we obtain for the \ctimes\ connection $\o_c$ 
\beq\label{gtrans2} 
\eqalign{ 
  \hatl^\ast\o_c(\partial_\m) &= 
  e^{-\l}\hatone^\ast\o_c(\partial_\m)e^\l + e^{-\l}\partial_\m e^\l 
  \;=\; \hatone^\ast\o_c(\partial_\m) + \partial_\m \l 
\\ 
  &= \frac{1}{4}\Con{c}\m{c} + \partial_\m\l \;, 
} 
\eeq 
which should be compared with the expression (\ref{extended6}).%
\footnote{Note that ${\hat{\l}}^\ast\o_c = {\hat{\l}}^\ast\o_c(\partial_\m) 
\cdot dx^\m$.} 
This is the \ctimes\ gauge transformation of the \ctimes\ potential 
$\frac{1}{4}\Con{c}\m{c}$. Since this transformation affects only 
quantities defined on or derived from the principal bundle \cxm, all other 
quantities on the complex Lorentz bundle \clplusm\ or on the complex spin 
bundle \cspinm\ remain unchanged. Thus, especially, the complex Lorentzian 
connection (\ref{extended5}) and the complex tetrad field $\s$ remain 
fixed. This would not hold true any longer if, in the diagram 
(\ref{sectionstructure}), the complex frame bundle $F_c(M)$ or \pgm\ are 
considered, see below. 

To study the gauge transformation of the Dirac spinor $\psi$, we use the 
gauge transformation property (\ref{gt2}) of the vector fields to get 
(cf.\ (\ref{extended10}) and (\ref{espinrep}))
\beq\label{gtrans3} 
\eqalign{ 
  \psi &= [(\hats,\hatone)\,,\,\psi_{(\hat{\s},\hat{1})} ] 
  \;=\;   [(\hats,\hatl)\,,\,\psi_{(\hat{\s},\hat{\l})} ] \;, 
\\ 
  \psi_{(\hat{\s},\hat{\l})} &= 
  \zeta_c\big(((\One,\One),e^\l)^{-1}\big) \psi_{(\hat{\s},\hat{1})} 
  \;=\; e^\l\psi_{(\hat{\s},\hat{1})} \;. 
} 
\eeq 
In summary, the \ctimes\ gauge transformation reads as follows: 
\beq\label{gtrans4} 
  e_a{}^{\!\m} \mapsto e_a{}^{\!\m} ,\quad 
  \frac{1}{4}\Con{c}\m{c} \mapsto \frac{1}{4}\Con{c}\m{c} + \partial_\m \l
  ,\quad  
  \psi \mapsto e^\l\psi \;. 
\eeq

\subsection{Further extension of the spinor derivative}

In the discussion following (\ref{esd}), we remarked that the extension of 
the spinor derivative was not unique. In \ref{theextension} we have 
exploited the remaining ambiguity to further extend the spinor derivative. 
To obtain the most general spinor derivative (\ref{xsdd}), only a slight 
change of the extended spin representation $\zeta_c$ (\ref{espinrep}) is 
necessary. We now define 
\beq\label{espinrep3} 
\eqalign{ 
  \zeta_\varepsilon : \cspin\times\ctimes &\longrightarrow \quad\glfourc
\\
  \qquad\left(\,(A,B)\,,\,c\right) &\longmapsto 
  \zeta((A,B))\!\cdot\! c^\varepsilon\;, 
} 
\eeq 
where $\vare \in \real$. Using this spin representation, it is easy to show 
that Dirac spinors now transform as (cf.\ (\ref{gtrans3})) 
\beq\label{gtrans5} 
  \psi_{(\hat{\s},\hat{\l})} = \zeta_\varepsilon\big(((\One,\One),e^\l)^{-1} 
  \psi_{(\hat{\s},\hat{1})} = e^{-\varepsilon\l}\psi_{(\hat{\s},\hat{1})} 
  \;. 
\eeq 
As in (\ref{extended8}), we can compute the spin connection 
corresponding to $\zeta_\varepsilon$ using an arbitrary cross section 
$(\hats,\hatl)$ to give the following result: 
\baq 
  \zeta_\vare\big((\hats,\hatl)^\ast\o_{sc}\big) &=& 
  (\zeta\circ\xi_o^{-1})\s^\ast\o_l \;+\; \vare\cdot\hatl^\ast\o_c 
\nonumber\\ 
  &=& 
  -\frac{1}{4}\c^b\c_a\cdot\frac{1}{2}(\Con{a}\m{b}-\Gamma_{b\m}{}^a)dx^\m 
  \;+\;  
  \vare\big(\frac{1}{4}\Con{c}\m{c}+\partial_\m \l\big)dx^\m \;.\quad 
\label{gtrans6} 
\eaq

\subsection{Restriction to \uone} 

So far we have dealt with the group \ctimes, which was needed to construct 
the bundle structure (\ref{bundlestructure}). In order to restrict \ctimes\ 
to its subgroup \uone, we first observe that the adjoint spinor%
\footnote{Adjoint spinors can be defined analogously to spinors as vector 
fields on an associated vector bundle of the extended spin bundle 
$(\cspin\!\times\!\ctimes)(M)$, where the representation of the extended 
spin group $\cspin\times\ctimes$ in \glfourc\ is taken to be the adjoint 
representation $(\c^0\zeta_c^\dagger\c^0)^T$. However, to avoid too much 
congestion in the exposition, we prefer to represent adjoint spinors 
only locally by simply taking the adjoint of an ordinary spinor.} 
transforms under the \ctimes\ gauge transformation (\ref{gtrans3}) according 
to 
\beq\label{gtrans7} 
  \PSI_{(\hat{\s},\hat{\l})} = 
  \big(\psi_{(\hat{\s},\hat{\l})}\big)^\dagger\c^0 = 
  e^{\overline{\l}}\,\PSI_{(\hat{\s},\hat{1})} \;, 
\eeq 
where $\overline{\l}$ means the complex conjugate of $\l$. Due to the 
``covariance'' of the covariant derivative in the local representation, see 
the second line of (\ref{invariance}), we have 
\beq\label{trans8} 
  \nabla_{\!\m}\psi_{(\hat{\s},\hat{\l})} = 
  e^\l\nabla_{\!\m}\psi_{(\hat{\s},\hat{1})} \;, 
\eeq 
so that the matter Lagrangian ${\cal L}_m$ in (\ref{lagrangian}) does not 
remain invariant under the whole \ctimes\ gauge transformation, but changes 
\beq\label{gtrans9} 
  {\cal L}_m \longmapsto \exp(\overline{\l}+\l){\cal L}_m \;. 
\eeq 
Since we {\em have to} demand the invariance of the Lagrangian, we conclude 
\baq 
  \overline{\l}+\l &=& 0 \quad\mbox{for all}\quad \l \quad \Longrightarrow 
\nonumber\\ 
  \exp(\l) &\in& \uone \;,   
\label{gtrans10} 
\eaq 
so that we must not consider the whole group \ctimes, but only its subgroup 
\uone\ of unit elements. As a consequence, instead of \cxm, its reduced 
bundle $\uone(M)$ must be considered. So, throughout this chapter, we 
subsequently 
\beq\label{gtrans11} 
  \mbox{replace every \ctimes\ by \uone.} 
\eeq

\subsection{Further properties of the gauge transformation}

In (\ref{gtrans2}), we discussed the gauge transformation of the \ctimes, 
or, because of (\ref{gtrans11}), of the \uone\ potential $\o_c$ on 
$\uone(M)$. We now want to study the same gauge transformation on the complex 
frame bundle $F_c(M)$. If we replace in the diagram (\ref{sectionstructure}) 
the trivial cross section \hatone\ by \hatl\ defined in (\ref{gtrans1}), 
we see that the cross section of the frame bundle becomes $e^\l\s$. The 
pull-back of the complex linear connection 1-form $\o$ via this cross 
section does not remain unchanged, but transforms according to 
\beq\label{gtrans12} 
  \big((e^\l\s)^\ast\o(\partial_\m)\big)^a{}_{\!b} = 
  e^{-\l}(\s^\ast\o(\partial_\m))^a{}_{\!b}e^\l + 
  \d^a{}_{\!b}e^{-\l}\partial_\m e^\l = 
  \Con{a}\m{b} + \d^a{}_{\!b}\partial_\m \l \;. 
\eeq 
Thus, the connection trace $\Con{c}\m{c}$ still transforms in a similar 
manner as in (\ref{gtrans2}). But now the cross section $e^\l\s$ is no longer 
an {\em orthonormal} tetrad field, but only {\em orthogonal\/}. So, unlike 
$\s$, this cross section can not be ``lifted'' to a cross section of the 
spin bundle, and, therefore, no local representation of Dirac spinors (cf.\ 
(\ref{gtrans3})) can be defined for $e^\l\s$. Even worse, any tangent vector 
$X$, written in the tetrad components $X = X^a(e_a{}^{\!\m}\partial_\m)$, 
becomes now charged, since $X^a$ is transformed to $e^{-\l}X^a$ due to the 
gauge transformation rule 
\beq\label{gtrans12a} 
  X = X^a(e_a{}^{\!\m}\partial_\m) 
    = e^{-\l}X^a(e^\l e_a{}^{\!\m}\partial_\m) \;. 
\eeq 
For these mathematical and physical reasons, it is {\em not allowed} to  
consider the \uone\ gauge transformation on the frame bundle $F_c(M)$ or, 
equivalently, on $\fff{G}(M)$, but only on the product bundles in the 
diagram (\ref{sectionstructure}). 

Stated differently, we must discard the right and the middle commutative 
rectangles in the diagram (\ref{sectionstructure}) and retain only the left 
rectangle. In this way, we detach the \uone\ potential $\o_c$ and its 
\uone\ gauge transformation completely from the basic complex linear 
connection $\o$ and also from the basic frame bundle geometry of the 
spacetime manifold $M$.

\subsection{Gauging the torsion trace}

Suppose now that we do {\em not} detach $\o_c$ from $\o$ but consider 
(\ref{gtrans12}) as the true \uone\ gauge transformation on $F_c(M)$, 
aiming at a gauge transformation of the torsion trace $T_\m$. In order to 
calculate the torsion trace from the transformed connection (\ref{gtrans12}) 
we ask about its coordinate components. 

Denoting the local coordinate frame by 
\beq\label{gtrans13} 
  k := (\partial_\m) = (\frac{\partial}{\partial x^\m}) \;, 
\eeq 
we reexpress it in terms of the transformed cross section $e^\l\s$ used to 
obtain (\ref{gtrans12}), 
\beq\label{gtrans14} 
\eqalign{ 
  k &= (e^\l\s)\cdot (e^{-\l}\Lambda) \quad \Leftrightarrow 
\\ 
  (\partial_\m) &= (e^\l e_a{}^{\!\n}\partial_\n)\cdot 
                   (e^{-\l}e^a{}_{\!\m}) \;, 
} 
\eeq 
where the expression $(e^{-\l}e^a{}_{\!\m})$ containing the reciprocal 
tetrad $e^a{}_{\!\m}$ plays the role of the gauge transforming matrix 
$\Lambda$, compare with (\ref{gt2}). With the help of the gauge 
transformation law (\ref{connectiongauge}) we obtain the desired coordinate 
components of (\ref{gtrans12}): 
\beq\label{gtrans15} 
\eqalign{ 
  \big(k^\ast\o(\partial_\m)\big)^\a{}_{\!\b} 
  &= 
  \big(e^\l\Lambda^{-1}\cdot(e^\l\s)^\ast\o(\partial_\m)\cdot e^{-\l}\Lambda 
  + (e^\l\Lambda^{-1})\cdot\partial_\m(e^{-\l}\Lambda) \big)^\a{}_{\!\b} 
\\ 
  &= 
  e_a{}^{\!\a}(\Con{a}\m{b}+\d^a{}_{\!b}\partial_\m\l)e^b{}_{\!\b} 
  + \d^\a{}_{\!\b}\partial_\m(-\l) + e_c{}^{\!\a}\partial_\m e^c{}_{\!\b} 
\\ 
  &= e_a{}^{\!\a}\Con{a}\m{b}e^b{}_{\!\b} 
  + e_c{}^{\!\a}\partial_\m e^c{}_{\!\b} 
\\ 
  &= \Con\a\m\b \;. 
} 
\eeq 
This result is totally independent from the \uone\ gauge function $\l$. Thus, 
we obtain the familiar result that the torsion trace $T_\m$ can not be gauged 
with \uone. 

Despite this undoubted result some authors like McKellar \cite{mck79} and 
Borchsenius \cite{bor76a} regarded the so-called $\l$-transformation, first 
introduced by Einstein \cite{ein55}, as the \uone\ gauge transformation for 
the torsion trace. This $\l$-gauge transformation reads 
\beq\label{gtrans16} 
  \Con\a\m\b \longmapsto \Con\a\m\b + \d^\a{}_{\!\b}\partial_\m \l \;,
\eeq 
where $\l$ is now an arbitrary complex valued function on the spacetime 
manifold $M$. It  was introduced from the purely formal reason, that the 
Ricci-scalar $R$ (\ref{riccic}) remains invariant under (\ref{gtrans16}). 
One might ask, if there is any sensible way to understand (\ref{gtrans16}) 
as a geometric feature? 

One suggestion might be to regard it as part of a conformal transformation 
of the coordinate reference frame $k$ (\ref{gtrans13}), that is, 
\beq\label{gtrans17} 
  k = (\partial_\m) \longmapsto (e^\l\partial_\m) \;, 
\eeq 
in analogy to the transformation of the tetrad field $\s \mapsto e^\l\s$. 
One can easily see, that this indeed results in the $\l$-transformation 
(\ref{gtrans16}) of the connection by using (\ref{connectiongauge}). But the 
problem is that $(e^\l\partial_\m)$ is no longer a coordinate reference 
frame:%
\footnote{This can be easily verified, since $[e^\l\partial_\m,e^\l 
\partial_\n] \neq 0$ unless $\l$ is constant everywhere.} 
Now the components of the connection on the right-hand side 
in (\ref{gtrans16}) are no longer coordinate components, forbidding 
their use for the ordinary covariant derivative $\nabla_{\!\m}$ 
in coordinate components. Instead, everything must now be represented in 
the special frame $(e^\l\partial_\m)$. For example, a vector field $X^\m$ 
in coordinate components would now read $e^{-\l}X^\m$, so that, if 
(\ref{gtrans17}) is regarded as the electromagnetic phase transformation, 
every covariant vector field would be charged. This situation is analogous 
to (\ref{gtrans12a}). 

Thus, it seems that there is no sound way to get a gauged torsion vector 
$T_\m$. We repeat, that the only way out of this problem is to detach the 
\uone\ potential $\o_c$ completely from the frame bundle geometry by 
considering only the left rectangle in the diagram (\ref{sectionstructure}). 
In this way, we obtain a consistent \uone\ gauge theory of electromagnetism 
and are able to interpret the vector $S_\m = \frac{1}{4}\Con{a}\m{a}$ as 
the true electromagnetic potential via (\ref{s-a}). The torsion vector 
$T_\m$ is related to the potential $A_\m$ only formally, as explained in 
\ref{torsionandem} on p.\ \pageref{torsionandem}.

                \chapter{Spin-Spin Contact Interaction}

One interesting consequence of the Einstein--Cartan theory is the 
prediction of a contact interaction between spinning particles. In the 
introduction, we have briefly discussed the case of Dirac particles, see 
(\ref{ecenergy}) and (\ref{ecdirac}). Since the contact interaction is 
coupled to the square%
\footnote{Note that in (\ref{ecenergy}) we have $\planck^4 / k = 
\planck^2 \times \hbar c$.} 
of the Planck length $\planck^2$, it is 
hopelessly too small to be detected in laboratory experiments \cite{sto85}. 

However, at high matter densities in the early universe, this tiny 
interaction can become even stronger than the mass effects of the 
interacting particles, see e.g.\ \cite{heh73,heh76}. And, as was remarked by 
Kanno \cite{kan88}, at the high temperature predominant in this early epoch, 
the contact interaction becomes much stronger than the weak interaction: At 
a first glance, the contact interaction in (\ref{ecenergy}) seems to be only 
a certain copy of the weak interaction, when this last interaction is written 
in the phenomenological Fermi contact form, i.e.\ without the gauge bosons. 
Since the Fermi coupling constant of the weak interaction is about 
$1.2\times 10^{-5}\mbox{GeV}^{-2}$, whereas the constant of the contact 
interaction is of the order $10^{-37}\mbox{GeV}^{-2}$, one may conclude that 
it does not make sense to look for an observable effect of the contact 
interaction in the presence of the weak interaction. However, it is 
well-known that the standard model of 
the electroweak interaction posseses a phase transition, where the broken 
symmetry is restored above a critical temperature of $100\mbox{GeV}$, see 
e.g.\ \cite{kir72,dol74,din92}. Above this temperature, the weak interaction 
becomes a long-range interaction of equal strength as the electromagnetic 
interaction, and the current-current terms are no longer appropriate to 
describe the electroweak forces. On the other hand, the contact interaction 
term in the Einstein--Cartan theory persists regardless of the energy scale 
considered, since it is directly induced by torsion without any 
intermediating bosons.%
\footnote{If the energy scale is as high as the quantum gravity scale, then 
this remark may become incorrect, since then the geometry of spacetime 
(including torsion) must be quantized.} 

In the early universe, when the density of spinning particles exceeded some 
critical value, the contact interaction also leads to pair creations, 
see \cite{ker75,rum79}. As was noted by Kerlick \cite{ker75}, 
the required mass density is more than thirty orders of magnitude smaller 
than the density required for pair creation via tidal forces 
caused by the curvature of spacetime \cite{zel70}. Thus, the torsion-%
induced pair creation effects are much stronger and more likely than the 
curvature effects, and must be taken into account in the discussion of the
scenario of the early universe \cite{ker75}. 

The contact interaction might also influence the singularity 
behaviour \cite{haw73} of the universe. Whereas Kerlick \cite{ker75} 
concluded that torsion enhances singularity, other authors came to the 
opposite conclusion, namely that the contact interaction prevents it, 
see e.g.\ \cite{heh74,kuc78,nur83,kan88}. 

We may say that the torsion-induced contact interaction has important 
consequences on the early stage of the universe. But so far, no prediction 
has been made which can be investigated by present astronomical 
observations. One reason for the uncertainty of the predictions is, of course, 
that the spin-spin contact interaction is very weak and takes place only 
in a small time interval during the early epoch of the universe. Another 
reason might be that quantum field theoretic investigations have been
completely left out in most cases (see however \cite{kan88} and 
\cite{gvo85}). One reason for the omission of quantization is that 
Einstein--Cartan theory, like other gravitational theories, can not be 
quantized rigorously, that is, in a renormalizable way. Therefore, any 
quantization of the contact interaction is necessarily incomplete as 
physical theory. 

In this chapter, we shall try to step towards a more realistic view of the 
spin-spin contact interaction by quantizing it in the first Born 
approximation. 

First of all we must find such a contact interaction in our theory developed 
in chapter 2. This is done by considering a many-particle theory. It turns 
out that the resulting spin-spin interaction differs from the one of the 
Einstein--Cartan theory in not containing any self-interactions of fermions. 
          
In the next section we discuss the works of Kerlick \cite{ker75} and of 
Rumpf \cite{rum79}. These authors studied the shift of the energy spectrum 
of a Dirac particle due to a constant background torsion field. They both 
concluded that the contact interaction is attractive for the opposite spin 
direction of interacting fields, but repulsive for aligned spins, and that 
it does not depend on whether one considers matter or anti-matter; thus, 
one may speak of a ``universal'' interaction \cite{ker75}. Here we will 
apply these considerations to the contact interaction of our theory. The 
resulting energy shifts differ significantly from the results of the 
Einstein--Cartan theory. 

In the third section, we investigate the new spin-spin contact interaction 
as well as the ordinary contact interaction of the Einstein--Cartan theory 
by quantizing both interactions in the first Born approximation. As a result,  
neither interaction is ``universal'' as first proposed by Kerlick in 
\cite{ker75} for the ordinary Einstein--Cartan contact interaction.

        \section{Many-particle theory}\label{many} 

\subsection{The missing contact interaction} 

In Einstein--Cartan theory the Lorentzian connection (\ref{lct}) is influenced 
by spinning particles. It possesses a non-vanishing contorsion part 
built from the axial current $\PSI\c^5\c^\m\psi$, see (\ref{ectorsion}). 
This contribution of Dirac fields to geometry results in the characteristic 
spin-spin contact interaction in the energy-momentum equation 
(\ref{ecenergy}) as well as in the Dirac equation (\ref{ecdirac}). 

In chapter \ref{maintheory} we have seen that the resultant connection 
(\ref{eqcon3}) also contains a non-vanishing contorsion, now built from 
both vector and axial currents. But there we could not observe a  
spin-spin contact interaction like in the Einstein--Cartan theory. Neither 
the energy-momentum equation (\ref{fieldd}) nor the Dirac equation 
(\ref{fielda}) contain contact interaction terms, this being in contrast to 
the Einstein--Cartan theory. 

But this does not mean that there is no contact interaction at all. The 
reason for the absence of the contact interaction is that so far we have 
treated a classical single particle field theory: In the Dirac 
equation (\ref{dirac}) the cubic self-interaction term 
\beq\label{cubic1} 
  (j_\m + j^{\sssty 5}_\m)\c^\m\psi = 0 
\eeq  
vanished due to the identity (\ref{pauli}). Now, let us consider the basic 
Dirac equation (\ref{dirac2}), 
\beq\label{contact1}   
  i\c^\m(\nabla^\ast_{\!\m}-S_\m)\psi - \frac{mc}{\hbar}\psi 
  +\big(\frac{3}{2}iU_\m+\frac{1}{8}V_\m\c^5\big)\c^\m\psi = 0 \;, 
\eeq 
which is valid without refering to the field equations for the connection, 
but uses only the 4-vector decomposition (\ref{decomposition}). The last term 
containing the vectors $U_\m$ and $V_\m$ vanishes due to the field equation 
(\ref{quvj}) and (\ref{pauli}). Now, if these two vectors have not only 
contributions from the same Dirac field $\psi$, but also from some other, 
different, Dirac field, say $\chi$, so that for example 
$U_\m = -i\planck^2/4 (\PSI\c_\m\psi + \CHI\c_\m\chi)$, then we would obtain 
\beq\label{cubic2}  
  (\CHI\c_\m\chi + \CHI\c^5\c_\m\chi\c^5)\c^\m\psi \neq 0 
\eeq 
instead of (\ref{cubic1}) in the Dirac equation (\ref{contact1}). 

Therefore, in order to observe the missing spin-spin contact interaction 
in our theory, we must consider a many-particle system.

\subsection{Many-particle system}

To discover the spin-spin contact interaction we discuss a many-particle 
system consisting of spinors $\psi_z$ with charges $\varepsilon(z) e$, 
$\vare(z) \in \real$, and masses $m_z$, where $z$ is a counting index. 
In (\ref{lagrangian}) only the matter Lagrangian ${\cal L}_m$ changes. This 
Lagrangian now becomes a sum of Lagrangians for each spinor $\psi_z$, its 
spinor derivative given by (\ref{xsdd}) with $\varepsilon = \varepsilon(z)$, 
thus, 
\beq\label{many1} 
\eqalign{  
  {\cal L} &= g\cdot\hbar c \sum_z                        
  \big[ i\PSI_z\c^\m( \partial_\m-\frac{1}{4}\con{a}\m{b}\s^{ba}  
                     +\frac{\varepsilon(z)}{4}\Con{a}\m{a})\psi_z 
       -\frac{m_z c}{\hbar}\PSI_z\psi_z \big] 
\\ 
  &\phantom{=}-\frac{g}{2k}R + \frac{g}{4k}l^2Y_{\m\n}Y^{\m\n} \;. 
} 
\eeq 
Instead of the field equation (\ref{eqcon1}), we now obtain quite analogously
\beq\label{many2} 
\eqalign{ 
  0 &= \frac{\d}{\d\Con{a}\m{b}}({\cal L}_m+{\cal L}_G+{\cal L}_Y) 
       \cdot\d^\c{}_{\!\m}e^{a\a} e_b{}^\b\cdot\frac{k}{g} 
\\ 
  &=\phantom{+}  
  \frac{1}{4}i\planck^2 
  \sum_z \big[\PSI_z\c^\c\s^{\b\a}\psi_z 
             +\frac{\varepsilon}{4}\PSI_z\c^\c\psi_z g^{\a\b}\big] 
\\ 
  &\phantom{=} 
   -\frac{1}{2}\big[ \Sigma^{\b\e}{}_\e g^{\a\c} 
                    +\Sigma^\e{}_{\!\e}{}^\a g^{\c\b} 
                    -\Sigma^{\b\a\c} -\Sigma^{\c\b\a}\big] 
\\ 
  &\phantom{=} 
   -l^2 g^{\a\b}\nabla^\ast_{\!\n}Y^{\n\c} \;. 
} 
\eeq 
This equation can be handled just in the same way as in the discussion 
following (\ref{eqcon1}) by using the 4-vector decomposition and contraction 
techniques. For example, if we use the expression (\ref{eq4v}) and contract 
(\ref{many2}) with $g_{\a\b}$, then we obtain now  
\beq\label{many3} 
  0 = i\planck^2\sum_z\varepsilon(z)\PSI_z\c^\c\psi_z 
     -4l^2\nabla^\ast_{\!\n}Y^{\n\c} 
\eeq 
instead of (\ref{eqcon2c}). In a similar fashion, we get for the vectors 
$U^\a$, $Q^\a$ and $V_\d$ 
\beq\label{many4} 
  -Q^\a = U^\a = -i\planck^2/4\sum_z\PSI_z\c^\a\psi_z \quad\mbox{and}\quad 
  V_\d = 3\planck^2\sum_z\PSI_z\c^5\c_\d\psi_z \;, 
\eeq 
to be compared with (\ref{quvj}). The resultant connection is formally the 
same as in (\ref{eqcon3}), but the vector and axial currents occuring in 
(\ref{eqcon4}) have to be replaced by the sums of individual currents via 
(\ref{many4}), thus, 
\eqlabel{many4a} 
\begin{eqroman} 
\baq 
  \Con{a}\m{b} &=& {\widehat{\Gamma}}^a{}_{\m b} + \d^a{}_b S_\m \;, 
\label{many4aa}\\ 
  {\widehat{\Gamma}}^a{}_{\m b} &=& 
  \Chr{a}\m{b} + \frac{1}{4}\planck^2\sum_z\big( 
  i\PSI_z\c^a\psi e_{b\m}-ie^a{}_\m\PSI_z\c_b\psi 
  -\eta^a{}_{\m bd}\PSI_z\c^5\c^d\psi_z\big) \;. 
\label{many4ab} 
\eaq%
\end{eqroman}%
  
We remark, that (\ref{many3}) is the correct inhomogeneous Maxwell equation 
for the many-particle theory: In view of (\ref{ymm}), (\ref{s-a}) and 
(\ref{l}), we can rewrite it as 
\beq\label{many5} 
  \sum_z\varepsilon(z)e\cdot\PSI_z\c^\c\psi_z = \nabla^\ast_{\!\n}F^{\n\c} \;. 
\eeq 
     
Let us now discuss the Dirac equation. It is easy to see that the Dirac 
equation (\ref{dirac2}) suffers minor changes only, 
\beq\label{many6} 
  i\c^\m(\nabla^\ast_{\!\m}+\varepsilon(z)S_\m)\psi_z 
  -\frac{m_z c}{\hbar}\psi_z 
  +\big(\frac{3}{2}iU_\m+\frac{1}{8}V_\m\c^5\big)\c^\m\psi_z = 0 \;, 
\eeq                   
where the vectors $U_\m$ and $V_\m$ are now given by (\ref{many4}). Thus, 
with the help of the identity (\ref{pauli}), this can be reexpressed as 
\beq\label{many7} 
  i\c^\m(\nabla^\ast_{\!\m}+\varepsilon(z)S_\m)\psi_z 
  -\frac{m_z c}{\hbar}\psi_z 
  -\frac{3}{8}\planck^2\sum_{z'\neq z}  
  (\PSI_{z'}\c_\m\psi_{z'}+\PSI_{z'}\c^5\c_\m\psi_{z'}\c^5)\c^\m\psi_z = 0 \;. 
\eeq 
This Dirac equation contains clearly a spin-spin contact interaction, which, 
however, differs from the interaction in the Einstein--Cartan theory, cf.\ 
(\ref{ecdirac}). The interaction term in the Dirac equation (\ref{many7}) 
contains besides the axial currents also the vector currents and allows 
therefore only interactions between distinct particles. So, at least 
on the classical level, both contact interactions differ significantly. 

The field equation for energy-momentum (\ref{eqtet}) gains a new spin-spin 
interaction term $W_{\a\b}$ on the right-hand side,%
\footnote{We leave out the detailed computations, since they are rather 
tedious.} 
\eqlabel{many8} 
\begin{eqroman} 
\baq
  W_{\a\b} &=& \frac{3}{k}(-U_\m U^\m+\frac{1}{12^2}V_\m V^\m) 
\label{many8a}\\ 
  &=& 
  \frac{3}{16k}\planck^4\big( \sum^{}_z \PSI_z\c_\m\psi_z 
  \sum^{}_{z'}\PSI_{z'}\c^\m\psi_{z'} 
  +\sum^{}_z\PSI_z\c^5\c_\m\psi_z\sum^{}_{z'}\PSI_{z'}\c^5\c^\m\psi_{z'} 
  \big)g_{\a\b} \qquad\;\;
\label{many8b}\\ 
  &=& 
  \frac{3}{8k}\planck^4\sum^{}_{z\neq z'}
  \left( \PSI_z\c_\m\psi_z\,\PSI_{z'}\c^\m\psi_{z'} 
        +\PSI_z\c^5\c_\m\psi_z\,\PSI_{z'}\c^5\c^\m\psi_{z'} \right) 
  g_{\a\b} \;,
\label{many8c} 
\eaq%
\end{eqroman}%
and also contains the energy-momentum tensors of the individual spinor fields
$\psi_z$. The result is  
\beq\label{many9} 
\eqalign{   
  \frac{1}{k}G^\ast_{\a\b} &= 
  \sum_z\frac{i\hbar c}{4} 
  \big[ \PSI_z\c_a(\nabla^\ast_{\!\b}+\varepsilon(z)S_\b)\psi_z 
       -(\nabla^\ast_{\!\b}-\varepsilon(z)S_\b)\PSI_z\c_\a\psi_z 
       +(\a\leftrightarrow\b) \big] 
\\ 
  &\phantom{=}+\frac{16}{k}l^2 
   \big[ S_{\a\c}S_\b{}^\c-\frac{1}{4}g_{\a\b}S_{\m\n}S^{\m\n} \big] 
\\ 
  &\phantom{=}+W_{\a\b} \;. 
} 
\eeq 
If we compare this equation with the corresponding energy-momentum equation 
(\ref{ecenergy}) of the Einstein--Cartan theory, then, besides the 
additional contributions from vector currents in (\ref{many8}), also the 
doubled factor $3/8$ instead of $3/16$ occures. This 
is due to the summation of the various contact interaction terms, where 
each interaction between two distinct Dirac fields was counted twice, when 
the basic expression (\ref{many8a}) is reexpressed through the individual  
currents via (\ref{many4}) as in (\ref{many8b}). 

As has been already noted in \cite{paper}, the vanishing of the 
self-interacting terms in (\ref{many8}) and also in the Dirac equation 
(\ref{many7}) are formally due to the identity (\ref{pauli}) and have 
their origin in our special choice of ${\cal L}_m$ 
in (\ref{lagrangian}), where the adjoint covariant derivative of 
$\PSI$ is missing. Usually, the matter Lagrangian is required to 
be real, necessitating the inclusion of both derivatives of $\psi$ and 
$\PSI$, cf.\ \cite{heh71}. Since in (\ref{lagrangian})  
the Lagrangians ${\cal L}_G$ and ${\cal L}_Y$ were 
already complex, there was no need to make ${\cal L}_m$ alone real 
valued by including the adjoint spinor derivative. 
Stated differently, if 
the Lagrangian of the Einstein--Cartan theory, which normally contains both 
dervivatives of $\psi$ and $\PSI$ (see for example \cite{heh71}), is 
changed by omitting the adjoint covariant derivative of $\PSI$, then 
the self-interaction terms in (\ref{ecenergy}) and in (\ref{ecdirac}) 
will change and become the same as in our theory. Thus, for the 
single-particle case, these interaction terms will  
vanish, and we must also consider in the Einstein--Cartan theory 
a many-particle theory to discover the spin-spin contact interaction.

        \section{Apparent universality of the contact interaction} 

\subsection{Einstein--Cartan theory} 

Kerlick \cite{ker75} and Rumpf \cite{rum79} concluded that the 
spin-spin contact interaction of the Einstein--Cartan theory is 
{\em universal}, that is, it does not depend on the matter type (whether 
particles or anti-particles) considered. It is attractive for Dirac fields 
with opposite spins and repulsive for aligned spins \cite{ker75}. 

To see how these authors argued in this context we briefly discuss 
the work of Rumpf 
\cite{rum79}, p.\ 649, using our notations. Consider the Dirac equation 
(\ref{ecdirac}) in a special Riemann--Cartan spacetime with 
flat metric $g_{\a\b} = \eta_{\a\b}$ like in Minkowski spacetime, but 
with non-vanishing torsion, thus, 
\beq\label{univ1} 
  i\hbar c\c^\m\partial_\m\psi - mc^2\psi 
  + \frac{3}{8}\planck^2\hbar cj^{{\sssty 5}\,\d}\c^5\c_\d\psi = 0 \;. 
\eeq 
Here the axial current $\PSI\c^5\c^\d\psi$ has been replaced by a 
{\em background} field $j^{{\sssty 5}\,\d}$, so that the spinor $\psi$ 
loses its cubic self-interaction. This replacement means that the 
totally antisymmetric torsion field in (\ref{ectorsion}) is solely caused 
by this background field. We may imagine that this axial current is due to 
a constant classical background Dirac field at rest, which in addition is 
polarized in the $z$ direction, 
\beq\label{univ2} 
  \psi_{bg} := \sqrt{n} 
  \left(\begin{array}{c} 1 \\ 0 \\ 0 \\ 0 \end{array}\right) 
  e^{-imc^2/\hbar\cdot t} \;, 
\eeq 
where $n$ is the constant particle density. Noting that 
\beq\label{univ3} 
  H := i\hbar c\partial_0 = i\hbar\partial_t 
\eeq 
is the Hamiltonian of the Dirac field $\psi$ in (\ref{univ1}), we can first 
compute the axial current $j^{{\sssty 5}\,\d}$ from $\psi_{bg}$ and then 
reexpress (\ref{univ1}) as follows (see (\ref{algebra3a})) 
\beq\label{univ4} 
  H\psi = -i\hbar c\c^0\vec{\c}\cdot\nabla\psi + mc^2\c^0\psi 
          +\frac{3}{8}\planck^2\hbar cn 
          \left(\begin{array}{cc} \s^3 & 0 \\ 0 & \s^3 \end{array}\right)\psi 
  \;. 
\eeq 
Here the symbol $\nabla$ denotes the ordinary gradient vector, and 
$\vec{\c}$ stands for $\vec{\c} = (\c^1,\c^2,\c^3)$. 
We can solve this eigenvalue equation by the ansatz of a free wave polarized 
in the positive (negative) $z$ direction ($N$ is a normalization constant) 
\beq\label{univ5} 
  \psi_\uparrow := 
  N\left(\begin{array}{c} 1 \\ 0 \\ 0 \\ 0 \end{array}\right) 
  \exp(-\frac{i}{\hbar}p_\m x^\m) 
  \qquad\mbox{and}\qquad 
  \psi_\downarrow := 
  N\left(\begin{array}{c} 0 \\ 1 \\ 0 \\ 0 \end{array}\right) 
  \exp(-\frac{i}{\hbar}p_\m x^\m) \;. 
\eeq 
The energy eigenvalues of these two solutions read 
\eqlabel{univ6} 
\begin{eqroman} 
\baq 
  E_{\uparrow\uparrow} &=& mc^2+\frac{3}{8}\planck^2\hbar cn \;;
\label{univ6a}\\ 
  E_{\uparrow\downarrow} &=& mc^2-\frac{3}{8}\planck^2\hbar cn \;,
\label{univ6b} 
\eaq%
\end{eqroman}%
where the arrow-subscipts at $E$ denote the spin directions of the background 
field and that of the test particle. It may be easily checked that this 
result remains true if we consider Dirac anti-particles rather than particles 
as test fields \cite{ker75}. Since the energy level is raised (lowered) if the 
spins are parallel (anti-parallel) we may conclude that the spin-spin contact 
interaction is repulsive for aligned spins and attractive for opposite spins. 
Since this feature does not depend whether one considers ordinary matter or 
anti-matter one may speak of a {\em universal} spin-spin contact interaction.

\subsection{The new spin-spin contact interaction} 

The situation encountered above changes if we consider the 
new spin-spin contact interaction in (\ref{many8}). Instead of 
(\ref{univ1}), we now have 
\beq\label{univ7} 
  i\hbar c\c^\m\partial_\m\psi - mc^2\psi 
  + \frac{3}{8}\planck^2\hbar c 
  (j^\d + j^{{\sssty 5}\,\d}\c^5)\c_\d\psi = 0 \;. 
\eeq 
Proceeding in exactly the same manner as above, we obtain for the energies 
of the test particles (see (\ref{algebra5})) 
\eqlabel{univ8} 
\begin{eqroman} 
\baq 
  E_{\uparrow\uparrow} &=& mc^2 \;; 
\label{univ8a}\\ 
  E_{\uparrow\downarrow} &=& mc^2-2\cdot\frac{3}{8}\planck^2\hbar cn \;. 
\label{univ8b} 
\eaq%
\end{eqroman}%
Contrary to the Einstein--Cartan theory discussed above, there is no 
observable force between aligned spins, whereas the attractive force between 
opposite spins is now twice as strong as before. Furthermore, now the energy 
shifts of a test field describing anti-matter in a background torsion are 
not equal to (\ref{univ8}), but are given by (see (\ref{algebra6}) and 
(\ref{algebra4b})) 
\eqlabel{univ9} 
\begin{eqroman} 
\baq 
  E'_{\uparrow\uparrow} &=& mc^2+2\cdot\frac{3}{8}\planck^2\hbar cn\;; 
\label{univ9a}\\ 
  E'_{\uparrow\downarrow} &=& mc^2 \;. 
\label{univ9b} 
\eaq%
\end{eqroman}%
We see here that opposite spins do not feel any force acting between them. 
However, the repulsive force between aligned spins turns out to be stronger 
than in (\ref{univ8}). From (\ref{univ8}) and(\ref{univ9}) it follows that 
the universality of the contact interaction is lost now: The energy shifts 
due to the contact interaction between an ordinary background matter field 
and a test particle describing ordinary matter differs from the case where 
the test particle describes an anti-matter field.

        \section{Quantizing the contact interaction}

In the last section we have obtained the energy shifts of a Dirac field 
caused by contact interaction terms. This was simply done by finding the 
energy eigen\-values of the modified Dirac equations. One drawback of this 
procedure is that the Dirac fields are not second-quantized, so that their 
Fermi-Dirac statistics are completely disregarded. This is particularly 
unsatisfactory, since {\it ``the only source of 
torsion is intrinsic fundamental-particle spin. ... Thus, torsion is 
fundamentally a microscopic, quantum mechanically related phenomenon''} 
\cite{sto85}. 

In this section we therefore quantize the contact interaction term and 
investigate, how the resulting interaction Hamiltonian acts on various 
quantum two-particle states. In this way, we shall obtain more detailed 
informations about the shifts of energy levels of Dirac particles. 
For example, the contact interaction will turn out to be 
{\em non-universal} even in the case of Einstein--Cartan theory. 

The theoretical method applied for this study is simply the first Born 
approximation. Thus, we only consider first-order 
reactions caused by the contact interaction Hamiltonian. It is well-known 
that four-fermion contact interactions as considered here, which are of the  
similar structure as the phenomenological Fermi interaction of the 
weak forces, lead to non-renormalizable theories. We will argue below 
why it is yet sufficient to study the contact interaction only in the 
first Born approximation.

\subsection{Interaction Hamiltonian} 

We begin with determining the effective Lagrangian density of the 
many-particle system considered in the first section. For the sake of 
simplicity, we take a two-particle system consisting of two arbitrary 
charged Dirac spinors $\psi_1 = \psi$ and $\psi_2 = \chi$, having masses 
$m_1 = m$ and $m_2 = n$ and charges $\vare_1$ and $\vare_2$, respectively. 
Now, if we insert the field equation (\ref{eqcon3}) for the connection 
together with (\ref{many4}) into the basic Lagrangian density 
(\ref{many1}), this Lagrangian density may be reexpressed (see eq.\ 
(\ref{algebra12}) to (\ref{algebra15})) as 
\beq\label{quant1} 
\eqalign{ 
  {\cal L} &= g\,i\hbar c[ \PSI\c^\m(\nabla^\ast_{\!\m}+\vare_1 S_\m)\psi
                          +\frac{imc}{\hbar}\PSI\psi ] 
             +g\,i\hbar c[ \CHI\c^\m(\nabla^\ast_{\!\m}+\vare_2 S_\m)\chi
                          +\frac{inc}{\hbar}\CHI\chi ] 
\\ 
  &\phantom{=}-\frac{g}{2k}R^\ast 
              +g\frac{l^2}{4k}S_{\m\n}S^{\m\n} 
              +g\frac{3}{8}\frac{\planck^4}{k} 
  [\PSI\c_\m\psi\,\CHI\c^\m\chi+\PSI\c^5\c_\m\psi\,\CHI\c^5\c^\m\chi] \;. 
} 
\eeq 
The last term is the {\em spin-spin interaction Lagrangian} denoted 
henceforth by 
\beq\label{quant2} 
  {\cal L}_I = g\frac{3}{8}\frac{\planck^4}{k} 
  [\PSI\c_\m\psi\,\CHI\c^\m\chi+\PSI\c^5\c_\m\psi\,\CHI\c^5\c^\m\chi] \;.  
\eeq 

Since we want to investigate only this contact interaction, we neglect 
the effects of gravity and electromagnetism. Thus, we set the charges to 
zero and employ from now on flat Minkowski spacetime with constant 
Minkowski metric 
\beq\label{quant3} 
  g_{\m\n} = \eta_{\m\n} = \mbox{diag}(1,-1,-1,-1) \;. 
\eeq 
Thus, now the density factor $g$ in (\ref{quant2}) equals 1 and therefore 
may be omitted in (\ref{quant1}). 

To obtain the interaction Hamiltonian $H_I$ to be quantized 
in the following, we must subject (\ref{quant2}) to the well-known 
Legendre transformation 
\beq\label{legendre} 
  H = \int d^3\!\!x[\sum_a\pi_a\partial_t\phi_a - {\cal L}] \;, 
\eeq
where ${\cal L}$ is an arbitrary Lagrangian density, depending on general 
fields, here denoted by $\phi_a$. The fields $\pi_a$ denote the conjugate 
fields, 
\beq\label{quant4} 
  \pi_a = \frac{\partial {\cal L}}{\partial[\partial_t\phi_a]} \;, 
\eeq 
see for example \cite{itz80}. Since in ${\cal L}_I$ in (\ref{quant2}) 
there are no derivative terms present, we obtain the corresponding 
interaction Hamiltonian simply by the change of sign, 
\beq\label{quant5} 
  H_I = -\int d^3\!\!x{\cal L}_I 
      = -\frac{3}{8}\frac{\planck^4}{k}\int_V d^3\!\!x 
  [\PSI\c_\m\psi\CHI\c^\m\chi+\PSI\c^5\c_\m\psi\CHI\c^5\c^\m\chi] \;.  
\eeq 
Here the space integration is to be performed only in a volume $V$ in 
order to obtain finite results later on.

\subsection{Quantization procedure} 

\subsubsection{Notation} 

We quantize the Dirac fields $\psi$ and $\chi$ using the usual canonical 
quantization procedure \cite{itz80}. Our notation is as follows: The 
{\em operators} $\psi$ and $\chi$ are expanded in terms of $c$-number 
plane wave solutions $u$, $v$, $s$, and $w$ of the ordinary Dirac equations 
in the Minkowski spacetime and operator-valued amplitudes $a$, $\adag$, $b$, 
$\bdag$, $c$, $\cdag$, $d$, and $\ddag$, 
\eqlabel{quant6} 
\begin{eqroman}%
\baq 
  \psi &=& \int\frac{d^3\!\!p}{(2\pi)^3}\frac{m}{p^0} 
           \big[\bdag_{rp}v_r(p)e^{ipx}+a_{rp}u_r(p)e^{-ipx}\big] \;, 
\label{quant6a}\\ 
  \chi &=& \int\frac{d^3\!\!p}{(2\pi)^3}\frac{n}{p^0} 
           \big[\ddag_{rp}w_r(p)e^{ipx}+c_{rp}s_r(p)e^{-ipx}\big] \;. 
\label{quant6b} 
\eaq%
\end{eqroman}%
We have suppressed the summation over the spin $r$, which takes the value 
$r = +\frac{1}{2}$ when the spin is parallel to the positive $x^3$-direction 
and the negative value $r = -\frac{1}{2}$ when the spin is anti-parallel. 
Further, $p$ denotes the 4-momentum with the condition $p^0 = 
\sqrt{m^2+\vec{p}^{\,2}} > 0$ in (\ref{quant6a}) and $p^0 = 
\sqrt{n^2+\vec{p}^{\,2}} > 0$ in (\ref{quant6b}), respectively. 
The plane wave solutions $u$, $v$, $s$, and $w$ are explicitly given in 
(\ref{algebra17}) and were taken from \cite{nac90}. The pleasant feature of 
these plane wave solutions is that the waves describing the anti-matters, 
$v$ and $w$, have the right spin directions: For example, 
$v_{+1/2}(p)$ describes an anti-matter wave solution with its spin in the 
positive $x^3$-direction. For our purposes here, we do not need the  
explicit expressions, but only the normalization conditions. These are 
given by (see \cite{nac90} and (\ref{algebra17}))
\eqlabel{quant8} 
\begin{eqroman} 
\baq 
  \ubar_r(p)u_{r'}(p) \;=\; -\vbar_r(p)v_{r'}(p) &=& V^{-1}\d_{rr'} \;;
\label{quant8a}\\ 
  \sbar_r(p)s_{r'}(p) \;=\; -\wbar_r(p)w_{r'}(p) &=& V^{-1}\d_{rr'} \;,
\label{quant8b} 
\eaq 
\end{eqroman} 
and 
\eqlabel{quant9} 
\begin{eqroman} 
\baq 
  \ubar_r(p)\c^\m u_{r'}(p) \;=\; \vbar_r(p)\c^\m v_{r'}(p) &=& 
  V^{-1}\frac{p^\m}{m}\d_{rr'} \;;
\label{quant9a}\\ 
  \sbar_r(p)\c^\m s_{r'}(p) \;=\; \wbar_r(p)\c^\m w_{r'}(p) &=& 
  V^{-1}\frac{p^\m}{n}\d_{rr'} \;.
\label{quant9b}\\ 
\eaq%
\end{eqroman}%
In calculating the energy shifts below, the following {\em axial vector}  
expression built from an arbitrary 4-momentum $p^\m$ with intrinsic 
rest mass $M:=\sqrt{p^\m p_\m}$  becomes useful: 
\beq\label{axialflow} 
  p^{5\m} := \left(\begin{array}{l}%
             \qquad p^3 \\ p^3 p^1/(p^0+M) \\ p^3 p^2/(p^0+M) \\ 
                        p^3 p^3/(p^0+M)+M \end{array}\right) \;. 
\eeq 
We can then employ this axial vector to express the axial currents of the 
plain wave solutions, see eqs.\ (\ref{algebra18}) to (\ref{algebra21}) 
\eqlabel{quanta} 
\begin{eqroman} 
\baq 
  \ubar_r(p)\c^5\c^\m u_r(p) \;=\; -\vbar_r(p)\c^5\c^\m v_r(p) &=& 
  \mp V^{-1}\frac{1}{m}p^{5\m} \quad\cdots r=\pm\frac{1}{2} \;;\quad
\label{quantaa}\\ 
  \sbar_r(p)\c^5\c^\m s_r(p) \;=\; -\wbar_r(p)\c^5\c^\m w_r(p) &=& 
  \mp V^{-1}\frac{1}{n}p^{5\m} \quad\cdots r=\pm\frac{1}{2} \;.\quad 
\label{quantab} 
\eaq%
\end{eqroman}%
Note that $p^{5\m}$ are not equal in the both formulae, since the rest 
masses of the particles are different in (\ref{axialflow}). Also, if we 
insert the explicit formula (\ref{axialflow}) into (\ref{quanta}), we see 
that the $p^3$ component of the 4-momentum $p^\m$ is not treated in the same 
way as the other two space-like components $p^1$ and $p^2$. The 
reason is simply that we have chosen the plane wave solutions to be 
polarized entirely in the $x^3$ direction and thus distinguished this space 
direction. 

The operator-valued amplitudes $a$, $\adag$, $b$, $\bdag$, $c$, $\cdag$, $d$, 
and $\ddag$ satisfy the anti-commutation relations \cite{nac90}
\eqlabel{quant10} 
\begin{eqroman} 
\baq 
  \{a_{rp},\adag_{r'p'}\} \;=\; \{b_{rp},\bdag_{r'p'}\} &=& 
  (2\pi)^3\frac{p^0}{m}\d^3(\vec{p}-\vec{p'})\d_{rr'} \;;
\label{quant10a}\\ 
  \{c_{rp},\cdag_{r'p'}\} \;=\; \{d_{rp},\ddag_{r'p'}\} &=& 
  (2\pi)^3\frac{p^0}{n}\d^3(\vec{p}-\vec{p'})\d_{rr'} \;, 
\label{quant10b} 
\eaq%
\end{eqroman}%
where all other anti-commutations vanish. The interpretations of 
these operators are as usual: For example, $\adag_{rp}$ generates a particle 
having 4-momentum $p$ and $x^3$-component of spin $r$, whereas $\bdag_{rp}$ 
generates the corresponding anti-particle state with the spin also in the 
$x^3$-direction. 

\subsubsection{Quantization} 

We quantize the interaction Hamiltonian (\ref{quant5}) according to the 
canonical quantization procedure. Thus we merely replace the fields $\psi$ 
and $\chi$ and their adjoint fields by the corresponding operator 
expressions (\ref{quant6a}) and (\ref{quant6b}), respectively. However, to 
yield finite results, we must {\em normal-order} the operator expressions. 
Denoting this normal-ordering by $:$ we obtain the following interaction 
Hamiltonian operator: 
\beq\label{quant11} 
  \widehat{H}_I = -\frac{3}{8}\frac{\planck^4}{k}\int_V d^3\!\!x 
  \big[ :\PSI\c_\m\psi\,\CHI\c^\m\chi: 
       +:\PSI\c^5\c_\m\psi\,\CHI\c^5\c^\m\chi:\big] \;.  
\eeq 
We remark that this operator expression does not vanish even if 
$\chi$ is replaced by $\psi$ and $\CHI$ by $\PSI$, that is, if the contact 
interaction is considered between the same kind of particles. This 
feature differs of course from the classical expression $H_I$ in 
(\ref{quant5}), which vanishes for identical Dirac fields due to the 
Pauli relation (\ref{pauli}). Note that, if we had written  
the interaction Hamiltonian operator as follows, 
\beq\label{quant11b} 
  \widehat{H}_I = -\frac{3}{8}\frac{\planck^4}{k}\int_V d^3\!\!x 
  \big[ :\PSI\c_\m\psi::\CHI\c^\m\chi: 
       +:\PSI\c^5\c_\m\psi::\CHI\c^5\c^\m\chi:\big] \;, 
\eeq 
we would have obtained infinite results, because the field operators $\psi$ and 
$\PSI$ on the one hand and $\chi$ and $\CHI$ on the other hand are 
taken at the same spacetime point, as can be verified by an explicit 
computation. 

Note also that the corresponding interaction Hamiltonian operator 
$\widehat{H}_{ECT}$ for the spin-spin contact interaction in the 
Einstein--Cartan theory reads 
\beq\label{quant12} 
  \widehat{H}_{ECT} = -\frac{3}{8}\frac{\planck^4}{k}\int_V d^3\!\!x 
  \big[ :\PSI\c^5\c_\m\psi\CHI\c^5\c^\m\chi:\big] \;.  
\eeq 
Whereas the vanishing of the vector current contribution in comparison with 
(\ref{quant11}) is clear from the 
classical interaction expression in (\ref{ecenergy}), the appearance of the 
same factor $3/8$ seems strange, if we compare the classical expressions
(\ref{ecenergy}) with 
(\ref{many8}). However, to study the spin-spin contact interactions of our 
theory and that of the Einstein--Cartan theory on the equal basis, we must 
consider in the Einstein--Cartan theory also a many-particle theory instead 
of a single-particle theory as presented in the introducing chapter. If this 
is done, then the torsion $T_{\a\b\c}$ in (\ref{ectorsion}) is no longer 
produced by only one Dirac field $\psi$ but by the sum of many different 
fields. In the same fashion as explained in connection with (\ref{many9}), 
this leads to the double factor $3/8$ in (\ref{ecenergy}) instead of $3/16$.

\subsection{Evaluation on two-particle states}

To investigate the energy shifts of the two-particle system (\ref{quant1}) 
due to the contact interaction we evaluate the expectation values of the
interaction Hamiltonian $\widehat{H}_I$ between the following two-particle 
states: 
\eqlabel{quant13} 
\begin{eqroman} 
\baq 
  |1\rangle &:=& \cdag_{rq}\adag_{r'q'}|0\rangle \;;
\label{quant13a}\\ 
  |2\rangle &:=& \ddag_{rq}\bdag_{r'q'}|0\rangle \;;
\label{quant13b}\\ 
  |3\rangle &:=& \cdag_{rq}\bdag_{r'q'}|0\rangle \;;
\label{quant13c}\\  
  |1o\rangle &:=& \adag_{rq}\adag_{r'q'}|0\rangle \;;
\label{quant13d}\\ 
  |2o\rangle &:=& \bdag_{rq}\bdag_{r'q'}|0\rangle \;;
\label{quant13e}\\ 
  |3o\rangle &:=& \bdag_{rq}\adag_{r'q'}|0\rangle \;.
\label{quant13f} 
\eaq%
\end{eqroman}%
The first state $|1\rangle$ consists of two particles of different kind, the 
second state $|2\rangle$ is built from two anti-particles, and the third 
state $|3\rangle$ contains one particle and one antiparticle. The last three 
states $|1o\rangle$, $|2o\rangle$ and $|3o\rangle$ describe corresonding 
two-particle states consisting of two identical particles (but of course in 
different states). 

The calculations of the expectation values of $\widehat{H}_I$ 
(\ref{quant11}) are standard, see \ref{expectation}. We introduce the symbol 
\beq\label{quant14} 
  \s(r,r') := \left\{\begin{array}{c}+1\mbox{ if }r=r' \\
                                     -1\mbox{ if }r\neq r' \end{array} 
              \right. 
\eeq 
and obtain 
\eqlabel{quant15} 
\begin{eqroman} 
\baq 
  \langle 1|\widehat{H}_I|1\rangle &=& -\frac{3}{8}\frac{\planck^4}{kVmn} 
  [\phantom{-}q^\m q'_\m + \s(r,r')q^{5\m}q'^5_\m] \;; 
\label{quant15a}\\ 
  \langle 2|\widehat{H}_I|2\rangle &=& -\frac{3}{8}\frac{\planck^4}{kVmn} 
  [\phantom{-}q^\m q'_\m + \s(r,r')q^{5\m}q'^5_\m] \;; 
\label{quant15b}\\ 
  \langle 3|\widehat{H}_I|3\rangle &=& -\frac{3}{8}\frac{\planck^4}{kVmn} 
  [-q^\m q'_\m + \s(r,r')q^{5\m}q'^5_\m] \;; 
\label{quant15c}\\ 
  \langle 1o|\widehat{H}_I|1o\rangle &=& -\frac{3}{2}\frac{\planck^4}{kVm^2} 
  [\phantom{-}q^\m q'_\m + \s(r,r')q^{5\m}q'^5_\m] \;; 
\label{quant15d}\\ 
  \langle 2o|\widehat{H}_I|2o\rangle &=& -\frac{3}{2}\frac{\planck^4}{kVm^2} 
  [\phantom{-}q^\m q'_\m + \s(r,r')q^{5\m}q'^5_\m] \;; 
\label{quant15e}\\ 
  \langle 3o|\widehat{H}_I|3o\rangle &=& -\frac{3}{2}\frac{\planck^4}{kVm^2} 
  [-q^\m q'_\m + \s(r,r')q^{5\m}q'^5_\m] \;. 
\label{quant15f} 
\eaq%
\end{eqroman}%
In the first three expectation values, which are taken for two different 
kinds of particles, the first term $p^\m p'_\m$ exactly corresponds to 
the vector-vector term in $\widehat{H}_I$ (\ref{quant11}), whereas the 
second summand $\s(r,r')q^{5\m}q'^5_\m$ is exactly the axial-axial term, cf.\ 
(\ref{quanta}). 
On the other hand, such a simple decomposition does not apply for the last 
three expectation values based on two identical particles: In order to  
obtain these simple expressions (\ref{quant15d}) to (\ref{quant15f}), 
one has to use the Fierz transformation rule (see \ref{expectation}) to order 
the entanglement of the various plain waves, which has its origin in the 
exchange degeneracy of identical fermions. Also, due to the greater 
statistical freedom of an identical particle system, the last three 
expectation values are 4 times the first three expressions.

The expectation values for the Hamiltonian $\widehat{H}_{ECT}$ 
(\ref{quant12}) of the Einstein--Cartan theory read (cf.\ \ref{expectation}) 
\eqlabel{quant16} 
\begin{eqroman} 
\baq 
  \langle 1|\widehat{H}_{ECT}|1\rangle &=& -\frac{3}{8}\frac{\planck^4}{kVmn} 
  [\s(r,r')q^{5\m}q'^5_\m] \;; 
\label{quant16a}\\ 
  \langle 2|\widehat{H}_{ECT}|2\rangle &=& -\frac{3}{8}\frac{\planck^4}{kVmn} 
  [\s(r,r')q^{5\m}q'^5_\m] \;; 
\label{quant16b}\\ 
  \langle 3|\widehat{H}_{ECT}|3\rangle &=& -\frac{3}{8}\frac{\planck^4}{kVmn} 
  [\s(r,r')q^{5\m}q'^5_\m] \;; 
\label{quant16c}\\ 
  \langle 1o|\widehat{H}_{ECT}|1o\rangle &=& 
  -\frac{3}{8}\frac{\planck^4}{kVm^2} 
  [2m^2+q^\m q'_\m + 3\s(r,r')q^{5\m}q'^5_\m] \;; 
\label{quant16d}\\ 
  \langle 2o|\widehat{H}_{ECT}|2o\rangle &=& 
  -\frac{3}{8}\frac{\planck^4}{kVm^2} 
  [2m^2+q^\m q'_\m + 3\s(r,r')q^{5\m}q'^5_\m] \;; 
\label{quant16e}\\ 
  \langle 3o|\widehat{H}_{ECT}|3o\rangle &=& 
  -\frac{3}{8}\frac{\planck^4}{kVm^2} 
  [2m^2-q^\m q'_\m + 3\s(r,r')q^{5\m}q'^5_\m] \;. 
\label{quant16f} 
\eaq%
\end{eqroman}%
The first three expressions (\ref{quant16a}) to (\ref{quant16c}) can be 
obtained simply by neglecting the vector-vector interacton parts in the 
corresponding results (\ref{quant15a}) to (\ref{quant15c}). This can be 
directly understood in terms of the underlying expressions for the 
interaction Hamiltonian $\widehat{H}_I$ and $\widehat{H}_{ECT}$, which 
differ only in the vector-vector interaction term. On the other hand, such 
a simple understanding can not be given for the last three expressions in
(\ref{quant16}).

  \subsection{Discussion}

Let us first discuss the contact interaction between different kinds of 
particles. Whereas (\ref{quant15a}) to 
(\ref{quant15b}) are obviously non-universal, the corresponding expectation 
values of the Einstein--Cartan theory (\ref{quant16a}) to (\ref{quant16c}) 
are universal, that is, they do not depend on whether one considers particles 
or anti-particles. Furthermore, if both interacting particles are at rest, 
then $q^{5\m}q'^5_\m = -mn$, as can be easily verified with the help of 
(\ref{axialflow}). 
Thus in this special case, the interaction energy increases (decreases) for 
aligned (opposite) spins in accordance with the result of the 
consideration in (\ref{univ6}).

But when two identical particles interact, then we see from 
(\ref{quant16d}) to (\ref{quant16f}), that the contact interaction fails to 
be universal also in the Einstein--Cartan case. Contrary to the new spin-spin 
contact interaction in (\ref{quant15d}) to (\ref{quant15f}), where merely a 
symmetry factor 4 is needed to adjust the formulae to the case of identical 
particles, the interaction energy in the Einstein--Cartan theory gains 
some miscellaneous contributions due to the Fierz transformation (see 
(\ref{fierz1})). For example, let us consider non-relativistic identical 
particles both having no momentum in the $x^3$-direction, that is, 
$|\vec{q}|, |\vec{q'}| \ll m$ and $q^3 = q'^3 = 0$. Then, from 
(\ref{algebra26}), we obtain for the Einstein--Cartan case 
\beq\label{quant17} 
  \langle 1o|\widehat{H}_{ECT}|1o\rangle =
  \frac{3}{8}\frac{\planck^4}{kV} 
  [3(\s(r,r')-1)-\frac{(\vec{q}-\vec{q'})^2}{2m^2}] \;, 
\eeq 
and, for the new contact interaction, a quite similar result: 
\beq\label{quant18} 
  \langle 1o|\widehat{H}_I|1o\rangle =
  \frac{3}{8}\frac{\planck^4}{kV} 
  [4(\s(r,r')-1)-4\frac{(\vec{q}-\vec{q'})^2}{2m^2}] \;. 
\eeq 
We observe, that in both cases the interaction energy is negative. Thus, 
it is possible that the contact interactions among identical particles 
with aligned spins could be attractive 
in contrast to the statement made by Kerlick, cf.\ (\ref{univ8}). 

We may say that the spin-spin contact interaction of the Einstein--Cartan
theory is in general {\em not universal} and it is not always true that 
aligned spins repell. 

Note that the new spin-spin contact interaction does not allow for 
self-interactions among spinors already on the classical level in contrast 
to the ordinary contact interaction of the Einstein--Cartan theory.

\subsection{Justification of the first Born approximation}

The Born approximation of the contact interaction can also be found in the 
work by Kanno \cite{kan88}. But contrary to our approach, he computed
the expectation value of the contact interaction energy for many-particle 
states with high matter density. Thereby he assumed that these states 
can be approximated by summing up the free wave states of each particles. He 
concluded that there occurs a 
matter--anti-matter segregation due to the contact interaction. 
In my opinion, his approach is not correct since at high densities, a 
quantum-mechanical many-particle system with torsion 
can not be approximated by a sum of 
plane wave states: At high matter density, the torsion (\ref{ectorsion}) 
becomes non-negligible and changes the Dirac equation (\ref{ecdirac}) 
significantly. Therefore, plane wave solutions of the ordinary Dirac 
equation (without the cubic interaction term) do not approximate solutions 
of the Dirac equation with torsion. Thus, it makes no sense to take an 
expectation value of the interaction Hamiltonian between free wave states, 
since no such states exist at high density. 

On the other hand, we have studied the contact interaction between two 
particles, so that the matter density is negligible and the plane wave 
solutions really approximate the solutions of the Dirac equation (cf.\ 
(\ref{many6})) very well. 

Let us now justify why it is legitimate to consider only the first Born 
approximation of the contact interaction. 
It is well known that the phenomenological Fermi contact interaction 
describes the weak interaction very well at low momenta. To be more precise,
the description of the weak force by the contact interaction is valid for 
energies up to the critical value $1/\sqrt{G_{Fermi}} \approx 300\mbox{GeV}$, 
see for example \cite{itz80}. If this value is exceeded, the phenomenological 
contact interaction violates the unitarity. Now, since the torsion-induced 
spin-spin contact interaction has a coupling constant, which is of the 
order of the squared Planck length%
\footnote{Note that in (\ref{quant3}) $\planck^4/k = \planck^2\cdot 
\hbar c$.} 
it is legitimate to consider the first Born approximation for energies up 
to the Planck energy $2.4\cdot 10^{18}\mbox{GeV}$. But at this enormously 
high energy, or, equivalently, at the Planck scale, we would need a quantum 
theory of gravity to describe the physics properly. If we restrict 
ourselves to energies below the Planck scale, then the first Born 
approximation of the contact interaction is physically sensible.

             \chapter{Summary and Outlook} 

\section{Summary} 

In the preceeding chapters I have reexamined and improved the unified field 
theory of gravity and electromagnetism developed in my diploma thesis 
\cite{diplom}. Furthermore, the special spin-spin contact interaction 
predicted by this theory was investigated in detail. 

Although the theory presented here was motivated by earlier works on unified 
field theories \cite{bor76a,mof77,kun79,mck79,fer82,jak85}, in which the 
torsion trace $T_\m$ of a real linear connection was identified with the 
electromagnetic vector potential $A_\m$, the new theory comes to completely 
different conclusions: 
In this new theory, the linear connection resulting from the field equations 
is complex-valued and it is not compatible with the metric, where this 
failure of compatibility is caused by a vector part $S_\m$ of the connection 
(the so-called non-metricity vector). According to the geometrical background 
of the new theory, this vector $S_\m$ can be unambiguously detached from the 
tangent frame bundle of the spacetime manifold and then identified with the 
electromagnetic vector potential on a trivial \uone\ bundle. Contrary to this 
truly geometric identification, the relation between the torsion trace and 
the vector potential on the tangent frame bundle can be obtained only if a 
special \uone\ gauge is chosen and held fixed on the \uone\ bundle. For this 
reason, the long-standing relation between the torsion trace and the 
electromagnetic potential is merely a formal consequence of the geometrical 
background underlying the new theory. 
Furthermore, due to this geometry the whole complex connection resulting 
from the field equations can be decomposed into the vector potential $S_\m$ 
and a Lorentzian connection compatible with the metric, this being done by 
means of pull-back techniques. If we consider the torsion trace of this 
Lorentzian connection part only, it is not related to electromagnetism even 
formally. Thus, in the end, the torsion trace is not related to 
electromagnetic phenomena at all. However, it is important to note that this 
conclusion can only be drawn with the help of an investigation of the special 
geometrical background of the new theory. 

This geometrical background has been explained in chapter 3 in every detail, 
thereby clarifying several difficult properties, which were not mentioned in 
the diploma thesis \cite{diplom}: First, the 
notion of a complex spin geometry and its relation to the usual spin geometry 
has been explained rigorously. Secondly, the pull-back procedure, by which an 
unique spinor derivative can be obtained from any complex linear 
connection, has been improved. Thereby it was shown why the \uone\ principal 
bundle accounting for the electromagnetic phase transformation is necessarily 
trivial in the geometrical framework of our theory. Also, the roles played 
by different ``intermediate'' bundles in this pull-back procedure has been 
clarified in detail. Thirdly, the decomposition principle of the linear 
connection, by means of which it is possible to obtain a meaningful theory 
of electromagnetism, has been elaborated. Forthly, the properties of the 
\uone\ gauge transformation in the geometrical framework 
has been investigated. From this gauge structure, we were able to see why it 
is necessary to detach the \uone\ vector potential from the tangent frame 
bundle and to pull it back onto a trivial \uone\ bundle: If, instead, the 
\uone\ gauge transformation is considered on the basic tangent frame bundle 
of the spacetime manifold, every covariant vector field gains a negative 
elementary charge, which is clearly unphysical. Also, the same reasons show 
why it is impossible to introduce a formal \uone\ gauge 
transformation, the so-called $\l$--transformation, for the torsion trace. 

Besides these electromagnetic and geometrical aspects, the new theory also 
incorporates a spin-spin contact interaction between spinning particles. This 
property is shared by none of the unified field theories proposed before and 
is one of the salient features of the new theory. The contact interaction 
is also a characteristic feature of the Einstein--Cartan theory, and has its 
origin in the spin-torsion coupling, by which the spacetime geometry can 
not only respond to mass-energy via the curvature, but also to spin via 
torsion. These properties of the spacetime together with the geometric 
interpretation of electromagnetism of our new gravitational field theory 
lead to the conclusion that the spacetime geometry is able to interact with 
three basic features of elementary particles: Mass-energy, spin, and 
electromagnetic charge. 

Contrary to the ordinary axial current contact interaction of the 
Einstein--Cartan theory, the new contact interaction has 
contributions from both the axial and vector currents of Dirac spinors. 
This has the effect that now there are no self-interactions among Dirac 
fields as was the case in the Einstein--Cartan theory. This feature 
respects the quantum nature (Fermi--Dirac statistics) of elementary fermions 
already on the classical, i.e.\ not second-quantized, level, and makes the  
new contact interaction more favourable than the 
ordinary one. By regarding the energy eigenvalues of 
test fields in an background torsion field, the new spin-spin interaction 
turns out to be non-universal in contrast to the ordinary one: Now the 
interacting force between particle fields is different from the corresponding 
force between a particle and an anti-particle field. This difference 
persists if the contact interaction is quantized. 

The contact interaction has been investigated further on the quantum level
by means of the first Born approximation, similar to the phenomenological 
Fermi contact interaction of the weak forces. It turned out that both the 
new and the ordinary contact interactions, upon quantization, are 
non-universal in the case of identical particles interacting with each 
other. And in this case, if the particle momenta are small, both contact 
interactions have similar structure and are attractive regardless of the 
spin directions of the interacting particles. This result is in sharp 
contrast to the common opinion \cite{ker75}, that the contact interaction
is attractive only if the spins are opposed, but repulsive if they are in 
alignment.

\section{Future research} 

\subsection{Weak interaction} 

Since the unified field theory considered in this work enables the spacetime 
geometry to interact with three fundamental properties of elementary 
particles, namely mass, spin, and charge, it is natural to ask whether it is 
possible to incorporate the weak forces into the geometrical framework 
provided by this theory. 

If we stay on the non-quantum level, this can be hardly achieved by the 
present theory itself, since the theory does not contain charged 
vector boson fields as required for the Weinberg--Salam theory. Thus, it 
seems necessary to further enlarge the spacetime geometry, using, for 
example, an arbitrary covariant spinor derivative instead of a spinor 
derivative built from a complex linear connection. This speculation is 
confirmed by a survey of unified theories of gravity and electroweak 
interaction based upon the ``geometry of the tangent bundle'' instead of a 
spin structure: Without exception these theories \cite{bor76b,nov85,bat84,%
yil89,bat90,nov92} are not acceptable as realistic physical theories. 

The idea of using an enlarged spin structure for the unification of gravity 
and electroweak forces is not new and has already been considered in many 
works, see e.g.\ \cite{nov73,tro87,chi87,chi89}. But in all of these 
works, there is one significant problem which could not be solved rigorously: 
In the Weinberg--Salam theory, the charged $W$-boson couples to electron and 
neutrino via the interaction term (cf.\ \cite{ren90}) 
\[ 
  \sim \PSI_\n\c^\m(1-\c^5)\psi_e W^+_\m \;. 
\] 
To obtain such an interaction between different spinor fields and a 
connection part using the concept of an enlarged spinor derivative, it is 
necessary to introduce a $\mbox{SU}(2)$ theory of the pair of spinors 
$(\psi_\n, \psi_e)$ explicitly or in a different, more indirect manner. 
But if this procedure is followed, the ``unified field theories'' of gravity 
and the weak forces are by no means superior to the standard Weinberg--Salam 
theory itself, because such a ``unified theory'' can also be obtained much 
more easily by embedding the standard model into the pseudo-Riemannian 
geometry of general relativity and adding the Einstein--Hilbert Lagrangian 
$-\frac{g}{2k}R^\ast$ to the Weinberg--Salam Lagrangian. 

Thus, the concept of an enlarged spin geometry alone would not lead to a 
satisfactory unification of gravity and weak interaction. 
In my opinion, it is necessary to consider a quantum field theoretic approach 
together with the enlarged spin geometry rather than a classical field theory 
alone. First hints in this direction are provided by the dynamical 
electroweak symmetry breaking (for a recent review see \cite{kin95}), where 
the electroweak symmetry is broken by a vacuum condensate of fermions. 
This fermion condensate has its origin in a four-fermion contact interaction 
(cf.\ \cite{lal92}) like in the Nambu--Jona-Lasinio model \cite{nam61}. 
Since the new spin-spin contact interaction of our theory is very similar 
to the contact interactions considered in the theories on dynamical symmetry 
breaking, it seems possible that the torsion of the spacetime geometry is 
related to electroweak symmetry breaking.

             \subsection{Contact interaction} 

So far, the effects of the spin-spin contact interaction on cosmology have 
been examined mainly on the classical, that is, non-quantized, level, see 
for example the references \cite{kop72,tra73,heh74,ker75,kuc76,kuc78,nur83}. 
Also, the quantum approach of \cite{kan88} does not seem to be consistent, 
as we have argued in 4.3.5. 

To study the effects of the spin-spin contact interactions in the early 
epoch of the universe, where the matter density was enormously high, we 
must evaluate the thermodynamical average of the interaction energies due 
to the contact interactions. This is not as straightforward as in the 
ordinary case, since the contributions from the non-flat metric and the 
spacetime curvature can not be neglected. 

Another interesting point is the following: The spin-spin contact interaction 
modifies the energy-momentum equation of ordinary general relativity by 
the tensor $W_{\a\b}$, which has a form similar to the contribution 
$\Lambda g_{\a\b}$ of the cosmological constant $\Lambda$ in the Einstein's 
field equation, cf.\ (\ref{many8}). If this similarity is taken seriously, 
then we would obtain a cosmological constant, which is proportional to the 
current-current interaction terms in $W_{\a\b}$. This would imply a time-%
dependent cosmological constant, whose value would have been very high in 
the early epoch of the universe, where the matter density has been very high, 
and whose actual value for the present universe is nearly zero. Such a 
time-dependent cosmological constant is supported by string theoretic 
considerations \cite{lop95}.

\begin{appendix}

              \chapter{4-Vector Decomposition} 
 
Let $\Sigma_{\a\b\c}$ be an arbitrary third rank tensor, which might be 
real or complex valued. Given $\Sigma_{\a\b\c}$ we can define four 
contravariant vectors  $Q_\a$, $S_\b$, $U_\c$, and $V_\d$ and a tensor rest 
$\Upsilon_{\a\b\c}$ in the following way: 
\baq
  \Sigma_{\alpha\beta\gamma} &=:& \frac{1}{18}\Big[\;\;
  (\, 5\Sigma_\alpha{}^\epsilon{}_{\!\epsilon}
  \;\;-\Sigma^\epsilon{}_{\!\alpha\epsilon}
  \;\;-\Sigma^\epsilon{}_{\!\epsilon\alpha}) g_{\beta\gamma} 
\nonumber\\
  && \qquad\;\; +
  (-\Sigma_\beta{}^\epsilon{}_{\!\epsilon}
   +5\Sigma^\epsilon{}_{\!\beta\epsilon} 
   -\Sigma^\epsilon{}_{\!\epsilon\beta}) g_{\alpha\gamma}
\nonumber\\
  && \qquad\;\; +
   (-\Sigma_\gamma{}^\epsilon{}_{\!\epsilon}
    -\Sigma^\epsilon{}_{\!\gamma\epsilon} 
    +5\Sigma^\epsilon{}_{\!\epsilon\gamma}) g_{\alpha\beta}
  \,\Big] + \Sigma_{[\alpha\beta\gamma]}
          + \Upsilon_{\alpha\beta\gamma} 
\label{ap4v1}\\
  &=:&\quad
  Q_\alpha\,g_{\beta\gamma}\;
  +\;S_\beta\,g_{\alpha\gamma}\;
  +\;U_\gamma\,g_{\alpha\beta}\;
  -\;\frac{1}{12}\eta_{\alpha\beta\gamma\delta}V^\delta\;
  +\;\Upsilon_{\alpha\beta\gamma} \;, 
\label{ap4v2} 
\eaq 
where $\Sigma_{[\a\b\c]}$ means the antisymmetrization of its indices, and 
$V^\d$ is given by $V_\d = 2\eta_{\a\b\c\d}\cdot\Sigma^{\a\b\c}$, 
$\eta_{\a\b\c\d}$ being the volume element (\ref{volume}). Note that this 
decomposition is possible if and only if a metric $g_{\m\n}$ is given.  
From (\ref{ap4v1}) we conclude
\baq 
  \Sigma_\alpha{}^\epsilon{}_{\!\epsilon} &=&\,\frac{1}{18}\Big[\,
   20\Sigma_\alpha{}^\epsilon{}_{\!\epsilon}
   -4\Sigma^\epsilon{}_{\!\alpha\epsilon}
   -4\Sigma^\epsilon{}_{\!\epsilon\alpha}
\nonumber\\
  && \qquad\;\;\,
   - \Sigma_\alpha{}^\epsilon{}_{\!\epsilon}
   +5\Sigma^\epsilon{}_{\!\alpha\epsilon} 
   - \Sigma^\epsilon{}_{\!\epsilon\alpha}
\nonumber\\
  && \qquad\;\;\,
   - \Sigma_\alpha{}^\epsilon{}_{\!\epsilon}
   - \Sigma^\epsilon{}_{\!\alpha\epsilon} 
   +5\Sigma^\epsilon{}_{\!\epsilon\alpha}
  \,\Big] + \Upsilon_\alpha{}^\epsilon{}_{\!\epsilon}
\quad\Leftrightarrow\quad
  \Upsilon_\alpha{}^\epsilon{}_{\!\epsilon}=0\;.
\label{ap4v3} 
\eaq 
Similarly, $\Upsilon^\epsilon{}_{\!\beta\epsilon}
=\Upsilon^\epsilon{}_{\!\epsilon\gamma}=0$. Furthermore from (\ref{ap4v1}) 
it also follows 
\beq\label{ap4v4} 
  \Sigma_{[\alpha\beta\gamma]}
  =
  \Sigma_{[\alpha\beta\gamma]}+\Upsilon_{[\alpha\beta\gamma]}
  \quad\Leftrightarrow\quad
  \Upsilon_{[\alpha\beta\gamma]} = 0\;.
\eeq 
This tensor rest $\Upsilon_{\a\b\c}$ has no trace part nor an antisymmetric 
part. Since the original tensor $\Sigma_{\a\b\c}$ has $4^3 = 64$ degrees 
of freedom (in the real case) and the four vectors take away only 16 
degrees, the tensor rest still has 48 degrees of freedom. An explicit 
example for such a tensor rest is given by 
\beq\label{ap4v5} 
  \Upsilon_{\a\b\c}= \nabla^\ast_{\!\a}B_{\b\c} 
                    -\frac{1}{3}(C_\b g_{\a\c}-C_\c g_{\a\b})\;, 
\eeq 
where $B_{\b\c}$ is an antisymmetric tensor given by the total differential 
of a vector field $B_\m$ by $B_{\b\c} = \partial_\b B_\c - \partial_\c 
B_\b$, and $C_\b$ is another vector field satisfying an ``inhomogeneous 
Maxwell equation'' $C^\m = \nabla^\ast_{\!\n} B^{\m\n}$. It is easy to 
verify the conditions (\ref{ap4v3}) and (\ref{ap4v4}). From this example we 
may conclude that the tensor rest possibly contains interesting structures. 
But these are of no relevance to our theory yet since $\Upsilon_{\a\b\c}$ 
does not couple to spinorial matter, see (\ref{dirac2}), but appears only 
in the gravitational Lagrangian part ${\cal L}_G$, see (\ref{eq4v}).

           \chapter{Computations in Chapter 4}

\section{Non-quantized Dirac field} 

Throuout this chapter we employ the Dirac representation defined by 
\beq\label{algebra1} 
  \c^0 = \left(\begin{array}{cc}\One & 0 \\ 0 & -\One \end{array}\right)\;, 
\quad 
  \c^a = \left(\begin{array}{cc} 0 &\s^a \\ -\s^a & 0 \end{array}\right)\;, 
\;\; a = 1,2,3\;,\quad 
  \c^5 = \left(\begin{array}{cc} 0 & \One \\ \One & 0 \end{array}\right)\;, 
\eeq 
where the Pauli matrices can be found in (\ref{paulimatrices}). For the 
special background Dirac spinor given in (\ref{univ2}) we immediately 
obtain the results 
\eqlabel{algebra2} 
\begin{eqroman} 
\baq 
  (j^d) &=& (\PSI_{bg}\c^d\psi_{bg}) 
       \;=\; n\cdot\left(\begin{array}{c}1\\0\\0\\0\end{array}\right)\;; 
\label{algebra2a}\\ 
  (j^{{\sssty5}d}) &=& (\PSI_{bg}\c^5\c^d\psi_{bg}) 
       \;=\; n\cdot\left(\begin{array}{c}0\\0\\0\\-1\end{array}\right)\;. 
\eaq%
\end{eqroman}%
Note that the vector current $j^d$ is time-like, whereas the axial current 
$j^{5d}$ is a space-like vector. Using these expressions, we calculate 
\eqlabel{algebra3} 
\begin{eqroman} 
\baq 
  -\c^0\cdot\frac{3}{8}\planck^2\hbar c \cdot j^{{\sssty5}d}\c^5\c_d &=& 
  -\c^0\cdot\frac{3}{8}\planck^2\hbar cn \cdot (-1)\c^5(-\c^3) 
\nonumber\\ 
 &=& 
   \frac{3}{8}\planck^2\hbar cn 
  \left(\begin{array}{cccc}%
        1&0&0&0\\0&-1&0&0\\0&0&1&0\\0&0&0&-1%
        \end{array}\right) 
\nonumber\\ 
   &=:& \frac{3}{8}\planck^2\hbar cn\cdot \t_{ECT} \;;
\label{algebra3a}\\ 
  -\c^0\cdot\frac{3}{8}\planck^2\hbar c(j^d\c_d+j^{5d}\c^5\c_d) &=& 
  \frac{3}{8}\planck^2\hbar cn 
  \left(\begin{array}{cccc}%
        0&0&0&0\\0&-2&0&0\\0&0&0&0\\0&0&0&-2%
        \end{array}\right) 
\nonumber\\ 
  &=& \frac{3}{8}\planck^2\hbar cn\cdot \t_I \;. 
\label{algebra3b} 
\eaq%
\end{eqroman}%
The minus sign in front of $\c^3$ comes from $\c_a = \eta_{ab}\c^b$. 
With the result (\ref{algebra3a}) it is straightforward to obtain 
(\ref{univ4}) from (\ref{univ1}). 

We do not compute the energy eigenvalues of the Einstein--Cartan theory 
given in (\ref{univ6}) but only those of the new contact interaction given 
in (\ref{univ8}) and (\ref{univ9}), since the computations are very similar. 
For this purpose, we replace in (\ref{univ4}) the interacting matrix 
$\t_{ECT}$ by $\t_I$ in (\ref{algebra3b}) and put in the special plane wave 
spinor $\psi_\uparrow$ (\ref{univ5}). Since we will only consider test 
particles at rest, we assume in $\psi_\uparrow$ that the 3-momentum 
$(p_1,p_2,p_3)$ vanishes. We then obtain 
\beq\label{algebra4}  
  cp^0\psi_\uparrow = 
  (mc^2\c^0+\frac{3}{8}\planck^2\hbar cn\t_I)\psi_\uparrow \;. 
\eeq 
Denoting the total energy by $E_{\uparrow\uparrow} = cp^0$, this eigenvalue 
equation is equivalent to 
\beq\label{algebra5} 
  E_{\uparrow\uparrow}\left(\begin{array}{c}1\\0\\0\\0\end{array}\right) = 
  \left(\begin{array}{cccc}%
        mc^2 & 0 & 0 & 0 \\ 
        0 & mc^2-\frac{3}{4}\planck^2\hbar cn & 0 & 0 \\ 
        0 & 0 & -mc^2 & 0 \\ 
        0 & 0 & 0 & -mc^2-\frac{3}{4}\planck^2\hbar cn \end{array}\right) 
  \left(\begin{array}{c}1\\0\\0\\0\end{array}\right) \;. 
\eeq 
Thus we obtain $E_{\uparrow\uparrow} = mc^2$ (\ref{univ8a}). Similarly, 
$E_{\uparrow\downarrow} = mc^2 - 2\cdot\frac{3}{8}\planck^2\hbar cn$ 
(\ref{univ8b}). 

To obtain (\ref{univ9}), we use the anti-matter plane waves given by 
\beq\label{algebra6} 
  \psi'_\uparrow = N \left(\begin{array}{c}0\\0\\0\\1\end{array}\right) 
                   \exp(+\frac{i}{\hbar}p_\m x^\m) 
  \qquad\mbox{and}\qquad                      
  \psi'_\downarrow = N \left(\begin{array}{c}0\\0\\1\\0\end{array}\right) 
                     \exp(+\frac{i}{\hbar}p_\m x^\m) \;, 
\eeq 
where the up-arrow and down-arrow indicate the spin directions of the 
plane waves cf.\ \cite{itz80}. Using these plane waves instead of 
$ \psi_\uparrow $ and $ \psi_\downarrow $ we immediately obtain instead of 
(\ref{algebra4}) 
\beq\label{algebra4b}  
  -cp^0\psi'_\uparrow = 
  (-mc^2\c^0+\frac{3}{8}\planck^2\hbar cn\t_I)\psi'_\uparrow \;, 
\eeq 
from which we obtain (\ref{univ9a}). The interaction energy for opposite 
spins (\ref{univ9b}) can be derived in a similar fashion.

\section{Interaction Hamiltonian} 

In this section we shall derive the effective Lagrangian (\ref{quant1}) in 
order to obtain the interaction Lagrangian ${\cal L}_I$ in (\ref{quant2}). 
The original Lagrangian is given by (\ref{many1}), which contains two 
spinor fields $\psi$ and $\chi$ and their adjoint fields. 

First, we introduce the abbreviations 
\beq\label{algebra7} 
  j^a = \PSI\c^a\psi\;,\quad k^a = \CHI\c^a\chi\;,\quad 
  j^{{\sssty5}a} = \PSI\c^5\c^a\psi\;,\quad 
  k^{{\sssty5}a} = \CHI\c^5\c^a\chi\;. 
\eeq 
Next, we write down the connection components obtained by considering its 
field equation. From (\ref{many7}), we obtain 
\beq\label{algebra8} 
  \Con{a}\m{b} = \Chr{a}\m{b} + 
  \big(Q^a e_{b\m}+U_b e^a{}_\m-\frac{1}{12}\eta^a{}_{\m bd}V^d\big) 
  + S_\m\d^a{}_b 
\eeq 
with 
\beq\label{algebra9} 
  -Q^a=U^a=-\frac{i\planck^2}{4}(j^a+k^a)\;,\quad 
  V^d =3\planck^2(j^{{\sssty5}d}+k^{{\sssty5}d}) \;. 
\eeq 
We now insert this result step by step into the three Lagrangian parts 
${\cal L}_m$, ${\cal L}_G$ and ${\cal L}_Y$ given by (\ref{many1}), see 
also (\ref{lagrangiana}). The matter Lagrangian ${\cal L}_m$ consists of the 
two individual Lagrangians for the spinors $\psi$ and $\chi$, denoted  
henceforth by ${\cal L}_{m\psi}$ and ${\cal L}_{m\chi}$, respectively. We 
have 
\baq 
  {\cal L}_{m\psi} &=& 
  g\cdot\hbar c\big[i\PSI\c^\m(\partial_\m-\frac{1}{4}\con{a}\m{b}\s^{ba} 
  +\frac{\vare_1}{4}\Con{a}\m{a})\psi - \frac{mc}{\hbar}\PSI\psi \big] 
\nonumber\\ 
  &=& 
  g\cdot\hbar c\big[i\PSI\c^\m(\nabla^\ast_{\!\m}+\vare_1S_\m)\psi - 
  \frac{mc}{\hbar}\PSI\psi \big] 
\nonumber\\ 
  & & 
  \mbox{}+g\cdot\hbar ci\big[-\frac{1}{4} 
  (Q_a\eta_{cb}+U_b\eta_{ca}-\frac{1}{12}\e_{acbd}V^d) 
  \PSI\c^c\s^{ba}\psi\big] \;. 
\label{algebra10} 
\eaq 
To evaluate the last term in (\ref{algebra10}), we use the following  
algebraic identity among $\c$-matrices (cf.\ \cite{heh71}, 
see also \cite{diplom}) 
\beq\label{algebra10b} 
  \c^c\c^b\c^a = \c^c\eta^{ba} + \c^a\eta^{bc} - \eta^{ac}\c^b 
                +i\c^5\c_d\e^{cbad} \;. 
\eeq                                                   
Remembering that $\s^{ba} = \frac{1}{2}(\c^b\c^a-\c^a\c^b)$, we get for the 
last term in (\ref{algebra10}) 
\baq 
  & & 
  -g\cdot\frac{\hbar ci}{4}\big[ 
  (Q_a\eta_{cb}-Q_b\eta_{ca}-\frac{1}{12}\e_{acbd}V^d) 
  (j^a\eta^{bc}-\eta^{ac}j^b+ij^{\sssty5}{}_d\e^{cbad}) \big] 
\nonumber\\ 
  &=& 
  -g\cdot\frac{\hbar ci}{4}\big[ 6Q_aj^a+\frac{i}{2}V_dj^{{\sssty5}d} \big] 
\nonumber\\ 
  &=& 
  -g\cdot\frac{\hbar ci}{4}\big[ 
  6\cdot\frac{i\planck^2}{4}(j_a+k_a)j^a 
  + \frac{i}{2}\cdot 3\planck^2(j^{\sssty5}{}_d+k^{\sssty5}{}_d) 
  j^{{\sssty5}d} \big] 
\nonumber\\ 
  &=& 
   +g\cdot\frac{3}{8}\frac{\planck^4}{k} 
  (k_aj^a+k^{\sssty5}{}_dj^{{\sssty5}d}) \;, 
\label{algebra11} 
\eaq 
where we have used $\planck^2 = \hbar ck$. Similar results hold for the 
other matter Lagrangian ${\cal L}_{m\chi}$. Adding both partial Lagrangians 
we obtain 
\beq\label{algebra12} 
\eqalign{ 
  {\cal L}_m &= 
  g\cdot\hbar c\big[i\PSI\c^\m(\nabla^\ast_{\!\m}+\vare_1S_\m)\psi - 
  \frac{mc}{\hbar}\PSI\psi \big] + 
  g\cdot\hbar c\big[i\CHI\c^\m(\nabla^\ast_{\!\m}+\vare_2S_\m)\chi - 
  \frac{nc}{\hbar}\CHI\chi \big] 
\\ 
  &\phantom{=} 
  + g\cdot\frac{3}{8}\frac{\planck^4}{k} 
  (k_aj^a+k^{\sssty5}{}_dj^{{\sssty5}d}) 
  + g\cdot\frac{3}{8}\frac{\planck^4}{k} 
  (j_ak^a+j^{\sssty5}{}_dk^{{\sssty5}d}) \;. 
} 
\eeq 
Next, we insert (\ref{algebra8}) into ${\cal L}_G$ and obtain 
\beq\label{algebra13} 
  {\cal L}_G = -\frac{g}{2k}R = -\frac{g}{2k}\big[ R^\ast + 
  3\nabla^\ast_{\!\m}(Q^\m-U^\m)+3(U_\m U^\m+Q_\m Q^\m+4U_\m Q^\m) + 
  \frac{1}{24}V_\m V^\m \big] \;. 
\eeq 
This result can be found in \cite{diplom}. It can be verified in a 
cumbersome computation that the derivative term $3\nabla^\ast_{\!\m} 
(Q^\m-U^\m)$ vanish, if we take into 
account the Dirac equations for $\psi$ and $\chi$. Then, by inserting 
(\ref{algebra9}) into (\ref{algebra13}), we get 
\beq\label{algebra14} 
  {\cal L}_G = -\frac{g}{2k}R^\ast 
  - g\cdot\frac{3}{8}\frac{\planck^4}{k} 
  (k_\m j^\m+k^{\sssty5}{}_\m j^{{\sssty5}\m}) \;.
\eeq 
For the third Lagrangian part ${\cal L}_Y$, we have simply 
\beq\label{algebra15} 
  {\cal L}_Y = g\frac{l^2}{4k}S_{\m\n}S^{\m\n} \;. 
\eeq 
Finally, adding the results (\ref{algebra12}), (\ref{algebra14}) and 
(\ref{algebra15}) yields exactly (\ref{quant2}).

\section{Spinorial algebra}

The plane wave spinors used in chapter 4 are taken from \cite{nac90} but 
with a little change in the normalization constant. We first define 
\beq\label{algebra16} 
  \chi_{\frac{1}{2}} = \left(\begin{array}{c}1\\0\end{array}\right)\;,\quad 
  \chi_{-\frac{1}{2}} = \left(\begin{array}{c}0\\1\end{array}\right)\;,\quad 
  \vare = \left(\begin{array}{cc}0&1\\-1&0\end{array}\right) \;. 
\eeq 
Then, 
\eqlabel{algebra17} 
\begin{eqroman} 
\baq 
  u_r(p)&=& \sqrt{\frac{p^0+m}{2mV}}\left(\begin{array}{c} 
            \chi_r \\ \frac{\vec{\s}\vec{p}}{p^0+m}\chi_r 
            \end{array}\right)\;;
\label{algebra17a}\\ 
  v_r(p)&=&-\sqrt{\frac{p^0+m}{2mV}}\left(\begin{array}{c} 
            \frac{\vec{\s}\vec{p}}{p^0+m}\vare\chi_r \\ \vare\chi_r 
            \end{array}\right)\;;
\label{algebra17b}\\ 
  s_r(p)&=& \sqrt{\frac{p^0+n}{2nV}}\left(\begin{array}{c} 
            \chi_r \\ \frac{\vec{\s}\vec{p}}{p^0+n}\chi_r 
            \end{array}\right)\;;
\label{algebra17c}\\ 
  w_r(p)&=&-\sqrt{\frac{p^0+n}{2nV}}\left(\begin{array}{c} 
            \frac{\vec{\s}\vec{p}}{p^0+n}\vare\chi_r \\ \vare\chi_r 
            \end{array}\right)\;.
\label{algebra17d} 
\eaq%
\end{eqroman}%
The use of the $2\times 2$-matrix $\vare$ results in the correct spin 
directions of the anti-matter waves: All spinors above have exactly the 
same spin direction determined by $p$ and $r$, see \cite{nac90}. 

We shall now derive the formula (\ref{quanta}). Using the Dirac 
representation (\ref{algebra1}) of the $\c$-matrices and (\ref{algebra17a}) 
we have 
\beq\label{algebra18} 
  \ubar_r(p)\c^5\c^\m u_r(p) = \frac{p^0+m}{2mV} 
  (-\chi_r^\dagger\frac{\vec{\s}\vec{p}}{p^0+m},\chi_r^\dagger)
  \c^\m \left(\begin{array}{c}%
  \chi_r^\dagger \\ \frac{\vec{\s}\vec{p}}{p^0+m}\chi_r^\dagger 
  \end{array}\right) \;. 
\eeq 
Note that there are no differences between the anholonomic $\c$-matrices 
$\c^a$ and the holonomic ones $\c^\m$, since we are working in flat 
Minkowski spacetime. We first derive the 0-th component 
\baq 
  \ubar_r(p)\c^5\c^0 u_r(p) &=& 
  \frac{p^0+m}{2mV} 
  (-\chi_r^\dagger\frac{\vec{\s}\vec{p}}{p^0+m},-\chi_r^\dagger) 
  \left(\begin{array}{c}%
  \chi_r \\ \frac{\vec{\s}\vec{p}}{p^0+m}\chi_r 
  \end{array}\right)  
\nonumber\\ 
  &=& 
  -\frac{1}{mV} \chi_r^\dagger 
  \left(\begin{array}{cc} p^3 & p^1-ip^2 \\ p^1+ip^2 & -p^3 \end{array} 
  \right) \chi_r 
\nonumber\\ 
  &=& 
  \mp\frac{p^3}{mV}\quad\cdots\quad r=\pm\frac{1}{2} \;. 
\label{algebra19} 
\eaq 
Next, we consider the other three components: 
\baq 
  \ubar_r(p)\c^5\vec{\c} u_r(p) &=& 
  \frac{p^0+m}{2mV} 
  (-\chi_r^\dagger\frac{\vec{\s}\vec{p}}{p^0+m},\chi_r^\dagger) 
  \left(\begin{array}{cc} 0 & \vec{\s} \\ -\vec{\s} & 0 \end{array}\right) 
  \left(\begin{array}{c}%
  \chi_r \\ \frac{\vec{\s}\vec{p}}{p^0+m}\chi_r 
  \end{array}\right)  
\nonumber\\ 
  &=& 
  -\chi_r^\dagger 
  \frac{(\vec{\s}\vec{p})\vec{\s}(\vec{\s}\vec{p})}{2mV(p^0+m)}\chi_r 
  -\frac{p^0+m}{2mV}\chi_r^\dagger\vec{\s}\chi_r \;. 
\label{algebra20} 
\eaq 
Using this identity we obtain for the first component 
\baq 
  (\vec{\s}\vec{p}) \s^1 (\vec{\s}\vec{p}) &=& 
  \left(\begin{array}{cc} p^3 & p^1-ip^2 \\ p^1+ip^2 & -p^3 \end{array} 
  \right)  
  \left(\begin{array}{cc} 0 & 1 \\ 1 & 0 \end{array}\right)  
  \left(\begin{array}{cc} p^3 & p^1-ip^2 \\ p^1+ip^2 & -p^3 \end{array} 
  \right)  
\nonumber\\ 
 &=& 
  \left(\begin{array}{cc} 2p^1 p^3 & -(p^3)^2+(p^1-ip^2)^2 \\ 
                          -(p^3)^2+(p^1-ip^2)^2 & -2p^1 p^3 \\ 
  \end{array}\right) 
  \quad\Rightarrow 
\nonumber\\ 
  \ubar_r(p)\c^5\c^1 u_r(p) &=& 
  \mp\frac{p^1 p^2}{mV(p^0+m)} \quad\cdots\quad r=\pm\frac{1}{2} \;. 
\label{algebra21} 
\eaq 
Similar considerations for the other two space-directions lead to the 
result (\ref{quanta}). 
                       
Note that $\PSI\c^5\c^\m\psi$ is always a space-like vector. Indeed, we 
obtain for the special expression (\ref{axialflow}) 
\beq\label{algebra22} 
\eqalign{ 
  p^{5\m}p^5_\m &= (p^3)^2-\frac{(p^3)^2\cdot\vec{p}^{\,2}}{(p^0+M)^2}-M^2 
                  -\frac{2M(p^3)^2}{(p^0+M)} 
\\ 
  &= (p^3)^2-(p^3)^2\frac{(p^0)^2-M^2}{(p^0+M)^2}-M^2 
                  -\frac{2M(p^3)^2}{(p^0+M)} 
\\ 
  &= -M^2 \;. 
} 
\eeq 

For the calculations of the interaction energies, the following identity 
has been be used (compare (\ref{algebra17})) 
\beq\label{algebra22b} 
  \ubar_r(p)\c^5 u_r(p) = 
  \frac{p^0+m}{2mV} 
  (\chi_r^\dagger,\chi_r^\dagger\frac{\vec{\s}\vec{p}}{p^0+m}) 
  \left(\begin{array}{cc}0&\One\\-\One&0\end{array}\right) 
  \left(\begin{array}{c}\chi_r \\ \frac{\vec{\s}\vec{p}}{p^0+m}\chi_r 
  \end{array}\right) = 0 \;. 
\eeq

\section{Expectation values}\label{expectation} 

We shall now calculate the expectation values of $\widehat{H}_I$ and 
$\widehat{H}_{ECT}$. First we consider the Einstein--Cartan case 
with identical particles (\ref{quant16d}): Introducing the abbreviation 
\beq\label{algebra23} 
  \widetilde{dp} := \frac{d^3\!\!p}{(2\pi)^3}\frac{m}{p^0} 
\eeq 
we obtain 
\baq 
  \langle 1o|\widehat{H}_{ECT}|1o\rangle &=& 
  -\frac{3}{8}\frac{\planck^4}{k}\int_V 
  \langle 0|a_{r'q'}a_{rq}:\PSI\c^5\c^\m\psi\,\PSI\c^5\c_\m\psi: 
  \adag_{rq}\adag_{r'q'}|0\rangle 
\nonumber\\ 
  &=& 
  -\frac{3}{8}\frac{\planck^4}{k}\int_V \int\!\!\int\!\!\int\!\!\int 
  \widetilde{dp_1}\cdots\widetilde{dp_4} 
\nonumber\\   
  & & 
  \langle 0|a_{r'q'}a_{rq}\, :
  \adag_{r_1p_1}\ubar_{r_1}(p_1)e^{ip_1x}\c^5\c^\m 
  a_{r_2p_2}u_{r_2}(p_2)e^{-ip_2x}\cdot 
\nonumber\\ 
  & & \phantom{\langle o|} 
  \adag_{r_3p_3}\ubar_{r_3}(p_3)e^{ip_3x}\c^5\c_\m 
  a_{r_4p_4}u_{r_4}(p_4)e^{-ip_4x} : \,  
  \adag_{rq}\adag_{r'q'}|0\rangle 
\nonumber\\ 
  &=& 
  +\frac{3}{8}\frac{\planck^4}{k}\int_V \int\!\!\int\!\!\int\!\!\int 
  \widetilde{dp_1}\cdots\widetilde{dp_4} 
\nonumber\\ 
  & & 
  \langle 0|a_{r'q'}a_{rq}\adag_{r_1p_1}\adag_{r_3p_3}
  a_{r_2p_2}a_{r_4p_4}\adag_{rq}\adag_{r'q'}|0\rangle\cdot 
\nonumber\\ 
  & &
  \ubar_{r_1}(p_1)\c^5\c^\m u_{r_2}(p_2) 
  \ubar_{r_3}(p_3)\c^5\c_\m u_{r_4}(p_4)\cdot 
  e^{i(p_1-p_2+p_3-p_4)x} 
\nonumber\\ 
  &=& 
   \frac{3}{8}\frac{\planck^4}{k}V 
  \big[ \phantom{-} 
        2\ubar_r(q)\c^5\c^\m u_{r'}(q')\ubar_{r'}(q')\c^5\c_\m u_r(q)  
\nonumber\\ 
  & & \phantom{\frac{3}{8}\frac{\planck^4}{k}V\big[ } 
       -2\ubar_r(q)\c^5\c^\m u_r(q)\ubar_{r'}(q')\c^5\c_\m u_{r'}(q')\big]\;. 
\label{algebra24} 
\eaq 
To obtain the last line, we have used the anti-commutation relations of the 
creation- and annihilation-operators (\ref{quant10}). We use the following 
special case of the Fierz transformation (see e.g.\ \cite{itz80}), 
\beq\label{fierz1} 
  \PSI\c^5\c^\m\chi\,\CHI\c^5\c_\m\psi = 
  -\PSI\psi\,\CHI\chi-\frac{1}{2}\PSI\c^\m\psi\,\CHI\c_\m\chi
  +\PSI\c^5\psi\,\CHI\c^5\chi 
  -\frac{1}{2}\PSI\c^5\c^\m\psi\,\CHI\c^5\c_\m\chi \;, 
\eeq 
and the identity (\ref{algebra22b}) to reexpress the last line 
(\ref{algebra24}) in the following way: 
\baq 
  \langle 1o|\widehat{H}_{ECT}|1o\rangle &=& 
  \frac{3}{4}\frac{\planck^4}{k}V 
  \big[ \ubar_r(q)\c^5\c^\m u_r(q)\ubar_{r'}(q')\c^5\c_\m u_{r'}(q') 
       -\ubar_r(q)u_r(q)\ubar_{r'}(q')u_{r'}(q') 
\nonumber\\ 
  & & \phantom{\frac{3}{4}\frac{\planck^4}{k}V \big[}        
       -\frac{1}{2}\ubar_r(q)\c^\m u_r(q)\ubar_{r'}(q')\c_\m u_{r'}(q') 
\nonumber\\ 
  & & \phantom{\frac{3}{4}\frac{\planck^4}{k}V \big[}        
       -\frac{1}{2}\ubar_r(q)\c^5\c^\m u_r(q) 
                   \ubar_{r'}(q')\c^5\c_\m u_{r'}(q') \big] 
\nonumber\\ 
  &=& 
  -\frac{3}{8}\frac{\planck^4}{k}V 
  \big[ 2\ubar_r(q)u_r(q)\ubar_{r'}(q')u_{r'}(q') 
       +\ubar_r(q)\c^\m u_r(q)\ubar_{r'}(q')\c_\m u_{r'}(q') 
\nonumber\\ 
  & & \phantom{-\frac{3}{8}\frac{\planck^4}{k}V\big[} 
       +3\ubar_r(q)\c^5\c^\m u_r(q)\ubar_{r'}(q')\c^5\c_\m u_{r'}(q')\big] 
\nonumber\\ 
  &=& 
  -\frac{3}{8}\frac{\planck^4}{k}V 
  \big[ 2V^{-2}+V^{-2}\frac{q^\m}{m}\frac{q'_\m}{m} 
       +3V^{-2}\frac{q^{5\m}}{m}\frac{q'^5_\m}{m}\s(r,r') \big] \;, 
\label{algebra25}  
\eaq 
where (\ref{quant8}) to (\ref{quanta}) were used in the last line. This 
expression is equal to (\ref{quant16d}). 

The computations of all other expectation values in (\ref{quant15}) and 
(\ref{quant16}) are similar to this example and are therefore left out. 

To obtain the expressions (\ref{quant17}) and (\ref{quant18}) we need the 
following relation valid for small momenta $|\vec{q}|,|\vec{q'}| \ll m$: 
\baq 
  q^\m q'_\m &=& 
  \sqrt{m^2+\vec{q}^{\,2}}\sqrt{m^2+\vec{q'}^{\,2}} - \vec{q}\vec{q'} 
\nonumber\\ 
  &\approx& 
  (m+\frac{\vec{q}^{\,2}}{2m})(m+\frac{\vec{q'}^{\,2}}{2m}) - \vec{q}\vec{q'} 
\nonumber\\ 
  &\approx& 
  m^2 + \frac{1}{2}(\vec{q}-\vec{q'})^2 \;. 
\label{algebra26} 
\eaq

\end{appendix}

\newpage
\thispagestyle{empty} 
\begin{center} 
{\Large Danksagung} 
\end{center}

Ich m\"ochte mich herzlich bei Herrn Prof.\ M.\ Kretzschmar f\"ur seine 
Unterst\"utzung der vorliegenden Arbeit bedanken. Ferner danke ich ihm f\"ur 
hilfreiche Anregungen und Ratschl\"age zu dieser Arbeit.\\[1mm] 

F\"ur Diskussionen zu dieser Arbeit bin ich Herrn Prof.\ N.\ A.\ 
Papadopoulos dankbar.\\[1mm]
 
Ich danke Herrn Dr.\ R.\ H\"au\"sling sehr f\"ur sein sorgf\"altiges und 
kritisches Korrekturlesen.\\[1mm]
 
Der Landesgraduiertenf\"orderung des Rheinland-Pfalz danke ich f\"ur die 
finanzielle Unterst\"utzung. 


\begin{thebibliography}{Einstein33}

\bibitem[Ash 91]{ash91} A.\ Ashtekar, {\it Lectures on Non--Perturbative
  Canonical Gravity}, World Scientific, Singapore, 1991

\bibitem[Bat 84]{bat84} N.\ A.\ Batakis, {\it The Gravitoweak Connection},
  Phys.\ Lett.\ {\bf 148B} 51 (1984)

\bibitem[Bat 90]{bat90} N.\ A.\ Batakis and A.\ A.\ Kehagias, {\it 
  Electroweak gauge boson masses from geometry}, Class.\ Quant.\ Grav.\ 
  {\bf 7} L63 (1990) 

\bibitem[Bau 81]{bau81} H.\ Baum, {\it Spin--Strukturen und 
  Dirac--Operatoren \"uber pseudeoriemannschen Mannigfaltigkeiten}, 
  Teubner--Verlag, Leipzig 1981

\bibitem[Ben 87]{ben87} I.\ M.\ Benn and R.\ W.\ Tucker, {\it An 
  Introduction to Spinors and Geometry with Applications in Physics},
  Adam Hilger, Bristol, 1987

\bibitem[Ber 91]{ber91} N.\ Berline, E.\ Getzler, and M.\ Vergne, {\it 
  Heat Kernels and Dirac Operators}, Springer, Berlin, 1991 

\bibitem[Bil 55]{bil55} B.\ A.\ Bilby, R.\ Bullough and E.\ Smith, {\it 
  Continuous distributions of dislocations, a new application of the
  methods of non--Riemannian geometry}, Proc.\ Roy.\ Soc.\ London
  {\bf A231} 263 (1955)

\bibitem[Bis 64]{bis64} R.\ L.\ Bishop and R.\ J.\ Crittenden, {\it
  Geometry of Manifolds}, Academic Press, New York, 1964 


\bibitem[Ble 81]{ble81} D.\ Bleecker, {\it Gauge Theory and Variational
  Principles}, Addison--Wesley, Massachusetts, 1981

\bibitem[Bon 54]{bon54} W.\ B.\ Bonnor, {\it The Equation of Motions in the
  Nonsymmetric Unified Field Theory}, Proc.\ Roy.\ Soc.\ London {\bf 226A}
  366 (1954)

\bibitem[Bor 76a]{bor76a} K.\ Borchsenius, {\it An Extension of the 
  Nonsymmetric Unified Field Theory}, Gen.\ Rel.\ Grav.\ {\bf 7} 527; {\it 
  Covariant Extensions and the Nonsymmetric Unified Field}, Gen.\ Rel.\
  Grav.\ {\bf 7} 709 (1976)

\bibitem[Bor 76b]{bor76b} K.\ Borchsenius, {\it Unified theory of gravitation,
  electromagnetism, and the Yang--Mills field}, Phys.\ Rev.\ {\bf 13}
  2707 (1976)

\bibitem[Bos 93]{bos93} M.\ Bosselmann, {\it Allgemeine Relativit\"atstheorie 
  als Eichtheorie vom Yang--Mills-Typ}, Diploma thesis, Mainz, 1993 

\bibitem[Bos 94]{bos94} M.\ Bosselmann and N.\ A.\ Papadopoulos, {\it 
  TG--Equivariance of Connections and Gauge Transformations}, preprint 
  Mainz, 1994


\bibitem[Buc 86]{buc86} I.\ K.\ Buchbinder, {\it Renormalization of Quantum
  Field Theory in Curved Space--Time and Renormalization Group Equations},
  Fortschr.\ Phys.\ {\bf 9} 605 (1986)

\bibitem[Cal 53]{cal53} J.\ Callaway, {\it The Equations of Motion in 
  Einstein's New Unified Field Theory}, Phys.\ Rev.\ {\bf 92} 1567 (1953)


\bibitem[Car 22]{car22} E.\ Cartan, {\it Sur une g\'en\'eralisation
  de la   notion de courbure de Riemann et les espaces \`a torsion}, Comptes
  Rendus   Acad.\ Sci.\ {\bf 174} 593 (1922);
  English Translation by G.\ D.\ Kerlick in {\it Cosmology and 
  Gravitation: Spin, Torsion Rotation and Supergravity}, Eds.: P.\
  G.\ Bergmann and V.\ De Sabbata, Plenum Press, New York, 1980

\bibitem[Car 23-25]{car23} Cartan, E., {\it Sur les vari\'et\'es \`a 
  connexion affine et la th\'eorie de la relativiste\'e g\'en\'eralis\'ee 
  I, II}, Ann.\ Ec.\ Norm.\ Sup., {\bf 40}, 325 (1923); {\bf 41} 1 (1924),
  {\bf 42} 17 (1925);
  English Translation by A.\ Magnon, A.\ Ash\-tekar and A.\ Trautmann,
  {\it On Manifolds with Affine Connection and the Theory of General 
  Relativity}, Bibliopolis, Naples, 1985  

\bibitem[Chi 87]{chi87} J.\ S.\ R.\ Chisholm and R.\ S.\ Farwell, {\it 
  Electroweak spin gauge theories and the frame field}, J.\ Phys.\ {\bf A20}
  6561 (1987)

\bibitem[Chi 89]{chi89} J.\ S.\ R.\ Chisholm and R.\ S.\ Farwell, {\it 
  Unified spin gauge theory of electroweak and gravitational interactions},
  J.\ Phys.\ {\bf A22} 1059 (1989)

\bibitem[Col 84]{col84} A.\ Coley, {\it A Note on the Geometric Unification 
  of Gravity and Electromagnetism}, Gen.\ Rel.\ Grav.\ {\bf 16} 459 (1984)

\bibitem[DeW 64]{dew64} B.\ DeWitt, {\it Dynamical Theory of Groups and 
  Fields} in: Les Houches 1963, Relativity, Groups, and Topology, Eds.:
  B.\ DeWitt, C.\ DeWitt, Gordon and Breach, New York, 1964

\bibitem[Din 92]{din92} M.\ Dine, R.\ G.\ Leigh, P.\ Huet, A.\ Linde and 
  D.\ Linde, {\it Towards the theory of the electroweak phase transitions}, 
  Phys.\ Rev.\ {\bf D46} 550 (1992) 

\bibitem[Dol 74]{dol74} L.\ Dolan and R.\ Jackiw, {\it Symmetry behavior 
  at finite temperature}, Phys.\ Rev.\ {\bf D9} 3320 (1974) 

\bibitem[D\"un 89]{due89} P.\ D\"unges, {\it Der Dirac--Operator \"uber 
  semiriemann'schen, raum- und zeitorientierbaren Spinmannigfaltigkeiten 
  von beliebiger Dimension und beliebigem Index}, Diploma thesis, Univ.\ 
  Mainz, 1989

\bibitem[Edd 21]{edd21} A.\ S.\ Eddington, {\it A generalisation of Weyl's 
  Theory of the Electromagnetic and Gravitational Fields}, Proc.\ Roy.\ Soc.\
  London, {\bf A99} 104 (1921)

\bibitem[Ein 55]{ein55} A.\ Einstein, {\it The Meaning of Relativity},  
  Appendix II of the 5.\ Edition, Princeton University Press, Princeton, 
  1955; A.\ Einstein and B.\ Kaufman, {\it A New Form of the General  
  Relativistic Field Equations}, Ann.\ Math.\ {\bf 62} 128 (1955)

\bibitem[Fer 81]{fer81} M.\ Ferraris and J.\ Kijowski, {\it General 
  Relativity is a Gauge--Type Theory}, Lett.\ Math.\ Phys.\ {\bf 5} 127
  (1981)

\bibitem[Fer 82]{fer82} M.\ Ferraris and J.\ Kijowski, {\it Unified 
  Geometric Theory of Electromagnetic and Gravitational Interactions},
  Gen.\ Rel.\ Grav.\ {\bf 14} 37 (1982)

\bibitem[Gam 93]{gam93} R.\ Gambini and J.\ Pullin, {\it Quantum 
  Einstein--Maxwell fields: A unified viewpoint from the loop 
  representation}, Phys.\ Rev.\ {\bf D47}, R5214 (1993) 

\bibitem[Ger 68]{ger68} R.\ Geroch, {\it Spinor Structure of Space-Times in 
  General Relativity. I}, J.\ Math.\ Phys.\ {\bf 9} 1739 (1968) 

\bibitem[Ger 70]{ger70} R.\ Geroch, {\it Spinor Structure of Space-Times in 
  General Relativity. II}, J.\ Math.\ Phys.\ {\bf 11} 343 (1970) 

\bibitem[Gre 72]{gre72} W.\ Greub, S.\ Halperin, and R.\ Vanstone, {\it 
  Connections, Curvature, and Cohomology}, vol I, II, III, Academic Press, 
  New York, 1972

\bibitem[Gvo 85]{gvo85} A.\ A.\ Gvozdev and P.\ I.\ Pronin, {\it Quantum 
  Statistics and Temperature Effects in the Theory of the Interactions of 
  Fermions with Torsion}, Moscow Univ.\ Phys.\ Bull.\ {\bf 40}, 30 (1985) 

\bibitem[Ham 89]{ham89} R.\ T.\ Hammond, {\it Einstein--Maxwell Theory 
  from Torsion}, Class.\ Quant.\ Grav.\ {\bf 6} L195 (1989)

\bibitem[Har 90]{har90} R.\ Harvey, {\it Spinors and Calibrations}, 
  Academic Press, Boston, 1990 

\bibitem[Haw 73]{haw73} S.\ W.\ Hawking and G.\ F.\ R.\ Ellis, {\it The  
  Large Scale Structure of Space-Time}, Cambridge University Press, 
  Cambridge, 1973 

\bibitem[Heh 65a]{heh65a} F.\ W.\ Hehl and E.\ Kr\"oner, {\it Zum
  Materialgesetz eines elastischen Mediums mit Momentspannungen}, Z.\
  Naturf.\ {\bf 20a} 336 (1965)

\bibitem[Heh 65b]{heh65b} F.\ W.\ Hehl and E.\ Kr\"oner, {\it \"Uber den
  Spin in der allgemeinen Rela\-tivit\"atstheorie, Eine notwendige 
  Erweiterung der Einsteinschen Feldgleichungen}, Z.\ Phys.\ {\bf 187}
  487 (1965)

\bibitem[Heh 66]{heh66} F.\ W.\ Hehl, {\it Der Spindrehimpuls in der   
  allgemeinen Relativit\"atstheorie}, Abh.\ Braunschweig.\ Wiss.\
  Ges.\ {\bf 18} 98 (1966)

\bibitem[Heh 71]{heh71} F.\ W.\ Hehl and B.\ K.\ Datta, {\it Nonlinear 
  Spinor Equation and Asymmetric Connection in General Relativity}, J.\ Math.\
  Phys.\ {\bf 12} 1334 (1971)

\bibitem[Heh 73]{heh73} F.\ W.\ Hehl and P.\ von der Heyde, {\it Spin and 
  the structure of space-time}, Ann.\ Inst.\ H.\ Poincar\'e, {\bf A19}  
  179 (1973) 

\bibitem[Heh 74]{heh74} F.\ W.\ Hehl, P.\ von der Heyde and G.\ D.\ 
  Kerlick, {\it General relativity with spin and torsion and its deviations 
  from Einstein's theory}, Phys.\ Rev.\ {\bf D10} 1066 (1974) 

\bibitem[Heh 76]{heh76} F.\ W.\ Hehl, P.\ von der Heyde, G.\ D.\ Kerlick 
  and J.\ M.\ Nester, {\it General Relativity with Spin and Torsion:
  Foundations and Prospects}, Rev.\ Mod.\ Phys.\ {\bf 48} 393 (1971)

\bibitem[Heh 91]{heh91} F.\ W.\ Hehl, J.\ Lemke, and E.\ W.\ Mielke, in: 
  {\it Geometry and Theoretical Physics}, ed.\ J.\ Debrus and A.\ C.\ 
  Hirshfeld, Springer, Berlin 1991 

\bibitem[Hla 57]{hla57} V.\ Hlavat\'y, {\it Geometry of Einstein's   
  Unified field Theory}, Nordhoff, Groningen 1975

\bibitem[Hor 94]{diplom} K.\ Horie, {\it Die Vereinheitlichung von 
  Gravitation und Elektromagnetismus durch die Torsion der Raum--Zeit},
  Diploma thesis, Mainz, 1994

\bibitem[Hor 95]{paper} K.\ Horie, {\it Geometric Interpretation of 
  Electromagnetism in a Gravitational Theory with Space--Time Torsion},
  submitted for publication

\bibitem[Inf 33]{inf33} L.\ Infeld and B.\ L.\ van der Waerden, {\it Die 
  Wellengleichung des Elektrons in der allgemeinen Relativit\"atstheorie}, 
  Sitz.\ Preuss.\ Akad.\ Wiss., {\bf 9} 380 (1933)

\bibitem[Inf 50]{inf50} L.\ Infeld, {\it The New Einstein Theory and the 
  Equations of Motion}, Acta Phys.\ Pol.\ {\bf X} 284 (1950)

\bibitem[Itz 80]{itz80} C.\ Itzykson and J.--B.\ Zuber, {\it Quantum Field
  Theory}, Mc Graw--Hill, New York, 1980

\bibitem[Jac 65]{jac65} J.\ D.\ Jackson, {\it Classical Electrodynamics},
  John Wiley \& Sons, New York, 1965

\bibitem[Jak 85]{jak85} A.\ Jakubiec and J.\ Kijowski, {\it On Interaction
  of the Unified Maxwell--Einstein Field with Spinorial Matter}, Lett.\
  Math.\ Phys.\ {\bf 9} 1 (1985)

\bibitem[Kae 76]{kae76} F.\ A.\ Kaempffer, {\it On a Possible Unification
  of Gravitational and Weak Interactions}, Gen.\ Rel.\ Grav.\ {\bf 7}
  327 (1976)

\bibitem[Kan 88]{kan88} S.\ Kanno, {\it Interaction between Fermions and 
  Gravitational Field in the Very Early Universe}, Prog.\ Theor.\ Phys.\
  {\bf 79} 1365 (1988)

\bibitem[Kat 92]{kat92} M.\ O.\ Katanaev and I.\ V.\ Volovich, {\it Theory 
  of Defects in Solids and Three-Dimensional Gravity}, Ann.\ Phys.\ {\bf 
  216} 1 (1992) 

\bibitem[Ker 75]{ker75} G.\ D.\ Kerlick, {\it Cosmology and Particle
  Pair Production via Gravitational Spin--Spin Interaction in the 
  Einstein--Cartan--Sciama--Kibble Theory of Gravity}, Phys.\ Rev.
  {\bf D12} 3004 (1975)

\bibitem[Kib 61]{kib61} T.\ W.\ B.\ Kibble, {\it Lorentz Invariance and 
  the Gravitational Field}, J.\ Math.\ Phys.\ {\bf 2} 212 (1961)

\bibitem[Kin 95]{kin95} S.\ King, {\it Dynamical Electroweak Symmetry 
  Breaking} Rep.\ Prog.\ Phys.\ {\bf 58} 263 (1995) 

\bibitem[Kir 72]{kir72} D.\ A.\ Kirzhnits and A.\ D.\ Linde, {\it 
  Macroscopic Consequences of the Weinberg Model}, Phys.\ Lett.\ {\bf 42B} 
  471 (1972) 

\bibitem[Kob 63]{kob63} S.\ Kobayashi and K.\ Nomizu, {\it Foundations of 
  Differential Geometry} Vol.\ I, John Whiley \& Sons, New York 1963

\bibitem[Kon 52]{kon52} K.\ Kondo, {\it On the geometrical and physical
  foundations of the theory of yielding} in Proc.\ of the 2nd Japan
  National Congress for Appl.\ Mech.\ 41 (1952)

\bibitem[Kop 72]{kop72} W.\ Kopczy\'nski, {\it A Non-singular Universe 
  with Torsion}, Phys.\ Lett.\ {\bf A39} 219 (1972) 

\bibitem[Kr\"o 64]{kro64} E.\ Kr\"oner, {\it Plastizit\"at und Versetzungen}
  in: A.\ Sommerfeld, Vorlesungen \"uber Theoretische Physik, 5. Edition,
  2, Chap.\ 9 (Akad.\ Verlagsges., Leipzig, 1964)

\bibitem[Kr\"o 81]{kro81} E.\ Kr\"oner, {\it Continuum theory of defects} in:
  Physics of defects, Les Houches 1980, Session XXXV (North--Holland, 
  Amsterdam, 1981)

\bibitem[Kuc 76]{kuc76} B.\ Kuchowicz, {\it Some Cosmological Models with 
  Spin and Torsion, I}, Astrophys.\ Space Sci.\ {\bf 39} 157 (1976) 

\bibitem[Kuc 78]{kuc78} B.\ Kuchowicz, {\it Friedmann-like Cosmological 
  Models without Singularity}, Gen.\ Rel.\ Grav.\ {\bf 9} 511 (1978)

\bibitem[Kun 79]{kun79} G.\ Kunstatter and J.\ W.\ Moffat, {\it 
  Conservation Laws in a Generalized Theory of Gravitation}, Phys.\ Rev.\
  {\bf D19} 1084 (1979)

\bibitem[Kur 52]{kur52} B.\ Kur{s}uno\u{g}lu, {\it Gravitation and 
  Electrodynamics}, Phys.\ Rev.\ {\bf 88} 1369 (1952)

\bibitem[Kur 74]{kur74} B.\ Kur{s}uno\u{g}lu, {\it Gravitation and 
  Magnetic Charge}, Phys.\ Rev.\ {\bf D9} 2723 (1974)

\bibitem[Kus 85]{kus85} A.\ N.\ Kushnirenko, {\it Unified Theory of Weak,
  Srong, Electromagnetic, and Gravitational Interaction}, Sov.\ Phys.\ J.\
  {\bf 28} 21 (1985)

\bibitem[Lal 92]{lal92} Z.\ Lalak, {\it Electroweak phase transition in NJL 
  models of symmetry breaking}, Phys.\ Lett.\ {\bf B278} 284 (1992) 

\bibitem[Law 89]{law89} H.\ B.\ Lawson and M.-L.\ Michelsohn, {\it
  Spin Geometry}, Princeton Univ.\ Press, Princeton, 1989

\bibitem[Lop 95]{lop95} J.\ Lopez and D.\ V.\ Nanopoulos, {\it A new 
  cosmological constant model}, preprint CERN-TH/95-6, 1995 

\bibitem[Mack 82]{mac82} G.\ Mack, {\it Allgemeine Relativit\"atstheorie},
  Lecture notes 1982, DESY T-82-03, 1982

\bibitem[McK 79]{mck79} R.\ J.\ McKellar, {\it Asymmetric connection 
  treatment of the Einstein--Maxwell field equations}, Phys.\ Rev.\
  {\bf D20} 356 (1979)

\bibitem[McC 92]{mcc92} J.\ T.\ McCrea, {\it Irreducible decompositions of 
  non-metricity, torsion, curvature and Bianchi identities in metric-%
  affine spacetime}, Class.\ Quant.\ Grav.\ {\bf 9} 553 (1992)

\bibitem[Mag 87]{mag87} A.\ M.\ R.\ Magnon, {\it Unification of Gravity and
  Electromagnetism in Dimension 4: Some Peculiar Aspects}, Nuovo Cim.\ {\bf 
  100B} 717 (1987)

\bibitem[Mis 73]{mtw} C.\ W.\ Misner, K.\ S.\ Thorne and J.\ A.\ Wheeler,
  {\it Gravitation}, W.\ H.\ Freeman and Company, 1973

\bibitem[Mof 75]{mof75} J.\ W.\ Moffat and D.\ H.\ Boal, {\it Solutions of 
  the Nonsymmetric Unified Field Theory}, Phys.\ Rev.\ {\bf D11} 1375 (1975)

\bibitem[Mof 77]{mof77} J.\ W.\ Moffat, {\it Space--Time Structure in a
  Generalization of Gravitation Theory}, Phys.\ Rev.\ {\bf D15} 3520 (1977)

\bibitem[Mof 79]{mof79} J.\ W.\ Moffat, {\it New Theory of Gravitation},
  Phys.\ Rev.\ {\bf D79} 3554 (1979)

\bibitem[Nac 90]{nac90} O.\ Nachtmann, {\it Elementary Particle Physics,
  Concepts and Phenomena}, Springer-Verlag, Berlin, 1990

\bibitem[Nak 90]{nak90} M.\ Nakahara, {\it Geometry, Topology and 
  Physics}, Adam Hilger, Bristol, 1990

\bibitem[Nam 61]{nam61} Y.\ Nambu and G.\ Jona-Lasinio, {\it Dynamical 
  Model of Elementary Particles Based on an Analogy with Superconductivity 
  I}, Phys.\ Rev.\ {\bf 122} 345 (1961) 

\bibitem[Nov 73]{nov73} M.\ Novello, {\it Weak and Electromagnetic Forces
  as a Consequence of the Self-Interaction of the $\c$ Field}, Phys.\ Rev.\
  {\bf D8} 2398 (1973)

\bibitem[Nov 85]{nov85} M.\ Novello, L.\ M.\ C.\ S.\ Rodrigues, {\it A 
  Unified Model for Gravity and Electroweak Interactions}, Lett.\ Nuovo 
  Cim.\ {\bf 43} 292 (1985)

\bibitem[Nov 92]{nov92} M.\ Novello and E.\ Elbaz, {\it Electrodynamics, 
  Gravity and the Corresponding Short-Range Fermi Forces}, Fortschr.\ 
  Phys.\ {\bf 40} 651 (1992) 

\bibitem[Nur 83]{nur83} I.\ S.\ Nurgaliev and W.\ N.\ Ponomariev, {\it The 
  Earliest Evolutionary Stages of the Universe and Space-Time Torsion}, 
  Phys.\ Lett.\ {\bf 130B} 378 (1983) 

\bibitem[Ora 65]{ora65} L.\ O'Raifertaigh, {\it Local Invariance and 
  Internal Symmetry}, Phys.\ Rev.\ {\bf 139B} 1052 (1965); 
  S.\ Coleman and J.\ Mandula, {\it All Possible Symmetries of the S 
  Matrix}, Phys.\ Rev.\ {\bf 159} 1251 (1967)

\bibitem[Pau 58]{pau58} W.\ Pauli, {\it Theory of Relativity}, 
  pages 224 to 227, Pergamon Press, London, 1958

\bibitem[Ren 90]{ren90} P.\ Renton, {\it Electroweak Interactions},
  Cambridge Univ.\ Press, Cambridge, 1990

\bibitem[Rom 69]{rom69} P.\ Roman, {\it Introduction to Quantum Field 
  Theory }, John Wiley \& Sons, New York, 1969

\bibitem[Ros 76]{ros76} D.\ K.\ Ross, {\it The Relationship of Weyl 
  Geometry to Quantum Electrodynamics}, Nuovo Cim.\ {\bf 33B} 449 (1976)

\bibitem[Rum 79]{rum79} H.\ Rumpf, {\it Creation of Dirac Particles in 
  General Relativity with Torsion and Electromagnetism I, II, III}, 
  Gen.\ Rel.\ Grav.\ {\bf 10} 509, 525, 647 (1979) 

\bibitem[Sch 54]{sch54} E.\ Schr\"odinger, {\it Space--Time 
  Structure}, Press Syndicate of the Univ.\ of Cambridge, Cambridge, 1954

\bibitem[Sci 62]{sci62} D.\ W.\ Sciama, {\it On the Analogy between Charge
  and Spin in General Relativity} in: Recent Developments in General
  Relativity, Pergamon Press, London, 1962

\bibitem[Sto 85]{sto85} W.\ Stoeger, {\it The Physics of Detecting Torsion
  and Placing Limits on Its Effects}, Gen.\ Rel.\ Grav.\ {\bf 17} 981
  (1985)
                                                     
\bibitem[Str 64]{str64} R.\ F.\ Streater and A.\ S.\ Whitman, {\it PCT, 
  Spin and Statistics, and All That}, Benjamin, New York, 1964 

\bibitem[Ton 55]{ton55} M.--A.\ Tonnelat, {\it La th\'eorie du champ 
  unifi\'e d'Einstein et quelques-uns des ses d\'eveloppements}, Les Grands 
  Probl\'emes des Sciences IV, Gauthier--Villars, Paris, 1955

\bibitem[Tra 71,72]{tra71} A.\ Trautman, {\it On the Einstein--Cartan 
  Equations I--IV}, Bull.\ Acad.\ Pol.\ Sci., Ser.\ Sci.\ Math.\ Astron.\ 
  Phys.\ {\bf 20} 185, 503, 895 (1971); {\bf 21} 345 (1972)

\bibitem[Tra 73]{tra73} A.\ Trautman, {\it Spin and Torsion may avert 
  Gravitatianal Singularities}, Nature (Phys.\ Sci.) {\bf 242} 7 (1973) 

\bibitem[Tro 87]{tro87} R.\ Trostel, {\it Higgs Potential and Spinor 
  Connection within Weinberg-Salam Model}, Prog.\ Theor.\ Phys.\ {\bf 78} 
  640 (1987) 

\bibitem[Wey 22]{wey22} H.\ Weyl, {\it Space Time Matter}, Dover Publ., 
  London, 1922

\bibitem[Yil 89]{yil89} A.\ Yildiz, M.\ K.\ Hinders, B.\ A.\ Rhodes and 
  G.\ V.\ H.\ Sandri, {\it Asymmetric Einstein Field Unification of Gravity 
  with Electroweak and Strong Forces}, Nuovo Cim.\ {\bf 102A} 1419 (1989) 

\bibitem[Zel 70]{zel70} Y.\ B.\ Zel'dovich, {\it Particle production in 
  cosmology}, JETP Lett.\ {\bf 12} 307 (1970) 

\bibitem[Zha 92]{zha92} C.-m.\ Zhang, G.-c.\ Yang, F.-p.\ Chen and X.-j.\ Wu,
  {\it Is There Evidence for Torsion?}, Gen.\ Rel.\ Grav.\ {\bf 24}
  359 (1992)

\end{thebibliography}
\end{document}